\documentclass[]{aa}

\usepackage{graphicx}
\usepackage[varg]{txfonts}
\usepackage[urlcolor=blue, colorlinks=true, citecolor=blue, linkcolor=blue]{hyperref}
\usepackage{natbib}
\bibpunct{(}{)}{;}{a}{}{,} 
\usepackage{cleveref}
\usepackage[super]{nth}
\usepackage{upgreek}
\usepackage{placeins}

\begin{document} 

\newcommand{\orcid}[1]{\href{https://orcid.org/#1}{\includegraphics[width=8pt]{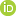}}}

   \title{The fundamental metallicity relation from SDSS ${ \left(z \sim 0 \right)}$ to VIPERS ${ \left(z \sim 0.7 \right)}$}

  \subtitle{ Data selection or evolution}


   \author{$^{\protect\orcid{0000-0003-1189-2617}}$F.~Pistis\inst{\ref{ncbj}}
          \and $^{\protect\orcid{0000-0003-3358-0665}}$A.~Pollo\inst{\ref{ncbj}, \ref{ju}}
          \and $^{\protect\orcid{0000-0002-2282-5850}}$M.~Scodeggio\inst{\ref{inaf_milan}}
          \and $^{\protect\orcid{0000-0002-9068-6215}}$M.~Figueira\inst{\ref{ncbj}, \ref{copernicus_torun}}
          \and $^{\protect\orcid{0000-0002-3818-8315}}$A.~Durkalec\inst{\ref{ncbj}}
          \and $^{\protect\orcid{0000-0003-3080-9778}}$K.~Ma\l{}ek\inst{\ref{ncbj}, \ref{marseille}}
          \and $^{\protect\orcid{0000-0001-6958-0304}}$A.~Iovino\inst{\ref{inaf_brera}}
          \and $^{\protect\orcid{0000-0003-0898-2216}}$D.~Vergani\inst{\ref{inaf_bologna}}
          \and $^{\protect\orcid{0000-0003-2342-7501}}$S.~Salim\inst{\ref{indiana}}
          }

   \institute{National Centre for Nuclear Research, ul. Pasteura 7, 02-093 Warsaw, Poland\\
              \email{francesco.pistis@ncbj.gov.pl}\label{ncbj}
              \and
              Astronomical Observatory of the Jagiellonian University, Orla 171, 30-001 Cracow, Poland\label{ju}
              \and
              INAF - Istituto di Astrofisica Spaziale e Fisica Cosmica Milano, via Bassini 15, 20133, Milano, Italy\label{inaf_milan}
              \and
              Institute of Astronomy, Faculty of Physics, Astronomy and Informatics, Nicolaus Copernicus University, Grudziądzka 5, 87-100 Toruń, Poland\label{copernicus_torun}
              \and
              Aix Marseille Univ. CNRS, CNES, LAM, Marseille, France\label{marseille}
              \and
              INAF - Osservatorio Astronomico di Brera, Via Brera 28, 20122 Milano, via E. Bianchi 46, 23807 Merate, Italy\label{inaf_brera}
              \and
              INAF - Osservatorio di Astrofisica e Scienza dello Spazio di Bologna, Via Piero Gobetti 93/3, I-40129 Bologna, Italy\label{inaf_bologna}
              \and
              Department of Astronomy, Indiana University, Bloomington, Indiana 47405, USA\label{indiana}
              }

\date{Received 14 October 2021 /
Accepted 25 May 2022}

 
  \abstract
  {Our knowledge of galaxy metallicity --- the result of the integrated star formation history and the evolution of the interstellar medium --- is important for constraining the description of galaxy evolution. As such, it has been widely studied in the local Universe, in particular, using data from the Sloan Digital Sky Survey (SDSS). The VIMOS Public Extragalactic Redshift Survey (VIPERS) allows us to extend such studies up to redshift of $z \sim 0.7$ and to quantify a possible evolution of the galaxy metallicity with high statistical precision.}
  {We focus on how to homogenize the comparison between galaxy samples having different characteristics. We check the projections of the fundamental metallicity relation (FMR) and the evolution of these projections between a sample selected at $z \sim 0$ (SDSS) and $z \sim 0.7$ (VIPERS). We check, in particular, whether and to what extent selection criteria can affect the results.}
  {We checked the influence of different biases introduced either by physical constraints (evolution of the luminosity function and differences in the fraction of blue galaxies) or data selection (the signal-to-noise ratio and quality of the spectra) on the FMR and its projections. To separate the differences occurring due to the physical evolution of galaxies with redshift from the false evolution mimed by these biases, we first analyzed the effects of these biases individually on the SDSS sample, and next, starting from the SDSS data, we built a VIPERS-equivalent $z \sim 0$ sample, replicating the main characteristics of VIPERS sample at $z \sim 0.7$ for a fair comparison.}
  {We found that the FMR projections are all sensitive to biases introduced by the selection on S/N and the quality flags of the emission line measurements in the spectra, especially the $\left[ \text{O{\,\sc{iii}}} \right]\lambda 4959$ line. The exception is the metallicity versus the sSFR plane which is insensitive to these biases. The evolution of the luminosity function introduces a bias only in the plane metallicity versus the star formation rate (SFR) while the fraction of blue galaxies has no impact on results.}
{With the applied methodology, the median metallicities estimated in each stellar mass-SFR bin of the samples at $z \sim 0$ and $z \sim 0.7$ agree within the uncertainties between SDSS and VIPERS samples ($ \Delta \log \left( \text{O}/\text{H} \right) \sim 0.6 \left< s_\text{VIPERS} \right> = 0.08$ dex, where $s_\text{VIPERS}$ stands for the metallicity standard deviation, without taking into account the biases). This difference can be reduced to $\sim 0.4 \left< s_\text{VIPERS} \right> = 0.06$ dex taking into account the biases, in particular the evolution of the luminosity function. We find a shift of the FMR projections towards lower metallicity which can be mimicked by a conservative selection on the S/N of emission lines. We also find either an overselection of high-metal galaxies at low stellar mass or an overestimation of the metallicity for the same sources at $z \sim 0.7$. Any bias taken into account in this study cannot mimic this overselection or overestimation at low redshift.
}


  \titlerunning{The Fundamental Metallicity Relation from SDSS ($z \sim 0$) to VIPERS ($z \sim 0.7$): selection or evolution?}
   \authorrunning{F. Pistis et al.}
   \maketitle
%

\section{Introduction}

The gas-phase galaxy metallicity, that is, the metal contents relative to hydrogen and helium, is an important tool in exploring the galaxy evolution. As the result of the integrated star formation history (SFH) and evolution of the interstellar medium (ISM), it gives insights into key galactic processes --- inflows, outflows, star formation, and quenching. Metallicity measurements at different epochs can be used to probe the early enrichment processes of both galaxies and the intergalactic medium (IGM) and help to constrain galaxy formation models  \citep{maiolino2019re}.

Metallicity is also tightly related with other galaxy properties, in particular with stellar mass ($\text{M}_\star$) --- one of the main galaxy features. Mechanisms such as inflows of metal-poor gas which ignite the star formation and dilute the metallicity of the ISM or the outflows of metal-rich gas which stop the star formation depend on this property. For this reason, it is crucial to assess observationally the relation between $\text{M}_\star$ and metallicity \citep{lequeux1979chemical}. Historically, it has been difficult to estimate the $\text{M}_\star$ of galaxies because of the inaccuracy of the measurements and the B-band luminosity was used instead \citep{garnett1987composition, skillman1989oxygen, skillman1989abundances, zaritsky1994h}.

Nowadays, we can derive galaxy physical properties from observations with a very good level of precision --- mostly due to the improvement of stellar population models \citep{bruzual2003stellar, kauffmann2003stellar}. Studies of the relation between $\text{M}_\star$ and metallicity, also referred to as the mass metallicity relation \citep[MZR, see e.g.,][]{savaglio2005gemini, kewley2008metallicity} have been performed with a high precision, for example, with the Sloan Digital Sky Survey \citep[SDSS, see e.g.,][]{tremonti2004origin, mannucci2010fundamental, curti2020mass} where a positive correlation was found. Additionally, these authors proposed various shapes for the MZR relation and distinct methods for metallicity measurements from different emission lines.

The origin of the MZR relation is still discussed. Different models have been proposed in the literature: ejection of metal-rich and infall of metal-poor gas models \citep{tremonti2004origin, finlator2008origin, dave2010nature, dave2011galaxy, chisholm2018metal}, so-called ``downsizing'' models, where star formation efficiency depends on galaxy mass \citep{spitoni2020connection, lian2018modelling, lian2018mass}, and models that include a variation of the initial mass function (IMF) with galaxy mass \citep{koppen2007possible, de2018effects}.

The MZR depends both on redshift and various galaxy properties. At high redshift, it was shown that, while metallicity decreases with redshift for a given $\text{M}_\star$, the MZR always shows a positive correlation \citep{erb2006mass, maiolino2008amaze, mannucci2009lsd}. Its scatter depends on different galaxy properties. \cite{ellison2007clues} first showed an anti-correlation between metallicity and specific star formation rate (sSFR) and \cite{mannucci2010fundamental, mannucci2011metallicity} observed an anti-correlation with the star formation rate (SFR). They first proposed that local galaxies are distributed on a surface defined by the $\text{M}_\star$, the metallicity, and the SFR, often referred to as the fundamental metallicity relation (FMR) because no evolution of this plane was observed up to $z \sim 3$. The same relation was observed by different authors \citep{yates2012relation, andrews2013mass, salim2014critical, hirschauer2018metal}. The absence of evolution up to $z \sim 0.8$ was confirmed by \cite{cresci2012zcosmos} using zCOSMOS data and by \cite{huang2019mass} using data from the extended Baryon Oscillation Spectroscopic Survey (eBOSS) of SDSS.

More recently, \cite{cresci2019fundamental} analyzed different FMR shapes proposed in the literature \citep[e.g.,][]{lara2010plane, hunt2012starburst, lara20123d} finding that the differences shown by various authors are mainly due to the different method used to estimate metallicity and SFR indicators used in their work \citep[similar studies were done on the MZR and chemical abundance, e.g.,][]{kewley2008metallicity, lopez2012eliminating}. It is especially noticeable if the comparison is done using calibrations based on $T_\text{e}$ or other methods based on non-identical diagnostics (i.e., strong line using photoionization models) are used \citep{kewley2008metallicity, lopez2012eliminating}. All the authors listed above agree on the close connection between the three galaxy properties that define the FMR. In the work of \cite{cresci2019fundamental}, the authors analyzed the data used to claim the nonexistence of the FMR at redshift $z \sim 2.5$ \citep[e.g.,][]{steidel2014kbss, wuyts2014metal, sanders2015mosdef} showing that the relation itself holds up to $z \sim 2.5$.

The only study expanding the FMR towards non-star-forming galaxies was recently carried out by \cite{kumari2021fmr} using the new empirical calibration based on nebular lines derived specifically for galaxies not classified as star-forming (SF) \citep{kumari2019metal}. Even these kinds of galaxies are in agreement with models that include an enhancement of star-formation due to the infall of metal-poor gas and starvation, which prevent the infall from halting the dilution process of metals.



Most of the studies claim that MZR evolved with redshift. For example, \cite{savaglio2005gemini} observed a shift of the MZR in their study of $\sim 60$ galaxies at redshift $0.4 < z < 1.0$ from the Gemini Deep Deep Survey (GDDS) compared with local studies. Using models that account for time-dependent metal outflow or time-dependent IMF, \cite{lian2018modelling, lian2018mass} found a two-phase evolution with a transition point at $z \sim 1.5$: the MZR evolves parallel to itself with no evolution of the slope and starts flattening until today. The models can reproduce the data from the literature from $z \sim 3.5$ to $z \sim 0$.

A physical galaxy model proposed by \cite{lilly2013fmr} linked i) the redshift evolution of the sSFR relative to the growth of dark matter halos; ii) the redshift evolution of gas-phase metallicities; and iii) the $\text{M}_\star$ to dark matter halo mass ratio. This model reproduces successfully the local FMR \citep{mannucci2010fundamental} and the MZR \citep{tremonti2004origin, mannucci2010fundamental} assuming a regulator system. This system is defined by the timescale of gas depletion ($\epsilon^{-1}$) and the mass loading $\lambda$ of the wind outflow $\lambda \cdot$SFR. An evolution of the FMR would lead to an evolution of these parameters.

Many authors do not take into account the effects of different methods used to estimate the FMR variables, such as slightly non-identical calibrations for the metallicity \citep[e.g.,][]{andrews2013mass, hirschauer2018metal}, non-homogeneous IMF used for $\text{M}_\star$ \citep[e.g.,][]{savaglio2005gemini, maiolino2008amaze, huang2019mass}; or intrinsic characteristics of the surveys, such as the different fraction of blue galaxies observed. This is mainly due to the limitations of previous surveys, with low statistics that do not allow for the study of these effects. We aim to check how different sample selections and observational effects of surveys can affect the shape of the FMR with a particular focus on the impact on the results and their interpretation.

The paper is divided as follows. In Sect.~\ref{sect:data}, we described the data used for our study: the intermediate redshift VIPERS survey ($z \sim 0.7$) and low-redshift SDSS survey ($z \sim 0$). Moreover, in this section, we discuss the initial selection of samples of SF galaxies for the study. In Sect.~\ref{sect:feat}, we described how the galaxy properties ($\text{M}_\star$, SFR, and metallicity) are estimated. In Sect.~\ref{sect:bias_sdss}, we describe additional biases (caused by physical constraints or data selection) introduced artificially to the SDSS sample and we study their effects on the shape of the FMR projections. In Sect.~\ref{sect:comparison}, we show the comparison between the two samples. Finally, in Sect.~\ref{sect:discussion}, we discuss the results of the paper and in Sect.~\ref{sect:concl} we report the conclusions. The cosmological parameters adopted in this paper are: $H_0 = 70\ \text{km} \, \text{s}^{-1} \, \text{Mpc}^{-1}$; $\Omega_\text{m} = 0.3$; $\Omega_\Lambda = 0.7$; and we assume a \cite{chabrier2003galactic} IMF.

\section{Data description and selection}\label{sect:data}
\subsection{VIPERS sample}\label{sec:VIPERS}
\subsubsection{Characteristics of the sample}

In this work, we use a sample of galaxies from the VIMOS Public Extragalactic Redshift Survey \citep[VIPERS,][]{guzzo2013vipers, garilli2014vimos, scodeggio2018vimos}, which is a spectroscopic survey made with the VIMOS spectrograph \citep{le2003proc} whose main purpose was to measure the redshift of $10^5$ galaxies in the range $0.5 < z <1.2$. The area covered by VIPERS is about $25.5\ \text{deg}^2$ on the sky and only galaxies brighter than $i_{AB} = 22.5$ were observed --- a pre-selection in the $(u-g)$ and $(r-i)$ color-color plane was used to remove galaxies at lower redshifts; see \cite{guzzo2014vimos} for a more detailed description of this survey. The Canada-France-Hawaii Telescope Legacy Survey Wide \citep[CFHTLS-Wide:][]{mellier2008cfhtls} W1 and W4 equatorial fields compose our galaxy sample, at $\text{R.A} \simeq 2$ and $\simeq 22\ \text{hours}$, respectively. The VIPERS spectral resolution ($R \sim 250$) allows for the study of individual spectroscopic properties of galaxies with an observed wavelength coverage of $5500$--$9500\ \text{\AA}$. The data reduction pipeline and redshift quality system are described in \cite{garilli2014vimos}. The final data release provides reliable spectroscopic measurements and photometric properties for $86\,775$ galaxies \citep{scodeggio2018vimos}.

This catalog was then cross-matched with the one used in \cite{turner2021unsupervised}, then analyzed with the SED fitting with the Code Investigating GALaxy Emission \citep[CIGALE,][]{burgarella2005star, noll2009analysis, boquien2019cigale}. This catalog is based on the photometric catalog built by \cite{moutard2016vipers, moutard2016photometry}, which is not complete in the W4 field, and it contains physical properties ($\text{M}_\star$, SFR, and magnitudes).

\subsubsection{Data selection}\label{sec:data_selection_vipers}

To ensure a high level of accuracy, we selected galaxies with highly secure redshift  \citep[$3.0 \leq z_\text{flag} \leq 4.5$, i.e. $> 90 \%$ confidence level;][]{scodeggio2018vimos} reducing the sample to $33\,785$ galaxies; of these, $14\,989$ have available measurements of all emission lines we need to estimate the metallicity. This effectively limits the redshift sample to $z_\text{max} \sim 0.9$. The cross-matching process further reduces the sample to $10\,126$ galaxies.


The spectroscopic redshift, the emission line fluxes, and their reliability flags have been estimated with the software Easy Redshift \citep[EZ]{garilli2010ez, garilli2014vimos}. This software is based on a decision tree, namely, the sequence of the operations to obtain the measurement, which can be customized according to the data studied. Additionally, all spectra and final spectroscopic redshift values were human-verified.

Regarding the measurement of the emission lines, the software looks for sharp peaks in the spectrum as possible emission lines \citep{garilli2010ez}. First, the software builds a peak list with the position of all pixels showing a flux above a user-defined threshold using continuum as a part of the decision tree (the significance of the peaks is estimated by subtracting the local continuum and computing the ratio between the peak height and the noise-weighted local continuum). Then, a Gaussian is fitted at each position and only the peaks for which the width of the fitted Gaussian is within defined limits are kept: the minimum and maximum width depends on the spectrum resolution, which (in the case of VIPERS) is equal to $250$.

After this initial automatic line identification to measure the redshift, each spectrum is then processed for a detailed fit of the emission lines. At the position of each identified line, a Gaussian fit is performed on the continuum-subtracted spectrum to obtain the spectral quantities. 
The total flux is computed by integrating the Gaussian function as resulting from the fit, in a range of $\pm 3$ times the width of the Gaussian fitted. The EW is instead given by the ratio of the line flux over the continuum mean value. In this way, both flux and EW of $\text{H}\alpha$, $\text{H}\beta$, $\left[ \text{O{\,\sc{ii}}} \right]\lambda 3727$, $\left[ \text{O{\,\sc{iii}}} \right]\lambda 4959$, and $\left[ \text{O{\,\sc{iii}}} \right]\lambda 5007$ have been computed. The errors on the line flux take into account the error on the continuum (computed as the root mean square around the fitted value), the Poissonian error on line counts, and the fit residuals; while the error on the EW is computed using the error propagation function.


Similar to the redshift flags, one can select lines based on the quality (reliability) flags \citep[computed again with the software EZ,][]{garilli2010ez}.
The flags of emission lines are in the form xyzt, where x, y, and z have values 1 or 0 and t has values 2, 1, or 0 if it satisfies or not particular conditions.

The x-value corresponds to the distance between the expected position and the Gaussian peak. It must be within $7\ \text{\AA}$ ($\sim 1$ pixel).
The y-values corresponds to the FWHM of the line. It must be between $7$ and $22\ \text{\AA}$ (equivalent to 1 and 3 pixels, respectively, in the instrumental configuration).
The z-value corresponds to the Gaussian amplitude and the observed peak flux. It must differ by no more than $30\%$.
The t-value is a condition on line flux and equivalent width. A value of $2$ means a strict condition, $\text{EW} \geq 3.5$ sigmas or $\text{flux} \geq 8$ sigmas; a value of $1$ a loose condition, $\text{EW} \geq 3$ sigmas or $\text{flux} \geq 7$ sigmas; while a value of $0$ corresponds to lower signal-to-noise ratios.

For VIPERS data, the use of a minimum flag equal to 1110 is recommended for each emission line in order to obtain a clean sample. At this step, we did not select any quality flag and it is important to check which kind of bias this selection can introduce into the analysis (see Sect.~\ref{sec:quality_flag}).

For this sample, the fluxes are corrected for attenuation following \cite{cardelli1989relationship} and assuming $R_V = 3.1$. The lines are corrected for both host galaxy and Milky Way extinctions. The former is estimated with the attenuation in the V-band ($A_\text{V}$) provided by the fit of the spectral energy distribution (SED) via the code Hyperz \citep{bolzonella2000sed, bolzonella2010zcosmos}. The latter is estimated with the color excess $\text{E} \left( \text{B} - \text{V} \right)$ derived from sky maps \citep{schlegel1998maps}.  In addition, the $\text{H}\beta$ line is corrected for stellar absorption via \cite{hopkins2003star} formula:
\begin{equation}
    S = \frac{\text{EW} + \text{EW}_C}{\text{EW}} F
,\end{equation}
where $S$ is the stellar absorption corrected line flux, $\text{EW}$ is the equivalent width of the line, $\text{EW}_C$ is the correction for stellar absorption, and $F$ is the line flux already corrected via \cite{cardelli1989relationship} from attenuation. We adopted the commonly used in the literature value of $\text{EW}_C = 2\ \text{\AA}$ \citep{miller2002stellar, goto2003stellar}.

\subsubsection{Star-forming galaxies selection}
We then need to select only SF galaxies. Usually one separates SF galaxies from other types of galaxies, that is, low ionization nuclear emission regions (LINERs) and Active Galactic Nuclei (AGNs) such as Seyfert galaxies (characterized by high ionization emission lines) using the Baldwin-Phillips-Terlevich diagrams \citep[BPT,][]{BPT81}.
The redshift range and the wavelength coverage of VIPERS do not allow for the detection of the $\left[ \text{N{\,\sc{ii}}} \right]\lambda 6584$ and $\text{H}\alpha$ lines. For this reason, we used the classification method proposed by \cite{lamareille2010spectral}, known as the blue BPT diagram. This diagram is built using a sample based on SDSS which is already corrected for stellar absorption \citep{brinchmann2004dr2, tremonti2004origin}. For this reason, we need to correct the emission lines (as explained above) for the VIPERS sample.

To distinguish between different sub-samples we use the equivalent widths (EWs) of the oxygen lines  ($\left[ \text{O{\,\sc{ii}}} \right]\lambda 3727$ and $\left[ \text{O{\,\sc{iii}}} \right]\lambda 5007$) relative to H${\beta}$ line. We used:
\begin{equation}
    \log \left( \left[ \text{O{\,\sc{iii}}} \right]/\text{H}\beta \right) 
    = \frac {0.11}{\log \left( \left[ \text{O{\,\sc{ii}}} \right]/\text{H}\beta \right) - 0.92}
    +0.85
\end{equation}
as a boundary line between AGN and SF galaxies, and
\begin{equation}
    \log \left( \left[ \text{O{\,\sc{iii}}} \right]/\text{H}\beta \right) 
    = 0.95 \times \log \left( \left[ \text{O{\,\sc{ii}}} \right]/\text{H}\beta \right) - 0.4
\end{equation}
to separate LINERs from Seyfert 2 type AGNs. Our classification method is reliable for all galaxies with $z<1$ because the change in ISM conditions in SF galaxies with redshift is not as strong as for galaxies at $z>1$ \citep{kewley2013theoretical, kewley2013cosmic}.

\begin{figure*}
    \centering
    \includegraphics[width=8.5cm]{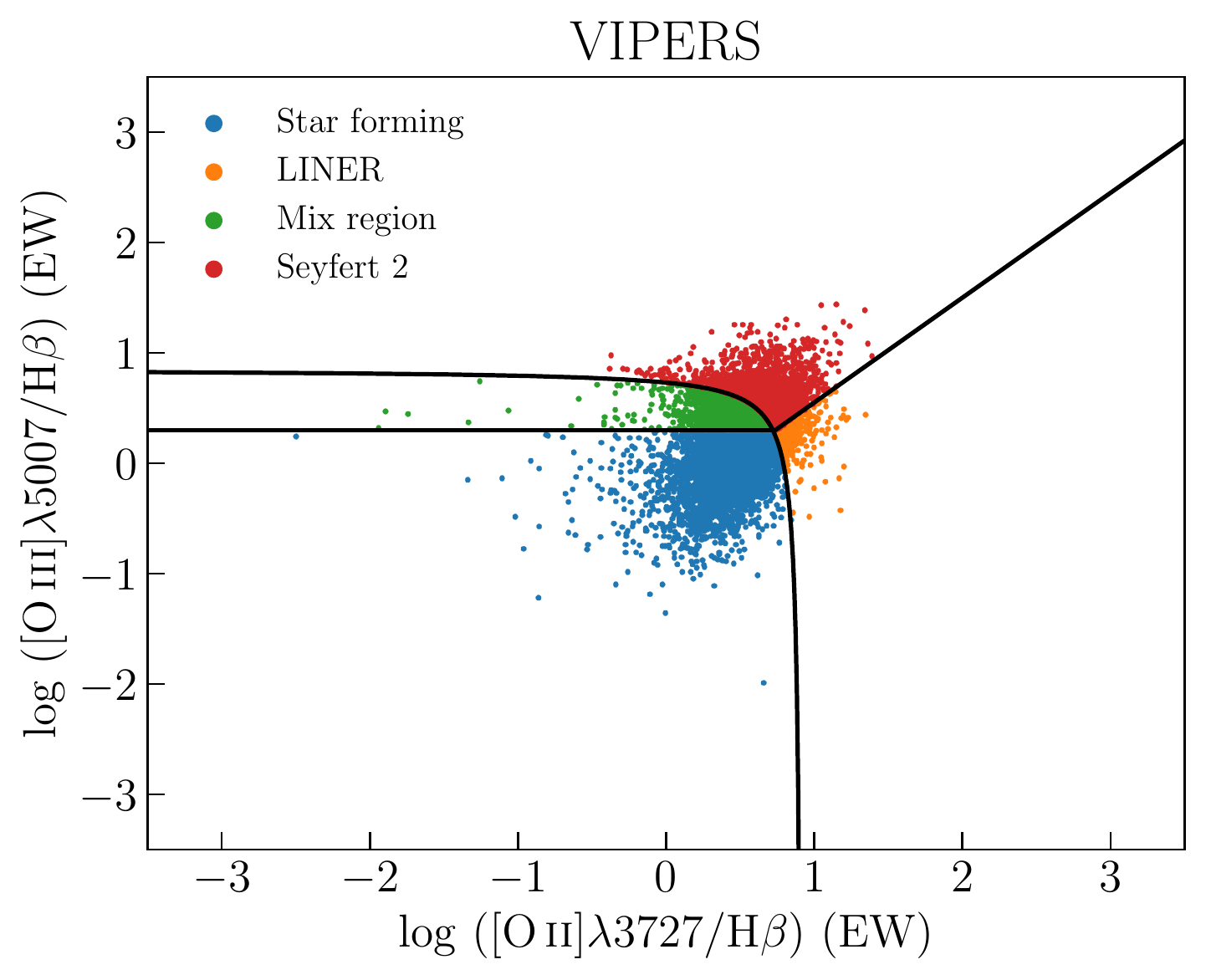}\includegraphics[width=8.5cm]{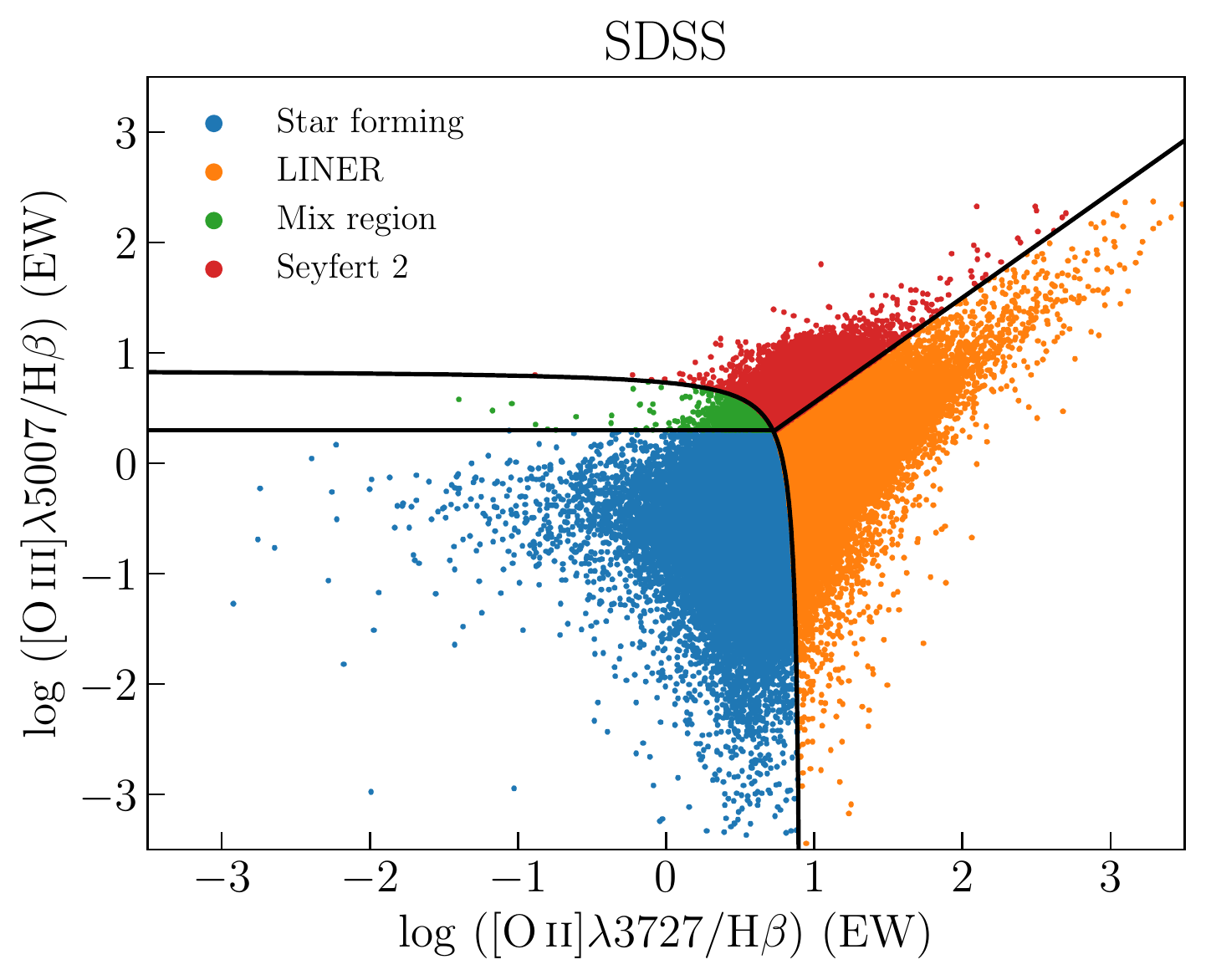}
    \caption{Diagnostic diagram using two line ratios: $\log \left( \left[ \text{O{\,\sc{iii}}} \right]\lambda 5007 / \text{H}\beta \right)$ vs $\log \left( \left[ \text{O{\,\sc{ii}}} \right]\lambda\lambda 3727 / \text{H}\beta \right)$ for VIPERS (left panel) and SDSS (right panel) samples. Solid lines show classification boundaries proposed by \cite{lamareille2010spectral}, blue points are the SF galaxies, orange points are the LINERs, green points are galaxies in the mix region, and red points are the Seyfert 2.}
    \label{fig:selection}
\end{figure*}

\cite{lamareille2010spectral} found a non-negligible number of Seyfert~2 galaxies falling into the SF region of the BPT diagram. They define the boundary of this ``mix'' region, located within SF galaxies, by the condition:
\begin{equation}
    \log \left( \left[ \text{O{\,\sc{iii}}} \right]/\text{H}\beta \right) 
    > 0.3.
\end{equation}
We exclude these galaxies from our study. They also define a composite region by the condition:
\begin{equation}\label{eq:bpt_composite}
    \begin{cases}
    \begin{split}
    \log \left( \left[ \text{O{\,\sc{iii}}} \right]/\text{H}\beta \right) &\leq - \left( \log  \left( \left[ \text{O{\,\sc{ii}}} \right]/\text{H}\beta \right) - 1.0 \right)^2 + \\
    &-0.1 \log  \left( \left[ \text{O{\,\sc{ii}}} \right]/\text{H}\beta \right) + 0.25
    \end{split}\\
    \log \left( \left[ \text{O{\,\sc{iii}}} \right]/\text{H}\beta \right) \geq \left( \log  \left( \left[ \text{O{\,\sc{ii}}} \right]/\text{H}\beta \right) - 0.2 \right)^2  - 0.6
    \end{cases}
,\end{equation}
where $\sim 85\%$ of galaxies are classified as SF and $\sim 15\%$ as LINERs \citep{lamareille2010spectral}. This contamination is the same for both samples (low redshift SDSS and intermediate redshift VIPERS) and it would just add a systematic to the metallicity. We check if the choice of the BPT diagram introduces a bias in the metallicity in Sect.~\ref{sec:bias_bpt} The results of our classification and selection boundaries are shown in Fig. \ref{fig:selection}, and we summarize in Table \ref{tab:subsamp_vipers} the numbers of galaxies in each sub-sample. The final sample of SF galaxies for VIPERS contains $5054$ galaxies ($\sim 34\%$ of the sample with a measure of all the emission lines we need).
\begin{table}
\caption{VIPERS sub-samples classification.}              
\label{tab:subsamp_vipers}      
\centering                                      
\begin{tabular}{c c c}          
\hline
\hline
 \noalign{\smallskip}
VIPERS sub-sample & Number of objects & Fraction ($\%$)\\
  \noalign{\smallskip}
\hline                                   
 \noalign{\smallskip}
    SF & $5054$ & $49.9$\\      
    Mix & $2455$ & $24.2$\\
    Seyfert & $2304$ & $22.8$\\
    LINER & $313$ & $3.1$\\
    Total & $10\,126$ & $100$ \\
     \noalign{\smallskip}
\hline          
\end{tabular}
\end{table}

\subsection{SDSS sample}
\subsubsection{Characteristics of the sample}

The best choice of the low-$z$ comparison sample is the SDSS sample. This sample was already used in many studies about MZR and FMR \citep[e.g.,][]{tremonti2004origin, mannucci2010fundamental, salim2014critical, curti2020mass}. This survey observed a large portion of the sky ($9380\ \text{deg}^2$) for the spectroscopic sample with a resolution $R \sim 1800$--$2200$ and wavelength coverage of $3800$--$9200\ \text{\AA}$ making it the most extended spectroscopic survey at low redshift.

We assemble this comparison sample by cross-matching two different samples.
A \textit{flux sample}: the MPA/JHU catalog\footnote{\url{https://wwwmpa.mpa-garching.mpg.de/SDSS/DR7/}} based on the SDSS DR7\footnote{\url{http://classic.sdss.org/dr7/}} \citep{abazajian2009sdss} composed of $927\,552$ galaxies with spectroscopic redshift and line fluxes \citep{kauffmann2003stellar, brinchmann2004physical, tremonti2004origin};
And a \textit{physical properties sample}: the A2.1 version of the GALEX-SDSS-WISE Legacy Catalog\footnote{\url{https://salims.pages.iu.edu/gswlc/}} \citep[GSWLC-2,][]{salim2016galex, salim2018dust} with $640\,659$ galaxies, based on SDSS DR10 \citep{anh2014sdss} with GALEX and WISE at $z < 0.3$. This catalog contains physical properties ($\text{M}_\star$, SFR, and magnitudes) obtained through SED fitting with CIGALE.
In the end, this cross-matched sample is composed of $601\,082$ galaxies.

\subsubsection{Data selection}
To avoid biases in the metallicity measurements, we select galaxies with a signal to noise ratio (S/N) equal to 15 for $\text{H}\alpha$, and equal to 3 for $\text{H}\beta$. Then, we corrected all emission lines for attenuation from the measured Balmer decrement, assuming the case B recombination ($\text{H}\alpha / \text{H}\beta = 2.87$) and adopting \cite{cardelli1989relationship} law assuming $R_V = 3.1$. Finally, we removed all galaxies with high extinction, that is, with values of $\text{E} \left( \text{B} - \text{V} \right)$ higher than $0.8$ \citep{curti2020mass}, reducing the sample to $259\,021$. The check on the comparison of the reliability of the emission lines between the two samples is left as a study of a possible bias in Sect.~\ref{sec:quality_flag}.

\subsubsection{SF galaxies selection}

We then proceeded with the selection of SF galaxies described in Sect. \ref{sec:VIPERS}. The BPT diagram is shown in the right panel of Fig. \ref{fig:selection} and the number of galaxies in each sub-sample is presented in Table \ref{tab:subsamp_gswlc}, showing the final sample of SF galaxies for SDSS contains $158\,416$ galaxies ($\sim 26\%$ of the initial cross-matched sample).
\begin{table}
\caption{SDSS sub-samples classification.}              
\label{tab:subsamp_gswlc}      
\centering                                      
\begin{tabular}{c c c}          
\hline                        
\hline
\noalign{\smallskip}
SDSS sub-sample & Number of objects & Fraction ($\%$)\\
 \noalign{\smallskip}
\hline                                   
\noalign{\smallskip}
    SF & $158\,416$ & $61.2$\\      
    Mix & $1578$ & $0.6$\\
    Seyfert & $9261$ & $3.6$\\
    LINER & $89\,766$ & $34.7$\\
    Total & $259\,021$ & $100$ \\
    \noalign{\smallskip}
\hline                                             
\end{tabular}
\end{table}


\section{Main galaxy physical properties}\label{sect:feat}

The studies of the FMR when using different samples need all three variables --- $\text{M}_\star$, SFR, and metallicity --- to be computed homogeneously. This allows us to avoid mistaking systematic differences for physical evolution in the relations.

\subsection{Stellar mass}

Since VIPERS and SDSS cover different wavelength ranges, we decided to use the estimation of the $\text{M}_\star$ made with the same tool with comparable parameters. The measurement for SDSS data is based on the integrated photometry to have the total $\text{M}_\star$ not suffering from the fiber effects.

The Bayesian analysis of the SEDs of both samples \citep{turner2021unsupervised} is based on simulated spectra generated with the SSP by \cite{bruzual2003stellar} based on a \cite{chabrier2003galactic} IMF. This SSP is then attenuated by dust assuming a specific dust attenuation law, the choice of which can strongly alter the derivation of $\text{M}_\star$es \citep[e.g., as shown in][]{lofaro2017dust, malek2018help, buat2019dust, hamed2021adonis_astarte}. The attenuation law used is a generalization of the \cite{calzetti2000dust} attenuation curve, modified to let its slope vary and to add a UV bump \citep{noll2009analysis}.

The templates are then combined with SFHs described by double exponentially declining events of star-formation, which produce an old and a young stellar population, for both samples and the second exponential models the burst. The difference between both samples is the length of the burst. For VIPERS short bursts ($10$--$1000 \text{ Myr}$) are allowed, while in the SDSS the bursts are longer ($100$--$5000 \text{ Myr}$). The old component in VIPERS has an additional, short e-folding time $\tau$ included in the models to allow us to reach a quiescent sSFR at $z \sim 1$.

\subsection{Star formation rate}

The SFR estimation of the SED fitting technique is more sensitive to the wavelength range covered by photometric data with a big influence from the infrared data compared to the estimation of $\text{M}_\star$ \citep{baes2020sed}. Since the VIPERS sample is missing observations at long wavelengths, we decided to measure the SFR from $\left[ \text{O{\,\sc{ii}}} \right]$ luminosity \citep[$\text{L}_{\left[ \text{O{\,\sc{ii}}} \right]}$,][]{kennicutt1998star} to obtain a comparable SFR between both samples and use all information inside the spectra. Among different SFR estimators for SF galaxies, the one from Figuera et al. (submitted) found good agreement (scatter but no bias) between $\left[ \text{O{\,\sc{ii}}} \right]$-based and SED-based SFR.
The $\left[ \text{O{\,\sc{ii}}} \right]$-based SFR is defined by:
\begin{equation}
    \text{SFR} \left( \text{M}_\sun \text{ yr}^{-1} \right) = \left( 1.4 \pm 0.4 \right) \times 10^{-41} \text{L}_{\left[ \text{O{\,\sc{ii}}} \right]} \left( \text{erg s}^{-1} \right)
,\end{equation}
where $\text{L}_{\left[ \text{OII} \right]} = 4 \pi \text{f}_{\left[ \text{OII} \right]} \text{d}_\text{L}^2$, $\text{f}_{\left[ \text{OII} \right]}$ is the flux of the emission line, and $\text{d}_\text{L}$ is the luminosity distance.

Because of the fiber system of the SDSS survey which measures the light with different spatial coverage, the SFR needs a correction to take into account the fiber aperture \citep{hopkins2003star}:
\begin{equation}
    \text{SFR}_\text{tot} \left( \text{M}_\sun \text{ yr}^{-1} \right) = \text{SFR} \left( \text{M}_\sun \text{ yr}^{-1} \right) \times 10^{-0.4 \left( u_\text{Petro} - u_\text{fiber} \right)}
,\end{equation}
where $u_\text{Petro}$ and $u_\text{fiber}$ are the modified form of the Petrosian magnitude \citep{petrosian1976surface} and the magnitude measured within the aperture of the spectroscopic fiber, respectively. The VIPERS sample does not need a spectroscopic fiber correction because the finite width of the slit is chosen to not overlap for different sources and it is expected to fully cover the galaxies on the CCD \citep{bottini2005mask, pezzotta2017structure, mohammad2018slit}. These values are then multiplied by the constant $0.63$ to pass from the IMF of \cite{salpeter1955imf} to the one of \cite{chabrier2003galactic} \citep{madau2014sfh}.

\subsection{Metallicity}

The goal of this sub-section is to check if different calibrations can introduce differences in the FMR and its projections, keeping the rest of the analysis independent of this choice. This is needed because there is no universal calibrator for the gas-phase metallicity measurement, so the MZR strongly depends on the method used to derive it. We use the oxygen abundance (defined as $\left[ 12 + \log \left( \text{O/H} \right) \right]$) to describe the metallicity.

Because direct comparisons are possible only between studies that used the same methods \citep{kewley2008metallicity}, we decided to check the impact of three different methods which use lines measured both in VIPERS and SDSS. Thanks to that we can make a comparable analysis at low and intermediate redshift ranges. All of them are based on the $R_{23}$ estimator \citep{pagel1979hii, mcgaugh1991hii, kewley2002hii} defined as:
\begin{equation}
    R_{23} = \frac{\left[ \text{O{\,\sc{ii}}} \right]\lambda 3727 + \left[ \text{O{\,\sc{iii}}} \right]\lambda\lambda 4959, 5007}{\text{H}\beta}
.\end{equation}
These methods are based on the following calibrations: i) \cite{tremonti2004origin} (thereafter T04); ii) \cite{pilyugin2001oxygen} (thereafter P01); and iii) \cite{zaritsky1994h} (thereafter Z94).

All these calibrations are valid for the upper branch of the double-valued $R_{23}$-abundance relation. Following \cite{nagao2006gas} we divided the upper-branch ($\left[ \text{O{\,\sc{iii}}} \right]\lambda 5007 / \left[ \text{O{\,\sc{ii}}} \right]\lambda 3727 < 2$) from the lower-branch. In this passage, we removed $\sim 1\%$ ($\sim 50$) of galaxies in the VIPERS sample and $\sim 1\%$ ($\sim 600$) in the SDSS sample. 

The T04 calibration estimates the metallicity from the theoretical model fits of strong emission-lines. The model fits are calculated by combining single stellar population (SSP) synthesis models from \cite{bruzual2003stellar} and CLOUDY photoionization models \citep{ferland1998cloudy}. The relation between metallicity and $R_{23}$ is given by:
\begin{equation}
    \left[ 12 + \log \left( \text{O/H} \right) \right]_\text{T04} = 9.185 - 0.313 x - 0.264 x^2 - 0.321 x^3
,\end{equation}
where $x \equiv \log R_{23}$. The precision of this calibration is around 0.09 dex and is valid for $12 + \log \left( \text{O/H} \right) \geq 8.4$.

The P01 is based on the comparison between direct measurement of the metallicity from the electron temperature $T_e$. The relation between metallicity and $R_{23}$ is given by:
\begin{equation}
    \left[ 12 + \log \left( \text{O/H} \right) \right]_\text{P01} = \frac{R_{23} + 54.2 + 59.45 P + 7.31 P^2}{6.07 + 6.71 P + 0.371 P^2 + 0.243 R_{23}}
,\end{equation}
where $P = R_3 / R_{23}$ and $R_3 = \left[ \text{O{\,\sc{iii}}} \right]\lambda\lambda 4959, 5007 / \text{H}\beta$. This calibration has a precision on the oxygen abundance determination of around $0.1$ dex. This calibration is valid for $12 + \nolinebreak \log \left( \text{O/H} \right) \geq \nolinebreak  8.2$.

The Z94 calibration is derived from the average of three previous calibrations made by: \cite{edmunds1984composition, dopita1986theoretical}; and \cite{mccall1985chemistry}. In this approach, the relation between metallicity and $R_{23}$ is given by:
\begin{equation}
\begin{split}
    \left[ 12 + \log \left( \text{O/H} \right) \right]_\text{Z94} &= 9.265 - 0.33 x - 0.202 x^2 +\\
    &- 0.207 x^3 - 0.333 x^4
\end{split}
,\end{equation}
where $x \equiv \log R_{23}$ and is valid for $12 + \log \left( \text{O/H} \right) \geq 8.4$.

After estimating the metallicity, we removed all galaxies outside the calibration range with $\left[ 12 + \log \left( \text{O/H} \right) \right]_\text{T04} < 8.4$, $\left[ 12 + \log \left( \text{O/H} \right) \right]_\text{P01} < 8.2$, and $\left[ 12 + \log \left( \text{O/H} \right) \right]_\text{Z94} < 8.4$. The VIPERS sample was reduced to $4431$ galaxies while no galaxies were removed from the SDSS sample.

\begin{figure*}
    \centering
    \includegraphics[width=8.5cm]{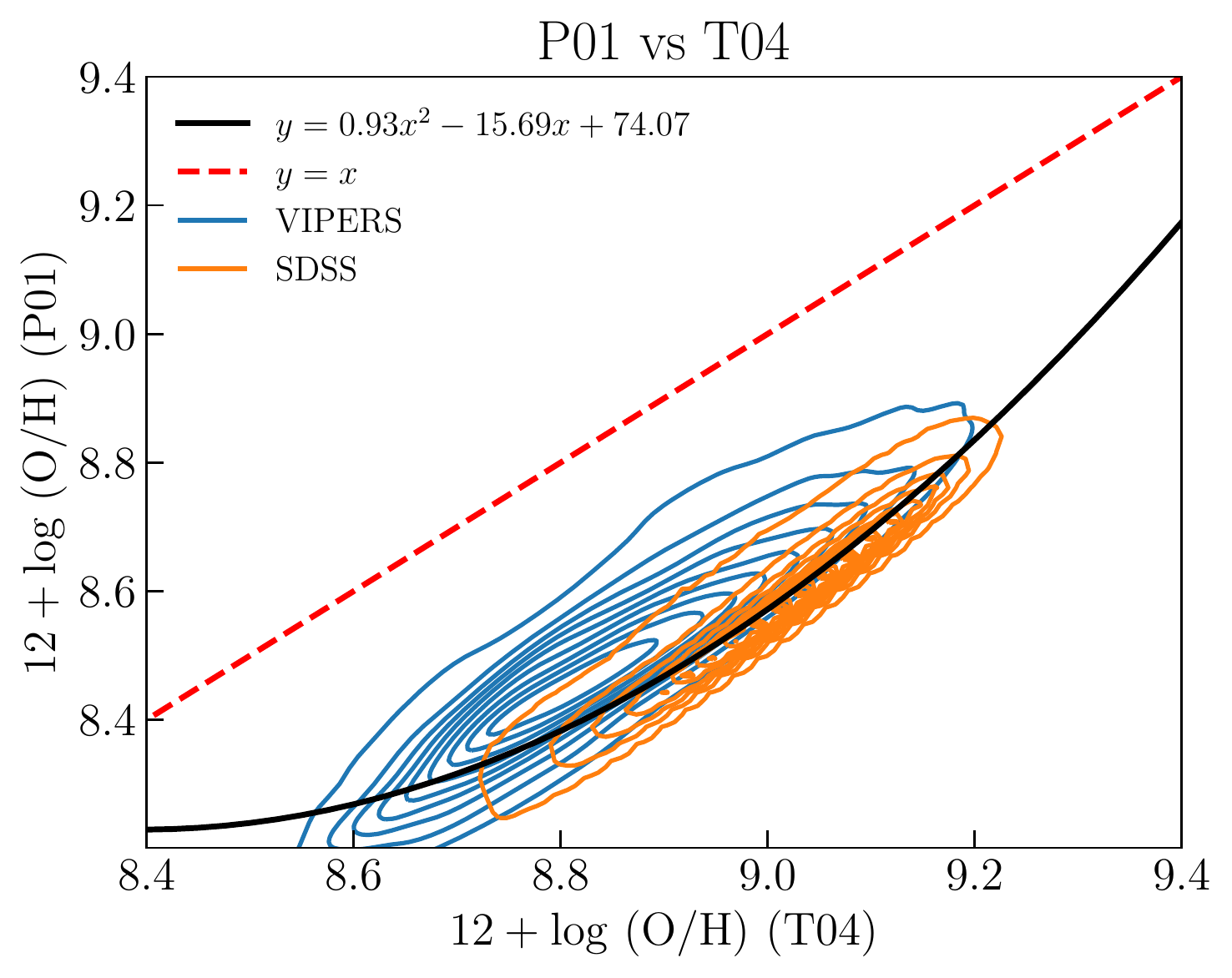}\includegraphics[width=8.5cm]{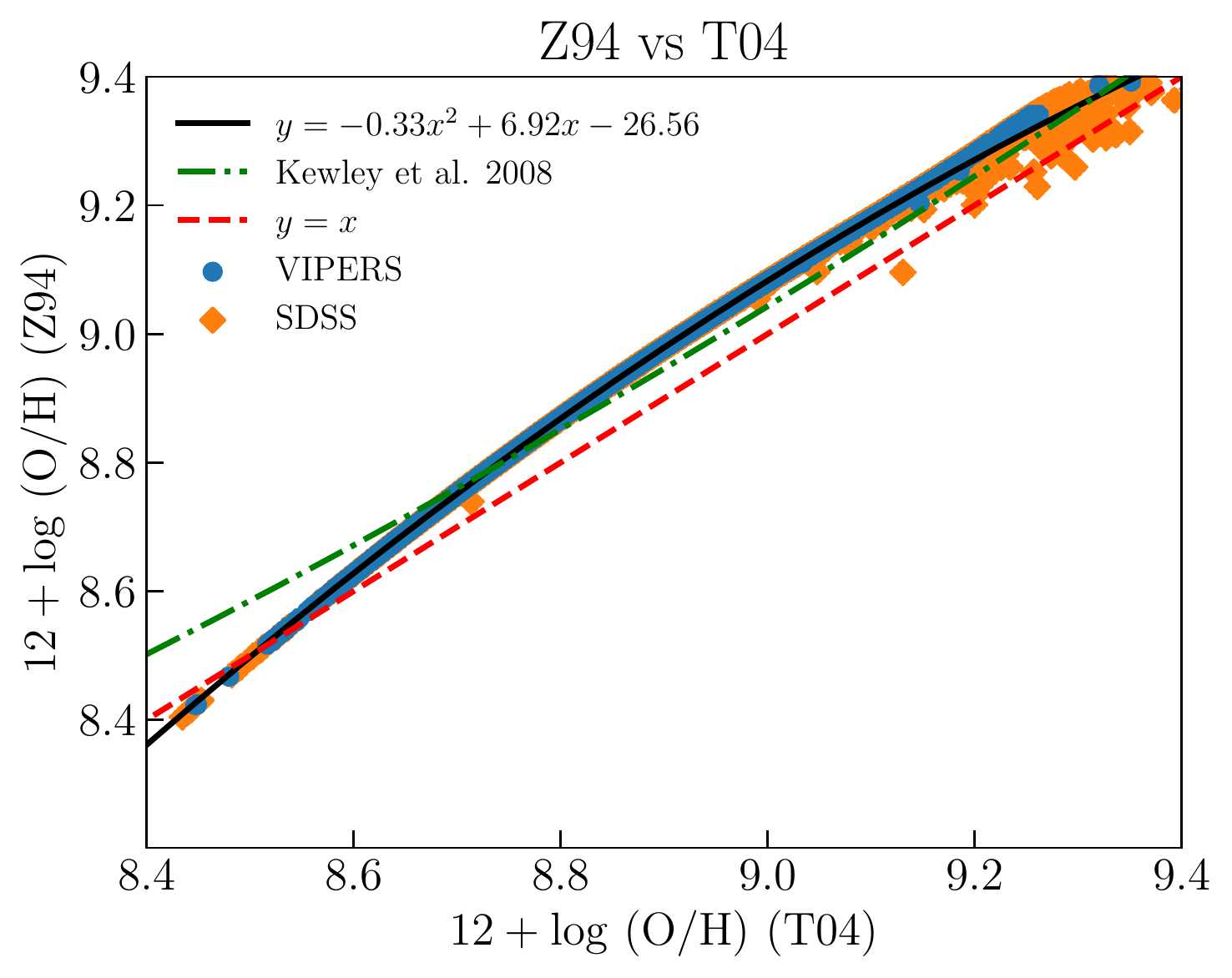}
    \caption{Comparion between different metallicity calibrators: P01 (left panel) and Z94 (right panel) vs T04 for VIPERS (blue) and SDSS (orange) samples. We show the $y=x$ line (dashed red line), our conversion polynomial from one calibrator to another (solid black line), and the conversion polynomial used by \cite{kewley2008metallicity} to convert T04 into Z94 (dash-dotted green line).}
    \label{fig:cal}
\end{figure*}

We then checked if these different methods can introduce significant biases to the analysis. We first removed all galaxies with metallicities outside the range of validity of the calibration, keeping $4772$ SF galaxies in the VIPERS sample and $155\,918$ in the SDSS sample. Figure~\ref{fig:cal} shows the difference between P01 and T04, and Z94 and T04 (T04 was chosen as the reference calibrator because it is among the most recent and most used ones among those considered here). We see that the galaxies are shifted from the $y=x$ line, and the distribution from P01 is spreader than Z94. The main difference between the calibrators is the dependence on the parameter $\mathbf{R_3}$ in P01, which can justify the higher dispersion compared to the other two calibrators. The description of the conversion between different methods can be found in Appendix~\ref{app:conversion}.

The VIPERS sample follows the same metallicity distribution as the SDSS sample (see Fig.~\ref{fig:cal}). This suggests that we are not underestimating or overestimating the metallicity of a sample compared with the other one. For simplicity, we decide to use only the T04 calibration.

\subsection{Property distributions and main sequence of the samples}\label{app:prop_dist}

From the SDSS sample, we removed all the sources with $\log \text{M}_\star \left[ \text{M}_\sun \right] < 7$, $\log \text{SFR} \left[\text{M}_\sun/\text{yr}\right] <-10$, and without a reliable rest-frame blue magnitude applying a cut at $\text{B} >-24$; reducing the total number of SF galaxies to $156\,018$. The selection on the rest-frame blue magnitude removes $\sim 700$ galaxies (less than $0.5\%$ of the total sample) with a wrong estimation of absolute $\text{B}$ magnitude\footnote{The $\text{B}$ magnitude of some galaxies in the GSWLC-2 catalog is flagged as $-99$, indicating some problem during the fitting procedure, and these objects were removed from the sample.}. The value $\text{B} >-24$ has been chosen with the aim of eliminating galaxies with wrong SED fitting results.

\begin{figure*}
    \centering
    \includegraphics[width=5.6cm]{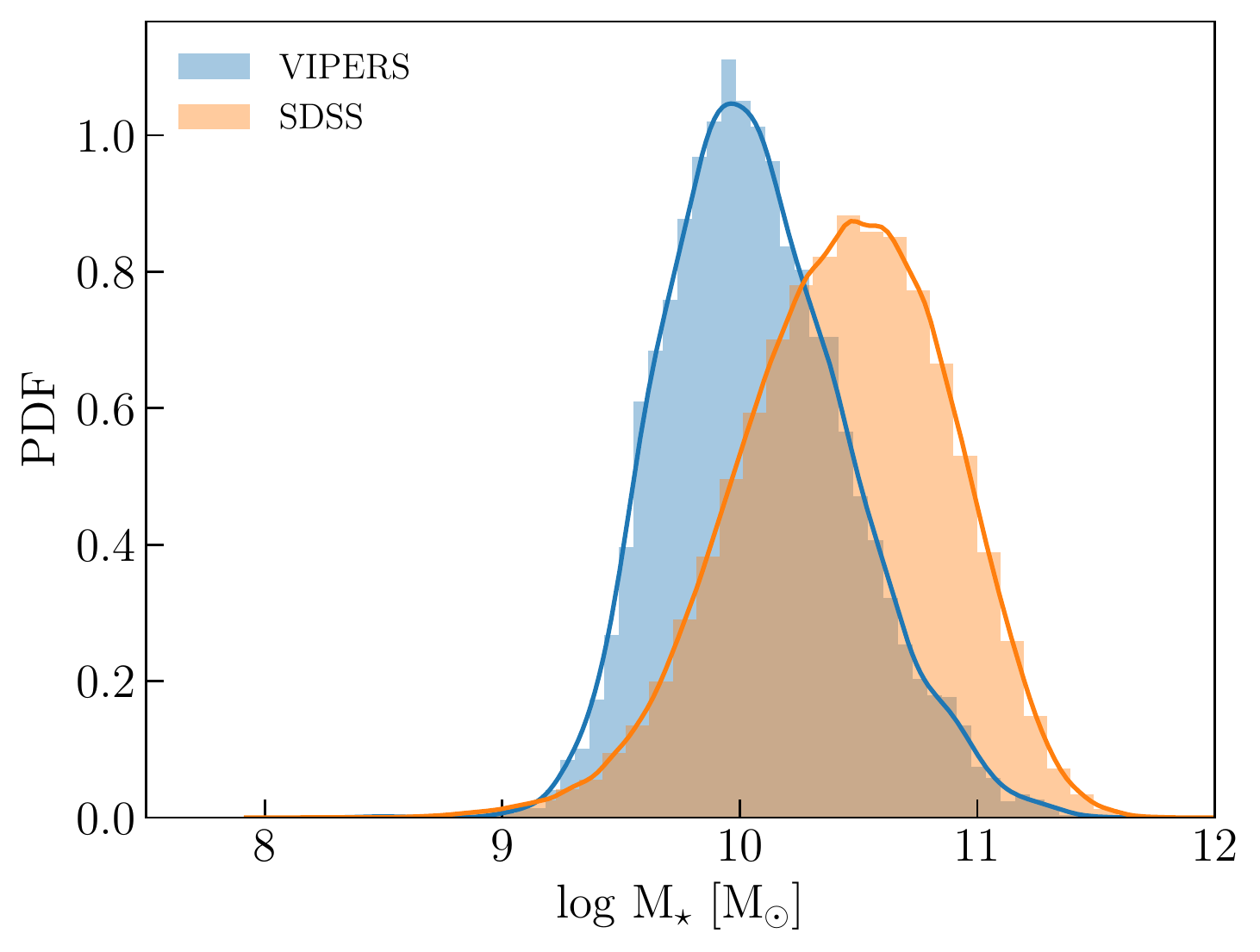}\includegraphics[width=5.6cm]{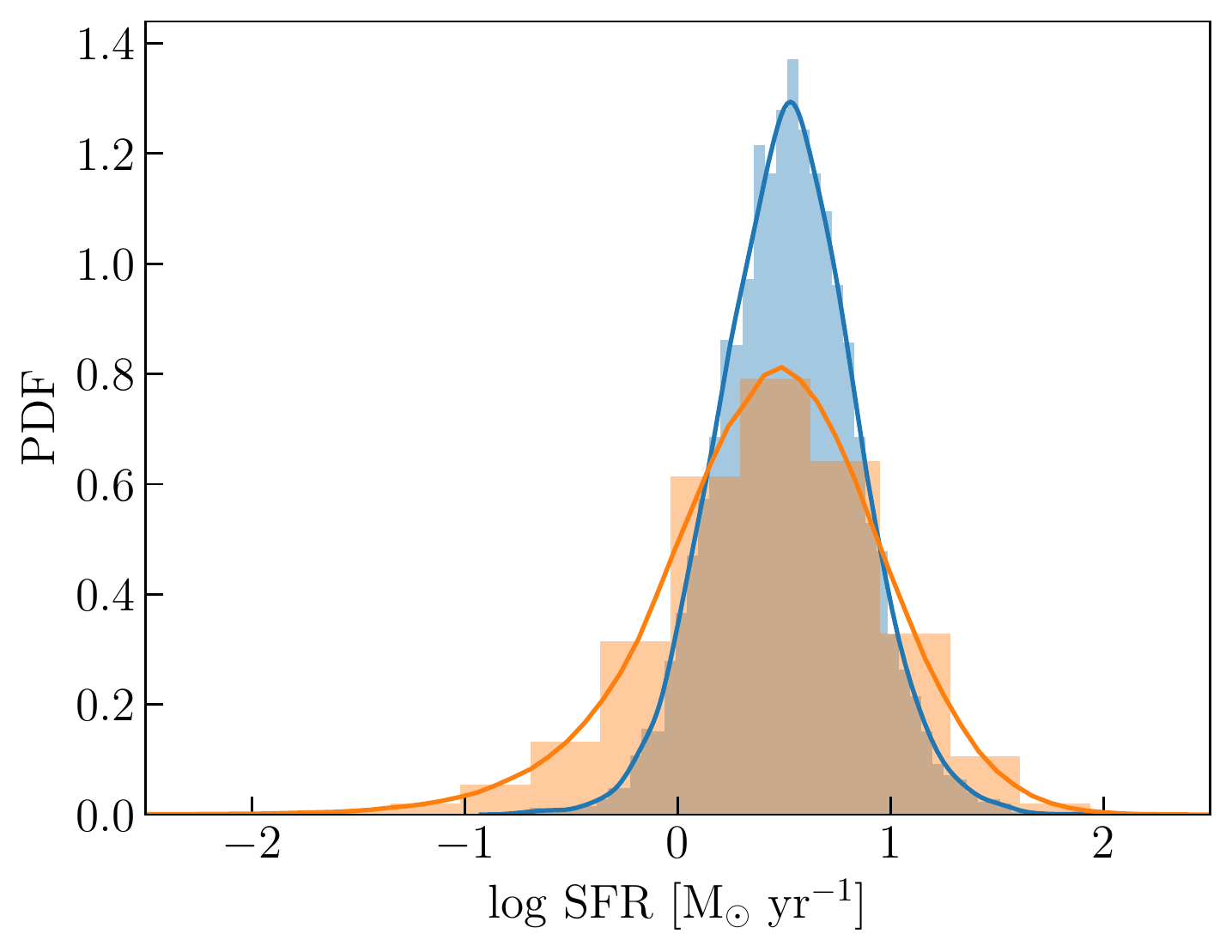}\includegraphics[width=5.6cm]{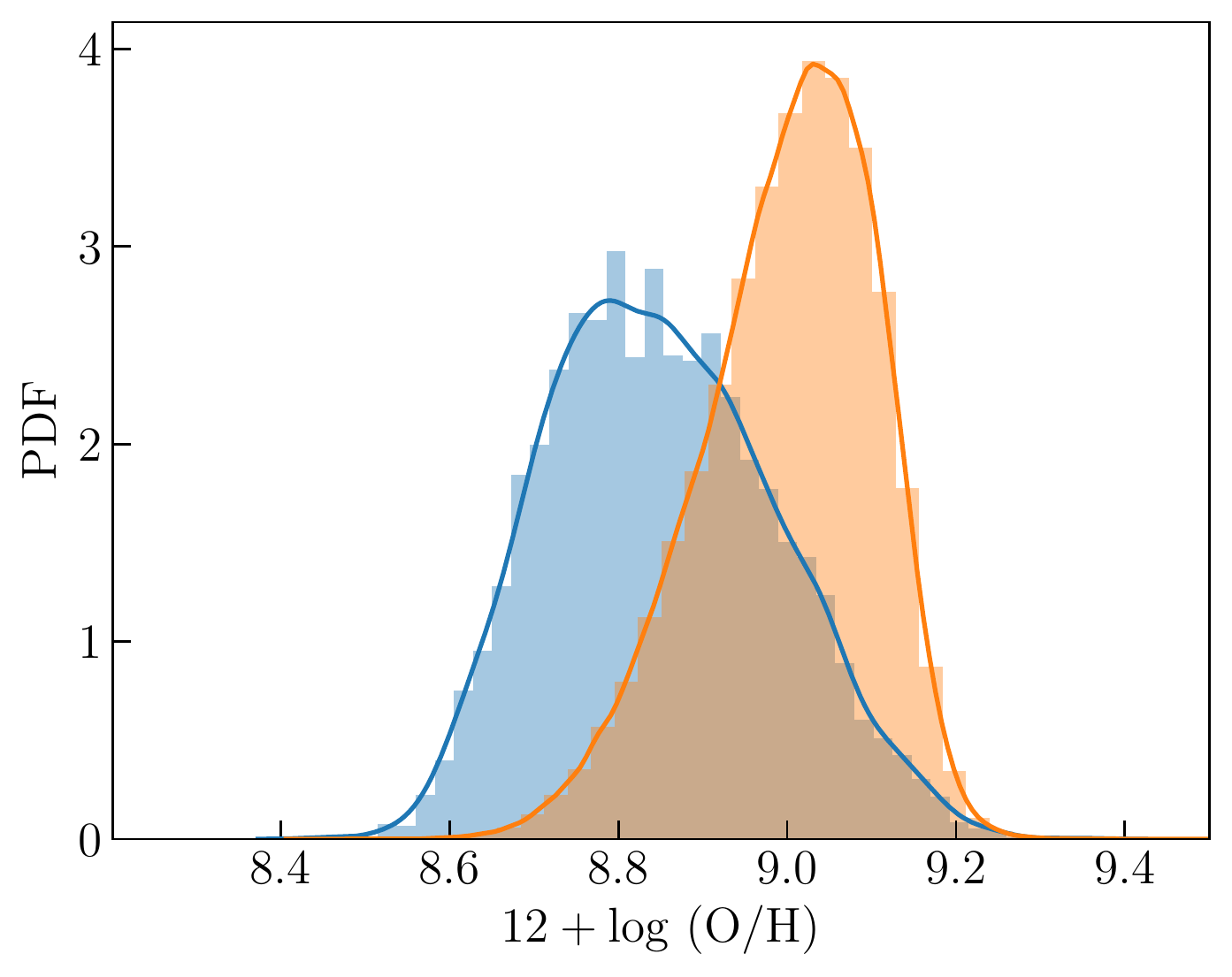}
    \includegraphics[width=5.6cm]{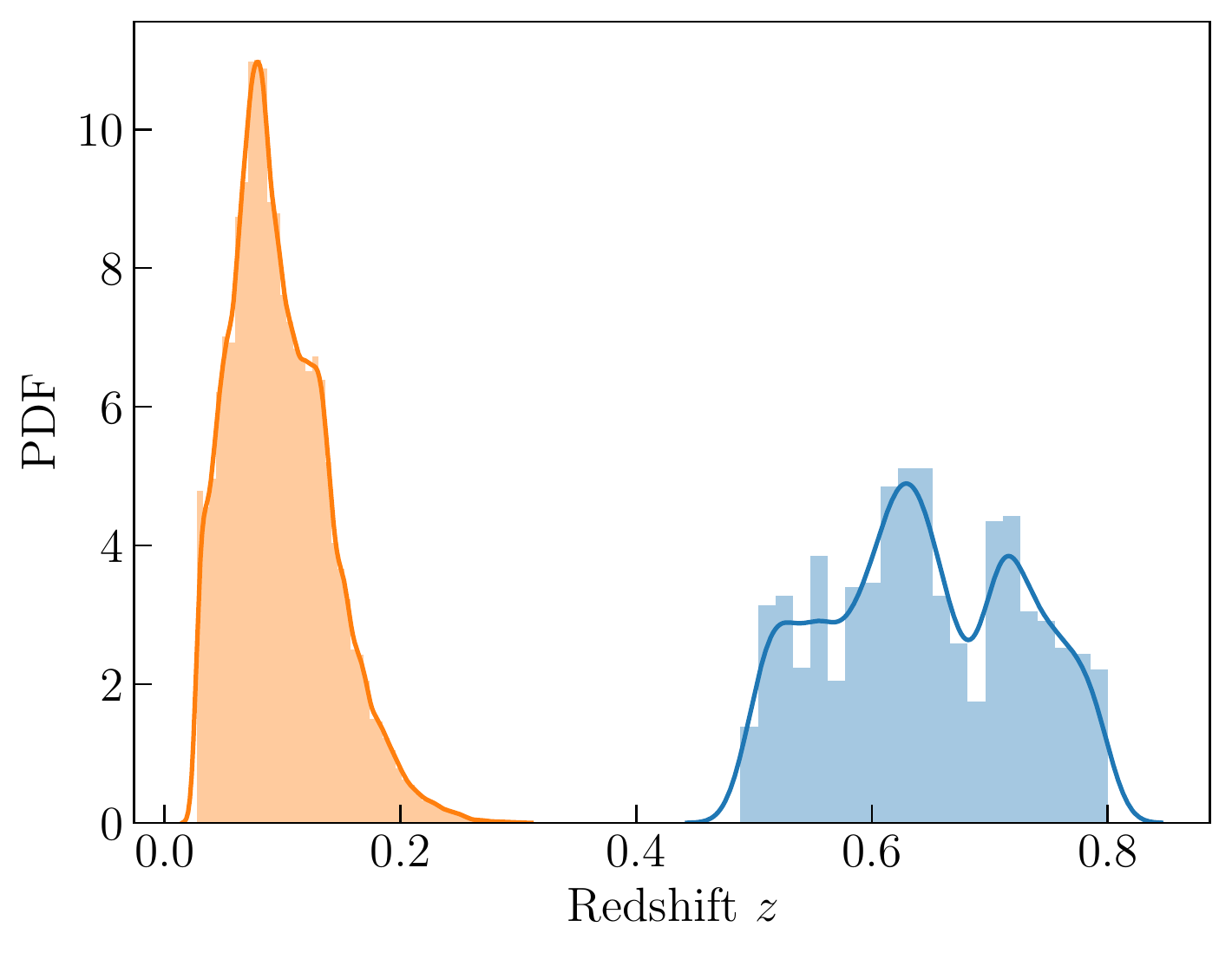}\includegraphics[width=5.6cm]{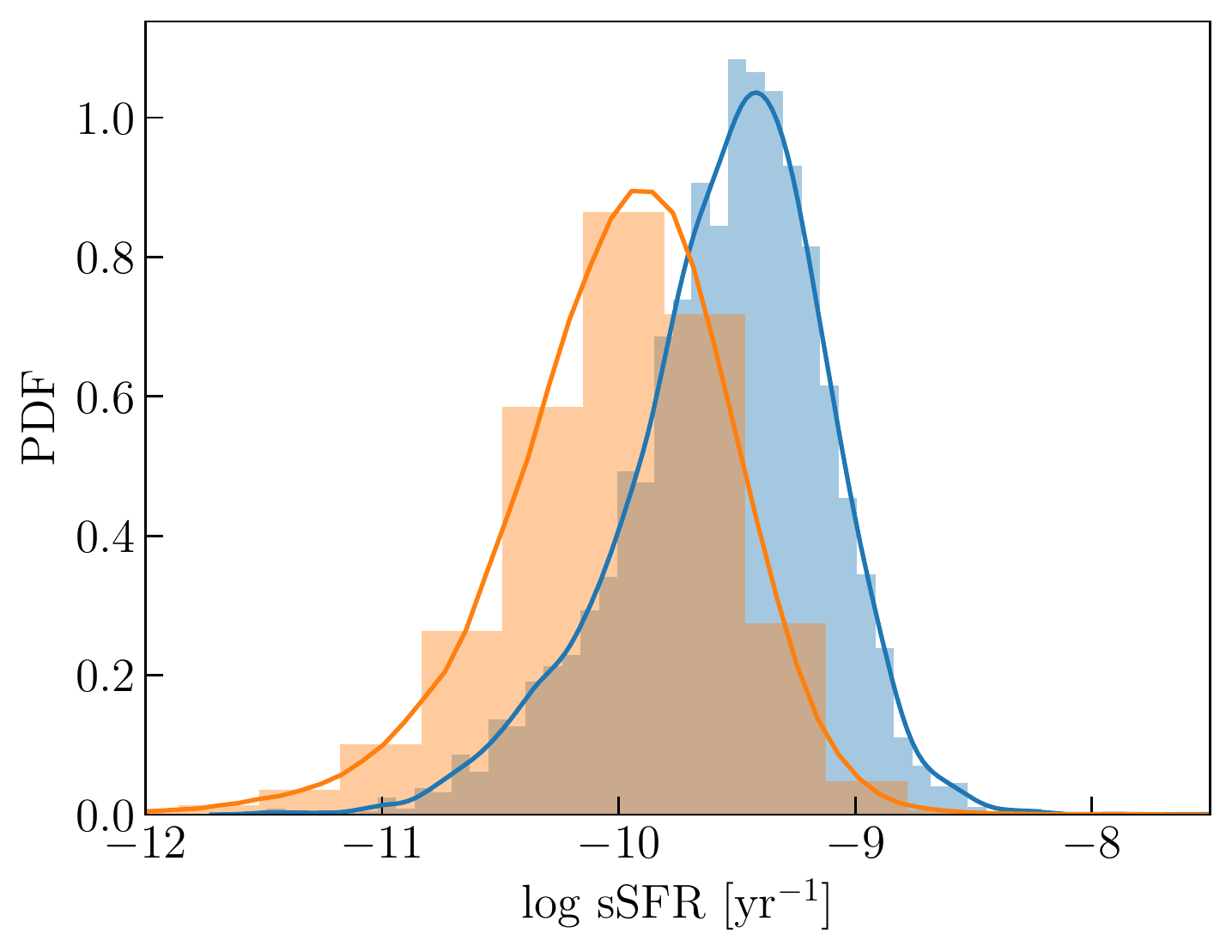}\includegraphics[width=5.6cm]{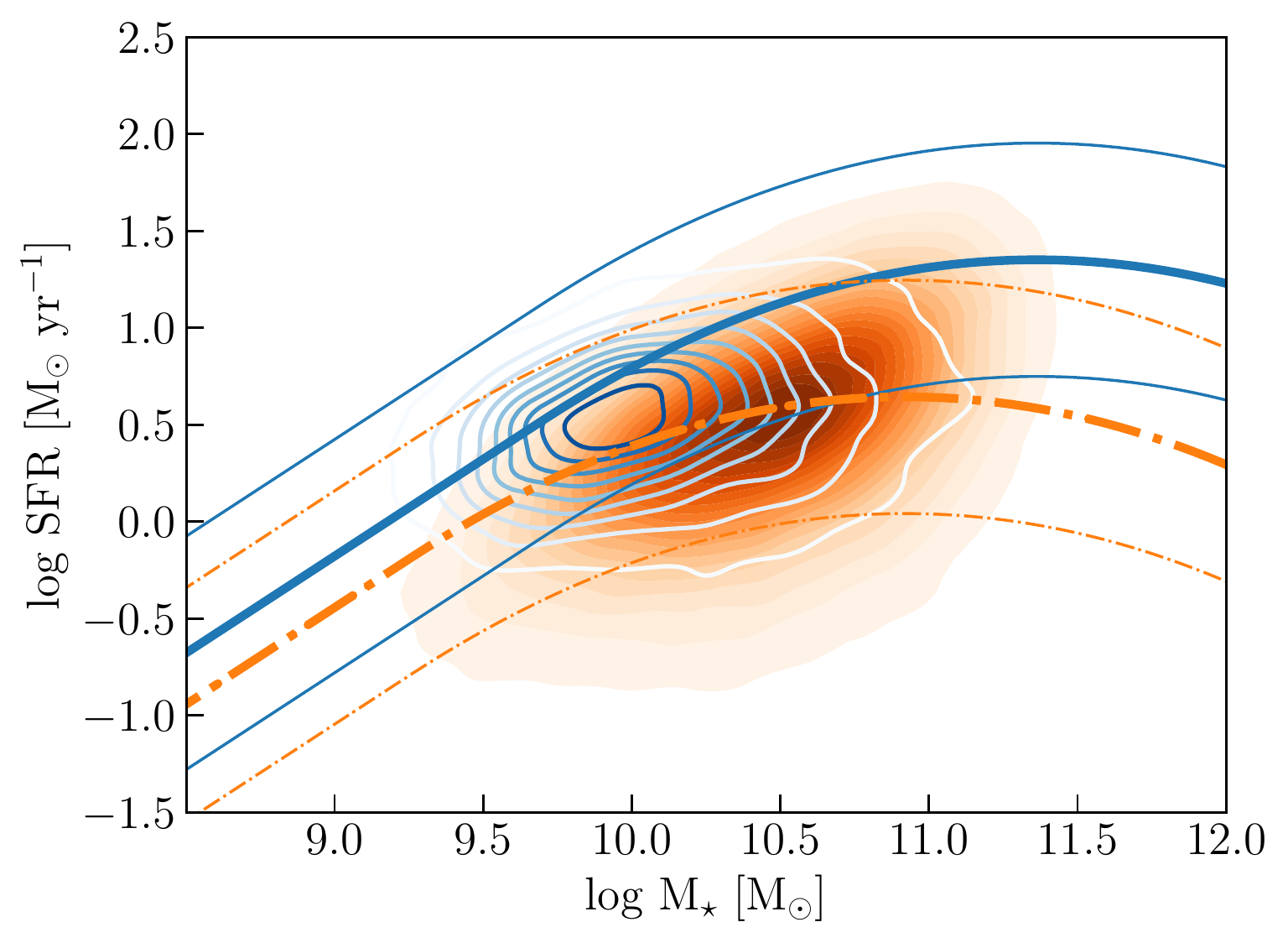}
    \caption{Histograms and KDE of $\text{M}_\star$ (top left), SFR (top mid), metallicity (top right), redshift (bottom left), sSFR (bottom mid), and main sequence (bottom right, with the comparison with the function described by \cite{schreiber2015ms} --- bold lines) for the VIPERS (blue) and SDSS (orange) samples. The thinner lines are the $\pm 4$ times the main sequence.}
    \label{fig:dists}
\end{figure*}

Figure~\ref{fig:dists} shows the histograms with the Kernel Density Estimates (KDEs) of the main galactic features. In particular, we report the $\text{M}_\star$, SFR, metallicity, redshift, sSFR, and the main sequence for the VIPERS and SDSS SF sub-samples used in this work.

The VIPERS sample has higher estimated values of sSFR and lower metallicity distributions as compared to the SDSS sample. From the galaxy SF main sequence (Fig.~\ref{fig:dists}), we observe that VIPERS is shifted towards higher SFR as compared to the SDSS sample. In the same plot, we show the so-called main sequence relation (and $\pm4 \times \text{MS}$) from \cite{schreiber2015ms}.
The observed shift of the VIPERS sample towards higher SFR agrees with the evolution of the main sequence.

\begin{figure}
    \centering
    \resizebox{\hsize}{!}{\includegraphics{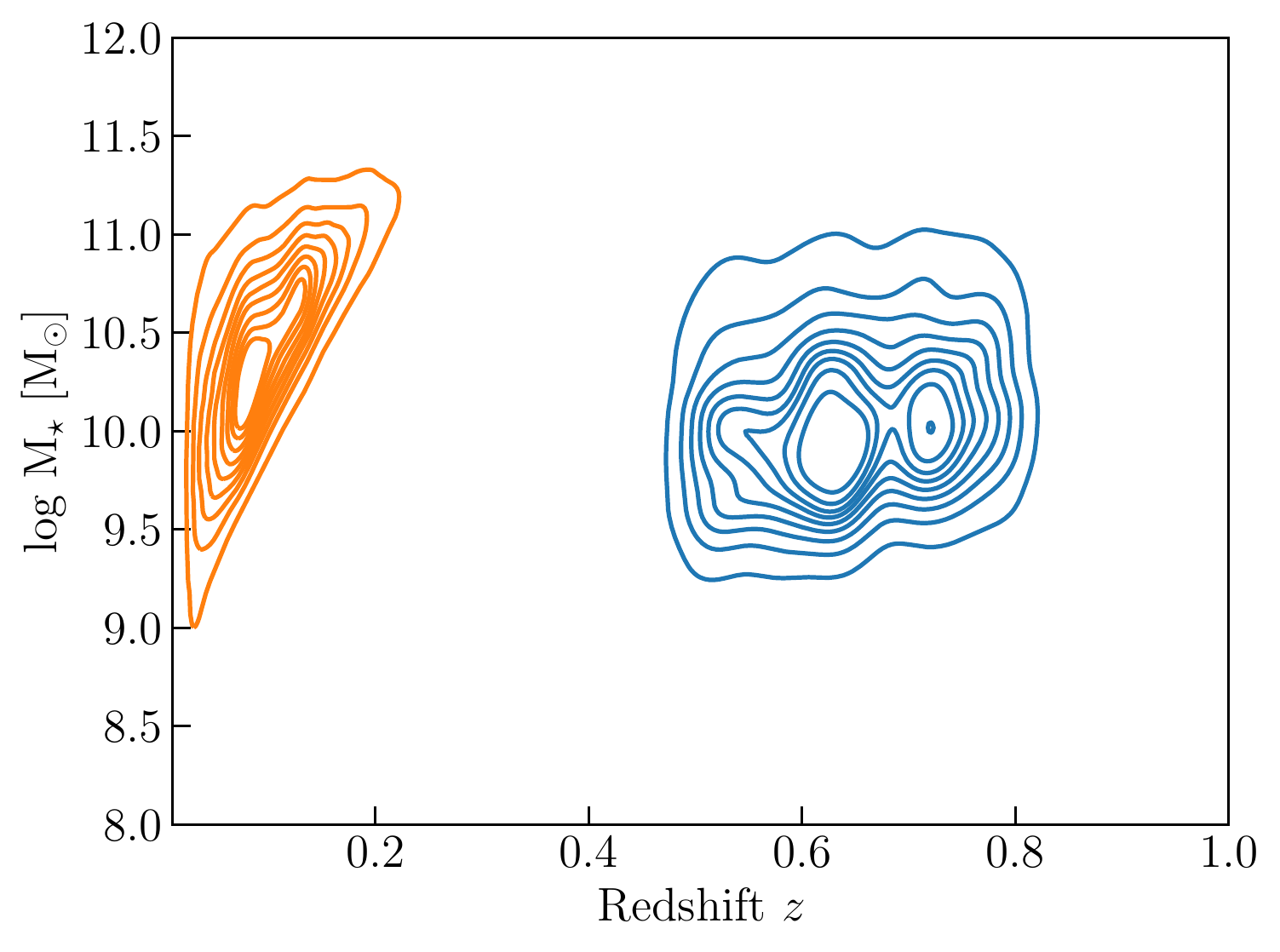}}
    \caption{Stellar mass vs redshift diagram for SDSS (orange) and VIPERS (blue) samples.}
    \label{fig:mass_redshift}
\end{figure}

Figure~\ref{fig:mass_redshift} shows the $\text{M}_\star$ versus redshift diagram where the characteristic cut of low mass galaxies due to the observational cutoff in the magnitude of the surveys (VIPERS: $i_{AB} = 22.5$; SDSS: $i_{AB} = 21.3$) is visible. This magnitude selection leads to a cut of galaxies with lower $\text{M}_\star$ with the redshift.
The VIPERS sample shows two separated populations in its redshift distribution at $z \sim 0.7$. We check whether this can add some bias in the analysis in Appendix~\ref{app:red}.

\section{Study the effects of biases}\label{sect:bias_sdss}

To do a fair comparison, we need to make SDSS as close as possible to the VIPERS SF selection and we need to check how different selections can affect the projections of the FMR. The characterization of a spectroscopic sample depends mostly on all the intrinsic characteristics of the spectrograph, of the sources at different redshift, for instance, the need to  reliably detect different emission lines that have different intrinsic luminosity, as well as on the spectra analysis. For these reasons, when different samples are compared (especially between significantly different redshift ranges), the sample selection criteria are generally different.

Observations at high redshift are usually more limited than local observations in the number of sources, at lower $\text{M}_\star,$ due to the rest-frame cut-off in magnitude. Therefore, a meaningful comparison between local and high redshift samples can be properly made only after taking into account the limitations that characterize both samples. By analyzing these constraints on the local data, we can say if the differences observed between different samples are physical or due to some selection effects. In the following subsections, we analyze different selection criteria individually: the S/N cut of emission lines used to compute the metallicity, the quality of the spectra (the only bias studied on both samples), the selection on the blue rest-frame absolute magnitude, and the fraction of blue galaxies. 

In the following analysis, the samples are divided into bins of $0.15$ dex width in $\text{M}_\star$, SFR, and sSFR. We keep only bins with as many galaxies inside as higher than 25 \citep{curti2020mass}.
The binning scheme is presented in Table~\ref{tab:bin}. We used a different method of binning the properties of SDSS in the range explored in this sample but not in VIPERS.
\begin{table*}
\caption{Binning scheme for VIPERS and SDSS samples according to the property to bin.}              
\label{tab:bin}      
\centering                                      
\begin{tabular}{l l r l r}          
\hline
\hline
 \noalign{\smallskip}
 & \multicolumn{2}{c}{VIPERS} & \multicolumn{2}{c}{SDSS}\\
  \noalign{\smallskip}
\multicolumn{1}{c}{Property} & \multicolumn{1}{c}{Range} & \multicolumn{1}{c}{Method} & \multicolumn{1}{c}{Range} & \multicolumn{1}{c}{Method}\\
   \noalign{\smallskip}
\hline                                   
 \noalign{\smallskip}
 $\log \text{M}_\star \left[ \text{M}_\sun \right]$ & Whole & Equal $0.15$ dex width bins & $\log \text{M}_\star < 9.0$ & Two bins with equal number of galaxies \\
 & & & $9.0 \leq \log \text{M}_\star \leq 11.5$ & Equal $0.15$ dex width bins \\
  & & & $\log \text{M}_\star > 11.5$ & Two bins with equal number of galaxies \\
  ${\log \text{SFR}  \left[ \text{M}_\sun \text{ yr}^{-1} \right]}$ & Whole &Equal $0.15$ dex width bins & ${\log \text{SFR} < -1}$ & Two bins with equal number of galaxies \\
  & & & ${\log \text{SFR} \geq -1}$ & Equal $0.15$ dex width bins \\  
$\log \text{sSFR} \left[ \text{yr}^{-1} \right]$ & Whole & Equal $0.15$ dex width bins & Whole & Equal $0.15$ dex width bins \\
     \noalign{\smallskip}
\hline          
\end{tabular}
\end{table*}
Finally, we estimate the median value inside each bin for both samples and their $1 s$ uncertainties are estimated from the $16$th and $84$th percentiles of the distributions inside the bin.

\subsection{Alternatives of the BPT diagram}\label{sec:bias_bpt}

The SDSS sample allows us to check if the use of the  version of the BPT diagram described by \cite{lamareille2010spectral} instead of the original diagram  can introduce a bias on the metallicity. First, we selected all sources in the composite region defined by Eq.~\ref{eq:bpt_composite} \citep{lamareille2010spectral}. Then, we separate SF galaxies and LINERs following \cite{kauffmann2003host}, according to which a galaxy is defined as non SF when\ \begin{equation}
    \log \left( \left[ \text{O{\,\sc{iii}}} \right]/\text{H}\beta \right) > \frac{0.61}{\log \left( \left[ \text{N{\,\sc{ii}}} \right]/\text{H}\alpha \right) - 0.05} + 1.3
.\end{equation}

\begin{figure*}
    \centering
    \includegraphics[width=8.5cm]{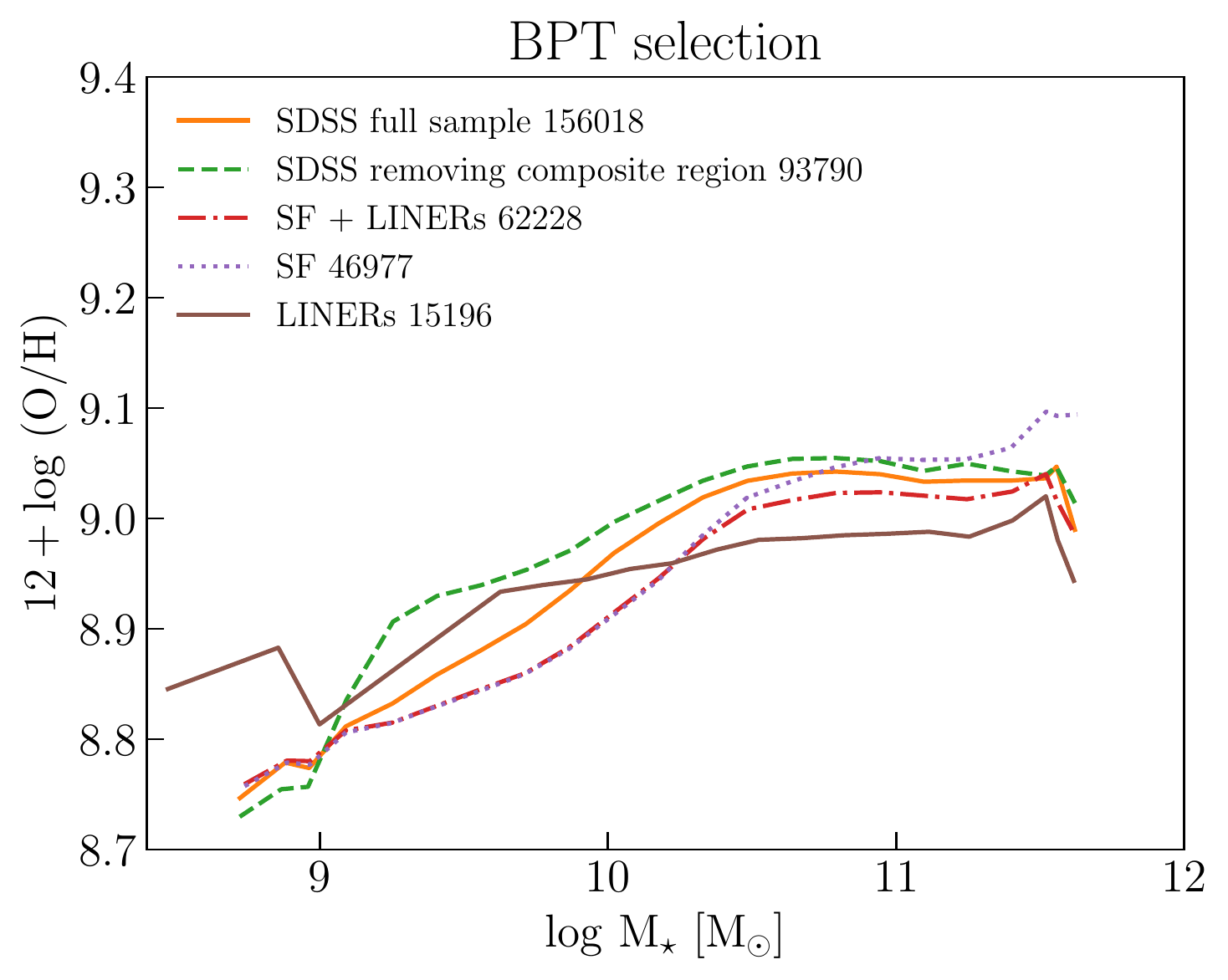}\includegraphics[width=8.5cm]{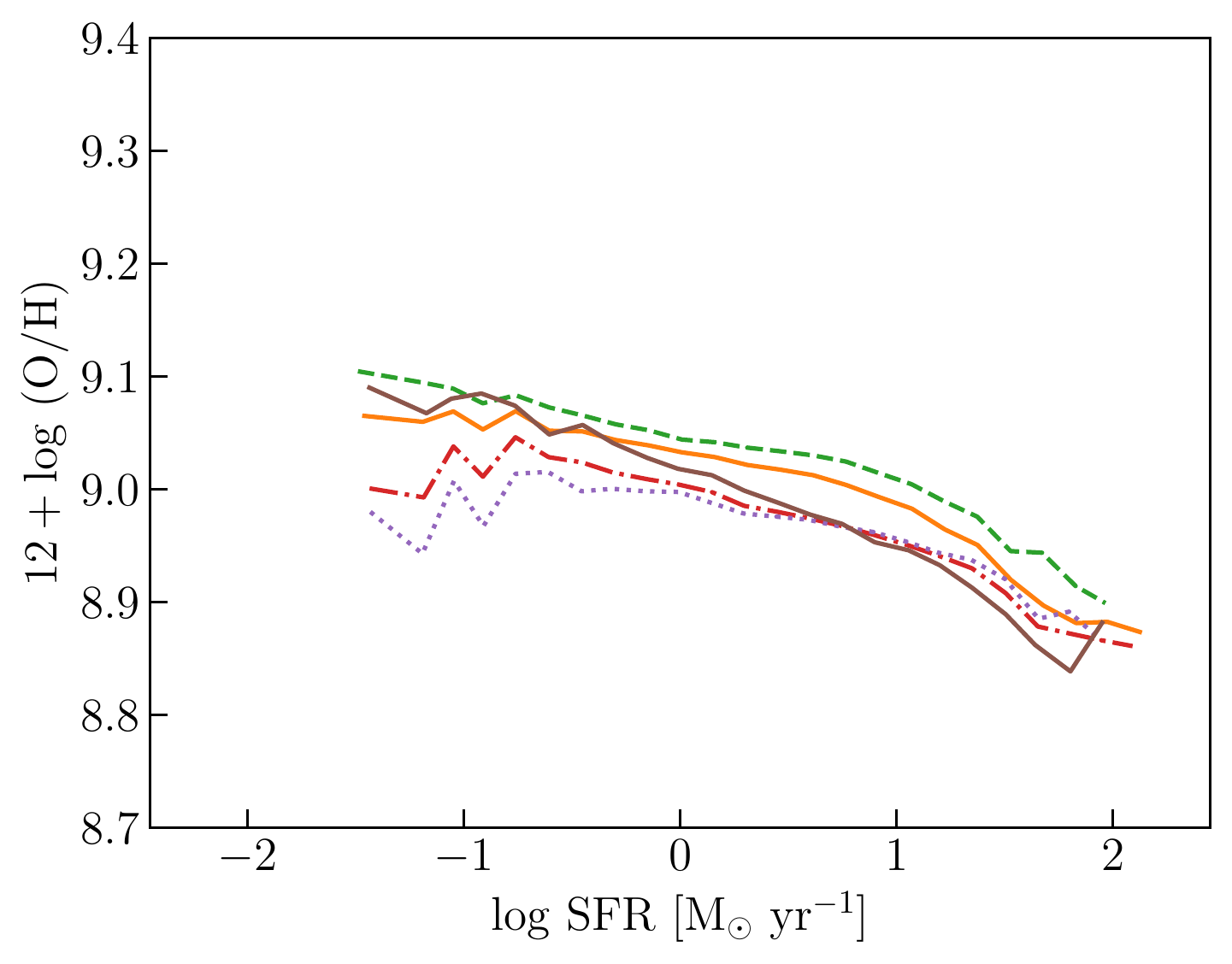}
    \includegraphics[width=8.5cm]{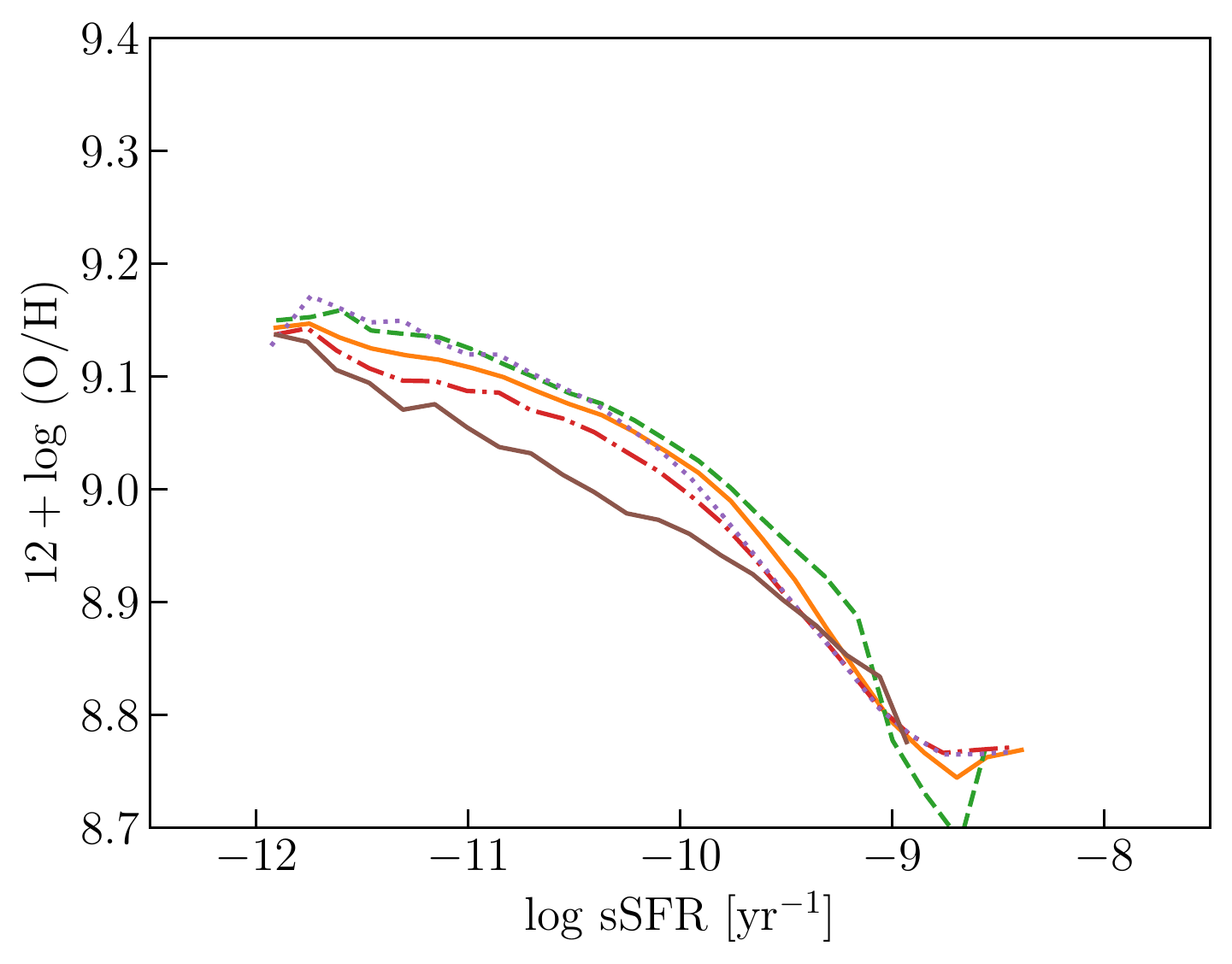}\includegraphics[width=8.5cm]{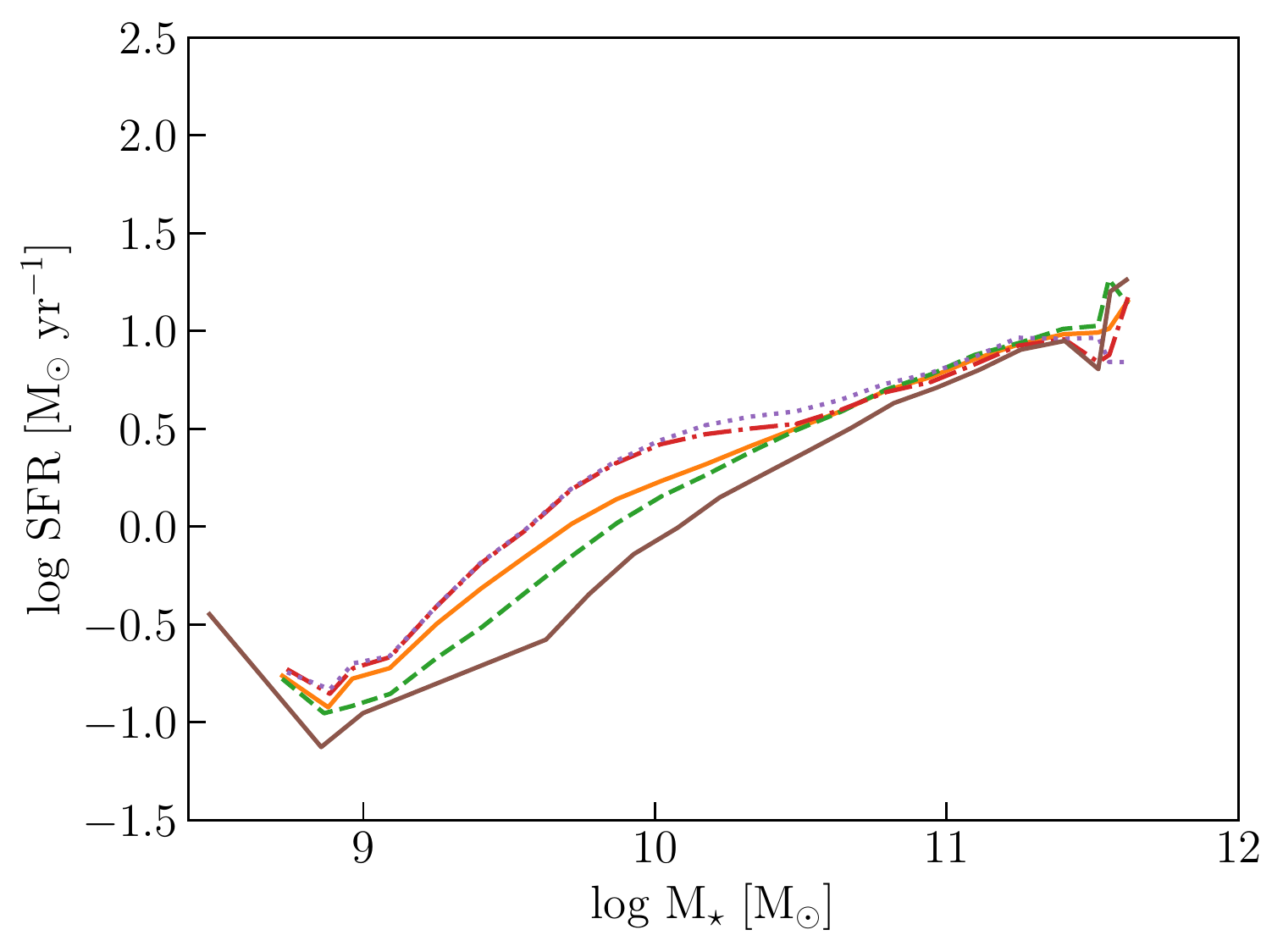}
    \caption{Effects of BPT diagram choice on the projections of the FMR (full sample: orange solid line); SDSS removing completely the composite region (green dashed line); composite region defined by Eq.~\ref{eq:bpt_composite} \citep{lamareille2010spectral} (SF + LINERs, red dash-dotted line); SF of the composite region (purple dotted line); and LINERs of the composite region (brown solid line).}
    \label{fig:bpt_bias}
\end{figure*}

Figure~\ref{fig:bpt_bias} shows the effects of the choice of the BPT diagram on the FMR projections. In these plots, we show in particular the effects of removing completely the composite region defined by \cite{lamareille2010spectral} and removing only the LINERs inside the same region.
Removing  the composite region completely ($62\,228$ SF plus LINERs galaxies, $\sim 40\%$ of the full sample) affects mainly the MZR at low $\text{M}_\star$, changing the overall shape of the MZR itself. This selection does not have a statistically significant effect on the other projections.

Regarding the effect of the LINERs contamination inside the composite region, $15\,196$ LINERs galaxies ($\sim 25 \%$ of the composite region and $\sim 10\%$ of the full sample), we checked the FMR projections for: i) the full composite region, ii) the SF galaxies, and iii) only LINERs. Figure~\ref{fig:bpt_bias} shows statistically negligible effects on the projections between the full composite region and only the SF galaxies inside this region. Since removing completely this region affects the results the most significantly, we decide to keep the LINERs contaminated area in our sample.

\subsection{S/N selection effect}

The first bias analyzed is related to different S/N cutoffs. We simultaneously applied a S/N cut on all emission lines used to compute the metallicities for different best-percentages ($10\%$, $25\%$, $50\%$; namely, the $X\%$ best-percentage means we keep the $X\%$ of the full SDSS sample with the highest S/N) and cut in the same range of S/N as the VIPERS sample. In Appendix~\ref{app:dist}, the distributions of S/N for the emission lines are shown.

\begin{figure*}
    \centering
    \includegraphics[width=8.5cm]{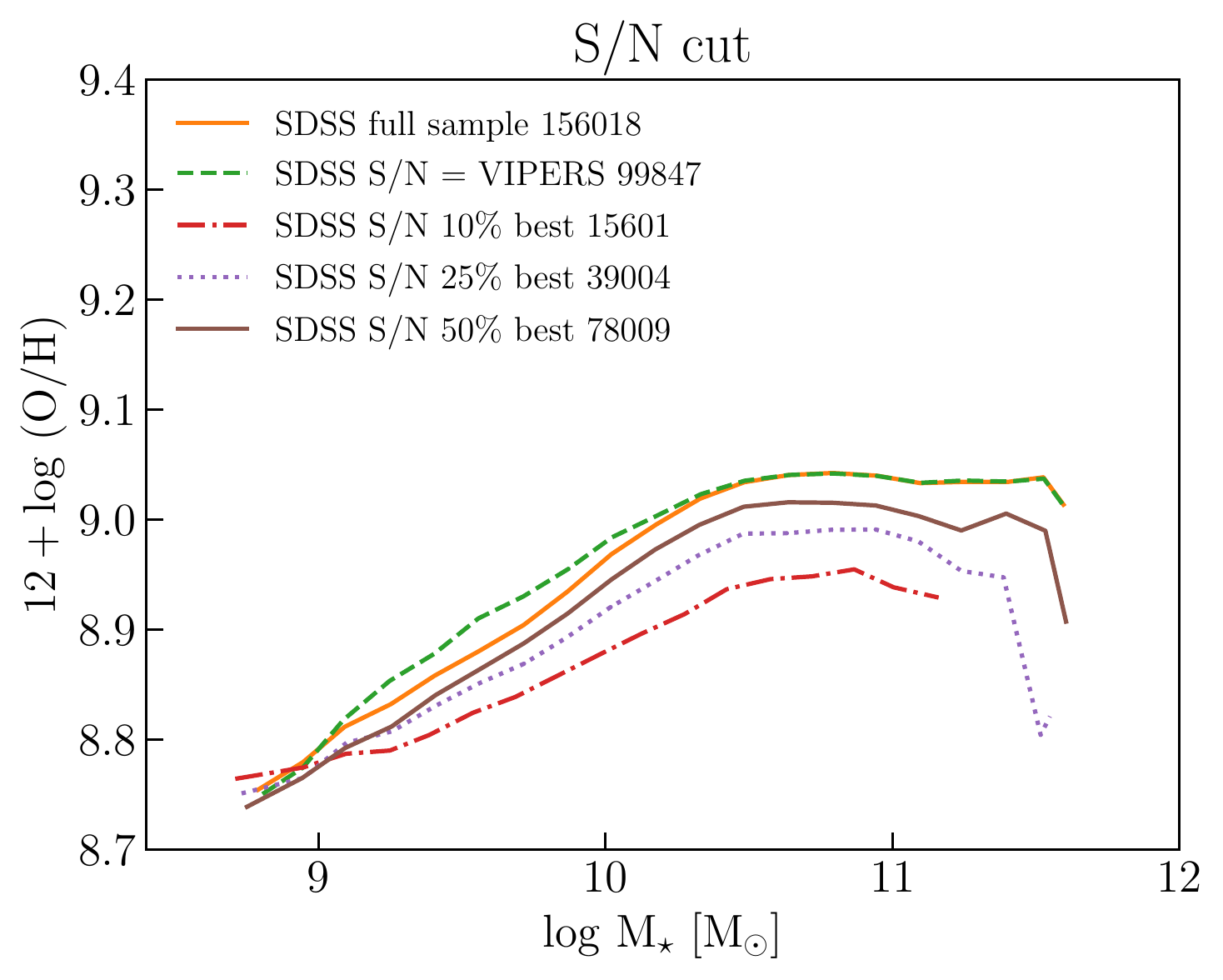}\includegraphics[width=8.5cm]{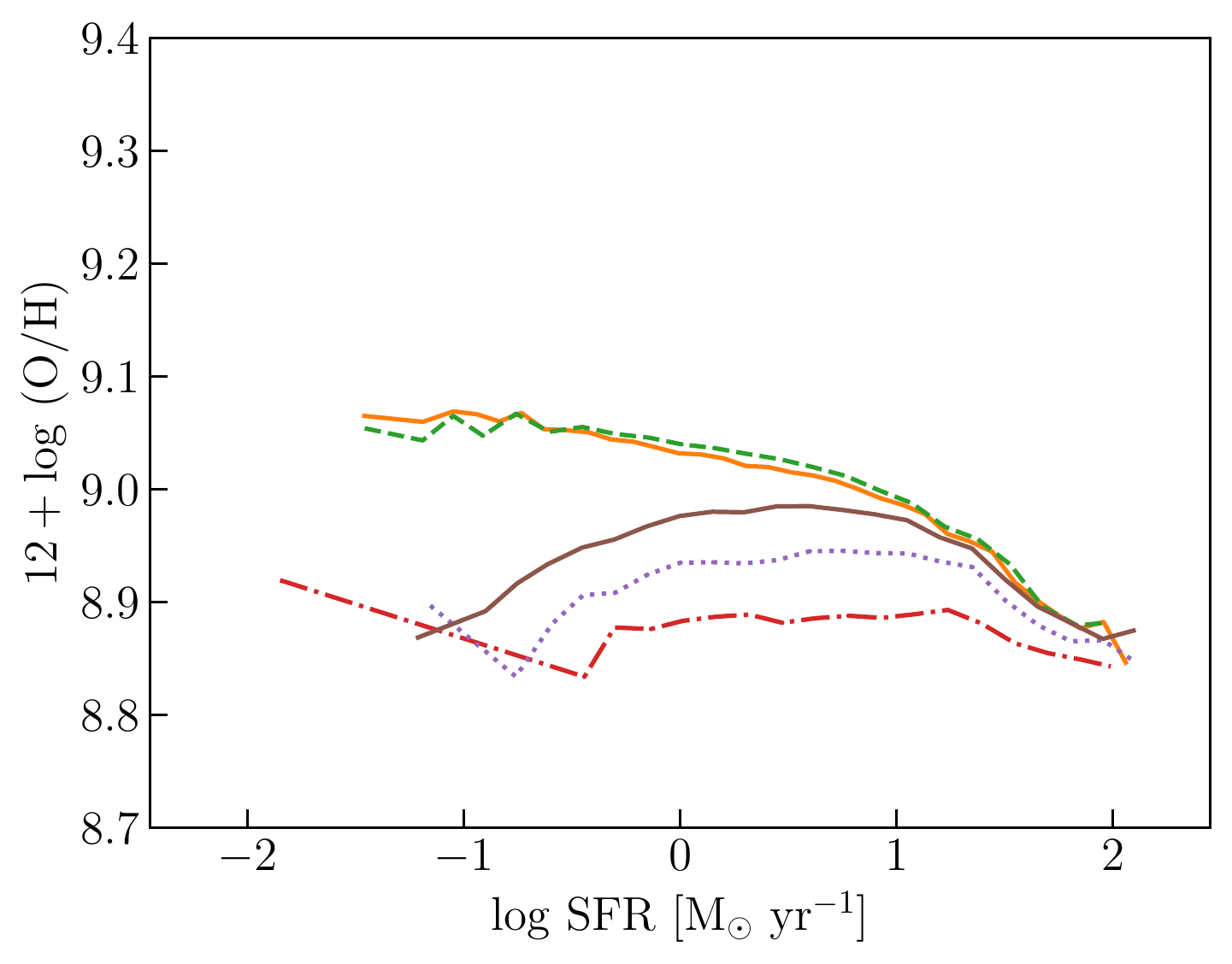}
    \includegraphics[width=8.5cm]{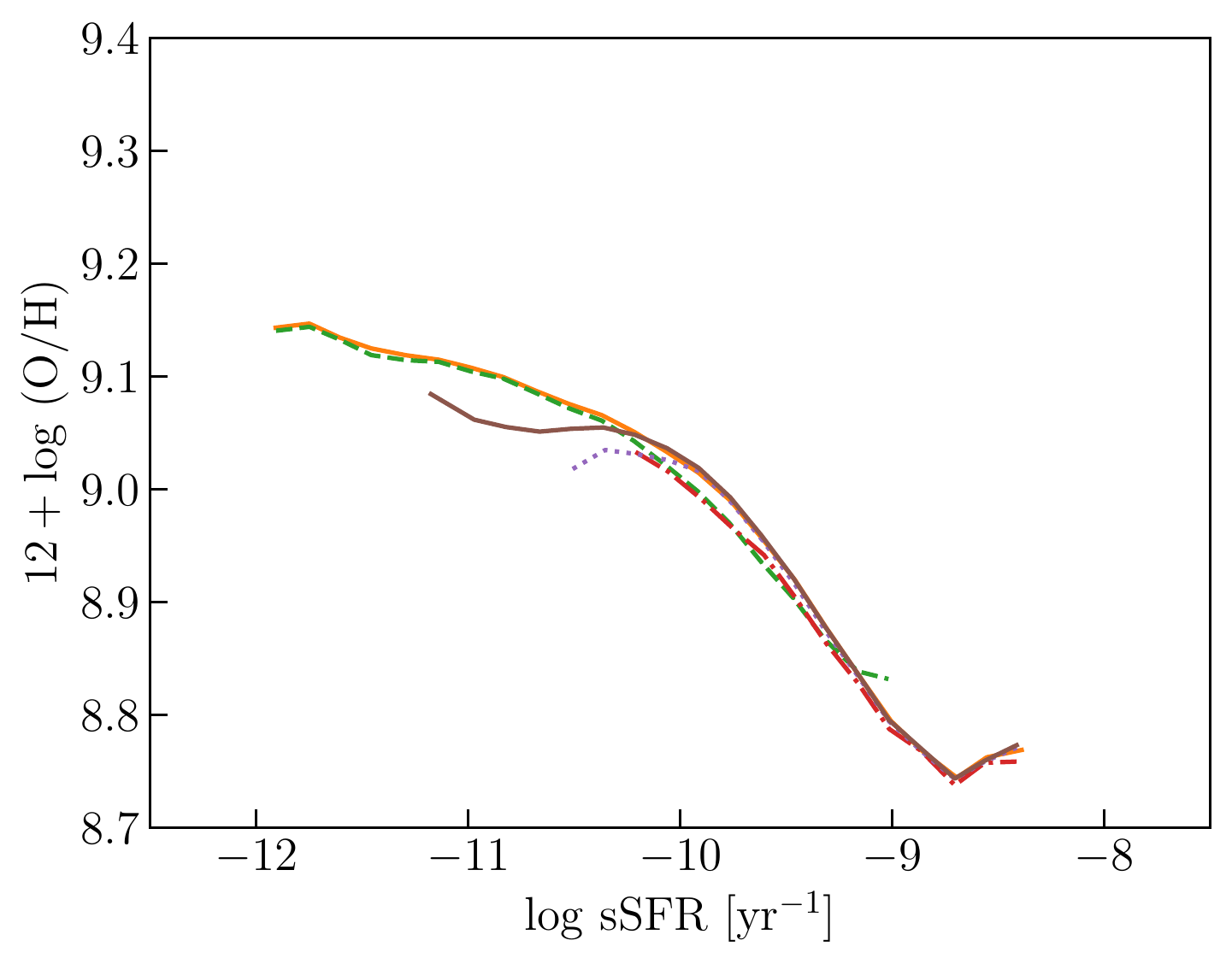}\includegraphics[width=8.5cm]{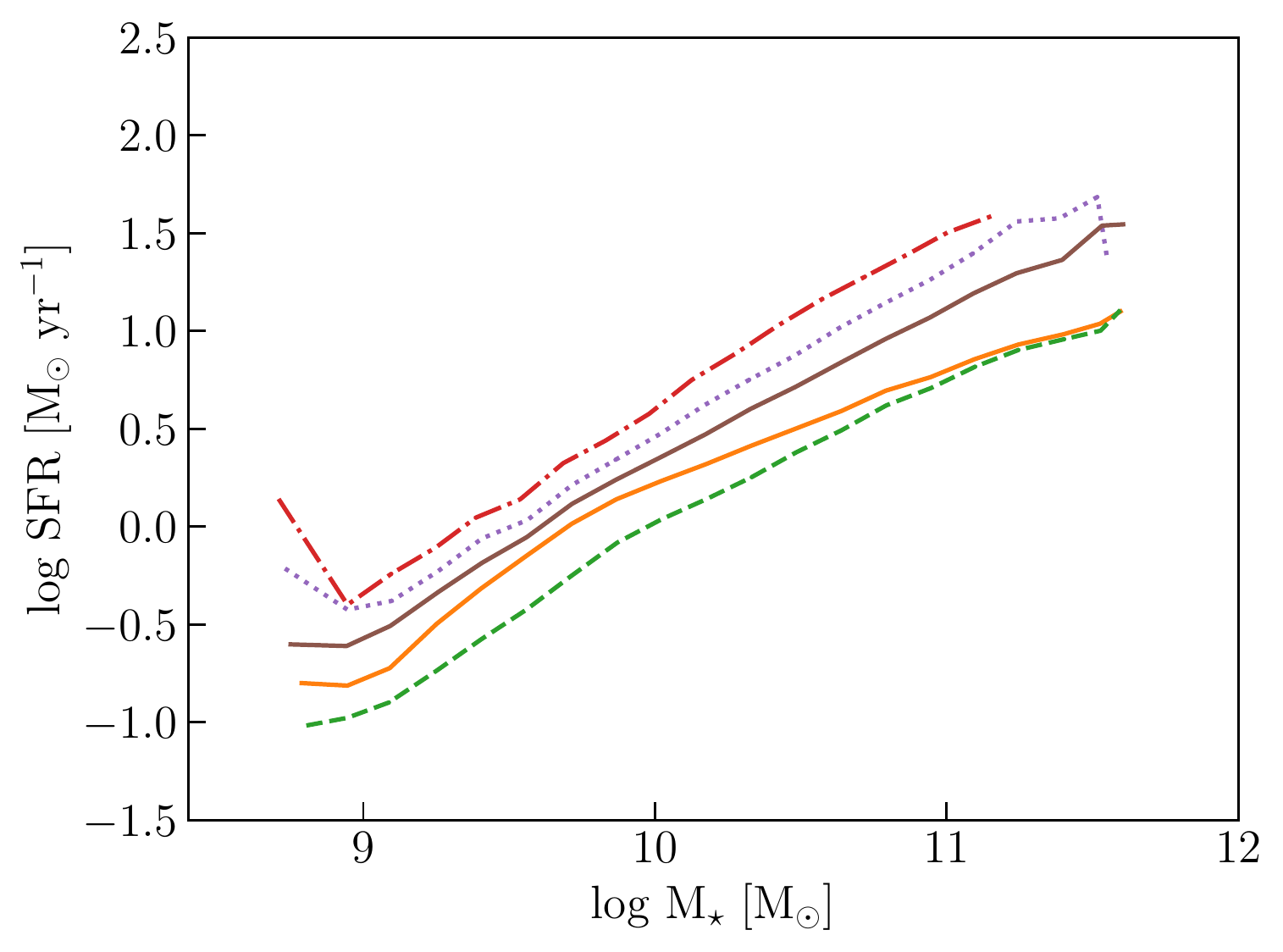}
    \caption{Effects of S/N cuts of the emission lines on the projections of the FMR (full sample: orange solid line; S/N interval equal to VIPERS: green dashed line; $10\%$ best: red dash-dotted line; $25\%$ best, purple dotted line; $50\%$ best: brown solid line).}
    \label{fig:s/n_bias}
\end{figure*}

Figure~\ref{fig:s/n_bias} shows the projection of the FMR after the application of the S/N selections. Decreasing the best-percentage of the S/N (i.e. increasing the cutoff on the lines) increases the flattening of the MZR and shifts it towards lower metallicities. A selection of higher S/N sources removes mainly metal-rich galaxies at higher $\text{M}_\star$es with a median $\Delta \log \left( \text{O} / \text{H} \right) = 0.09$ dex and a maximum $\Delta \log \left( \text{O} / \text{H} \right) = 0.11$ dex --- $\Delta \log \left( \text{O} / \text{H} \right)$ is defined as the difference in metallicity between the SDSS and VIPERS samples in each bin. This means that metal-poor galaxies at a given $\text{M}_\star$es have intrinsically higher S/N of emission lines, an effect that is in agreement with the one described by \cite{curti2020mass}. This type of galaxies also has a higher SFR, with a younger stellar population that can emphasize the emission lines increasing the photo-ionization state of the gas-phase metals. It is important to note that the sample with S/N in the same range of the VIPERS data has higher metallicity in the range $9.0 \leq \log \text{M}_\star \left[ \text{M}_\sun \right] \leq 10.25$ than the full sample.

In the metallicity versus the SFR plane, the cut is mainly at low SFR for sources with higher metallicities and flatten the curve. In this plane, the median $\Delta \log \left( \text{O} / \text{H} \right) = 0.17$ dex and a maximum $\Delta \log \left( \text{O} / \text{H} \right) = 0.23$ dex, showing a much more important dependence on this selection than the MZR. The curves at low $\text{M}_\star$es and high SFRs are almost invariant with these cuts.
It is interesting to note that the metallicity versus the sSFR plane, is almost insensitive to the cut on S/N. This means that the effects of the cut in function of the $\text{M}_\star$ and SFR cancel each other out.

Increasing the S/N cut removes the galaxies from the bottom part of the main sequence from the analysis. The same part with galaxies containing a higher abundance of metals. This bias moves the distribution of the main sequence towards the top part of the diagram.

\subsection{Quality of spectra: Flag selection}\label{sec:quality_flag}

The next step is to check which kind of bias can be introduced by a selection on the quality flag of the emission lines described in Sect.~\ref{sec:data_selection_vipers}. The flag definition is more complicated than the simple S/N selection. In particular, the first three values need a specific analysis of the spectra. In this part, we checked the effects of the flag selection on all the lines at the same time for the VIPERS sample; while on the SDSS sample, we checked the effects on the selection to fulfill the loose condition ($\text{EW} \geq 3 \sigma$ or $\text{flux} \geq 7 \sigma$) of the flag t-value of VIPERS for each emission line separately.

\begin{figure*}
    \centering
    \includegraphics[width=8.5cm]{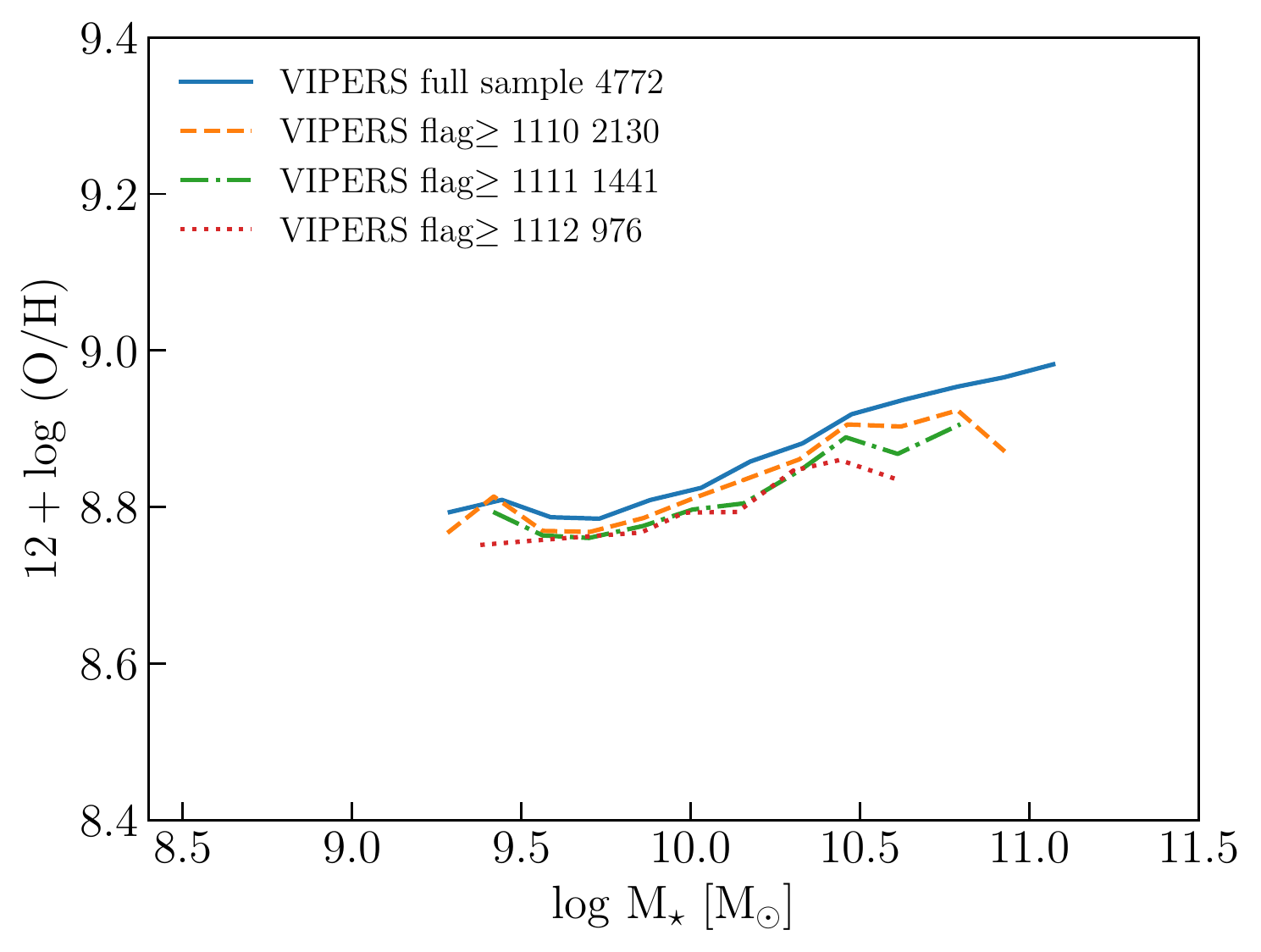}\includegraphics[width=8.5cm]{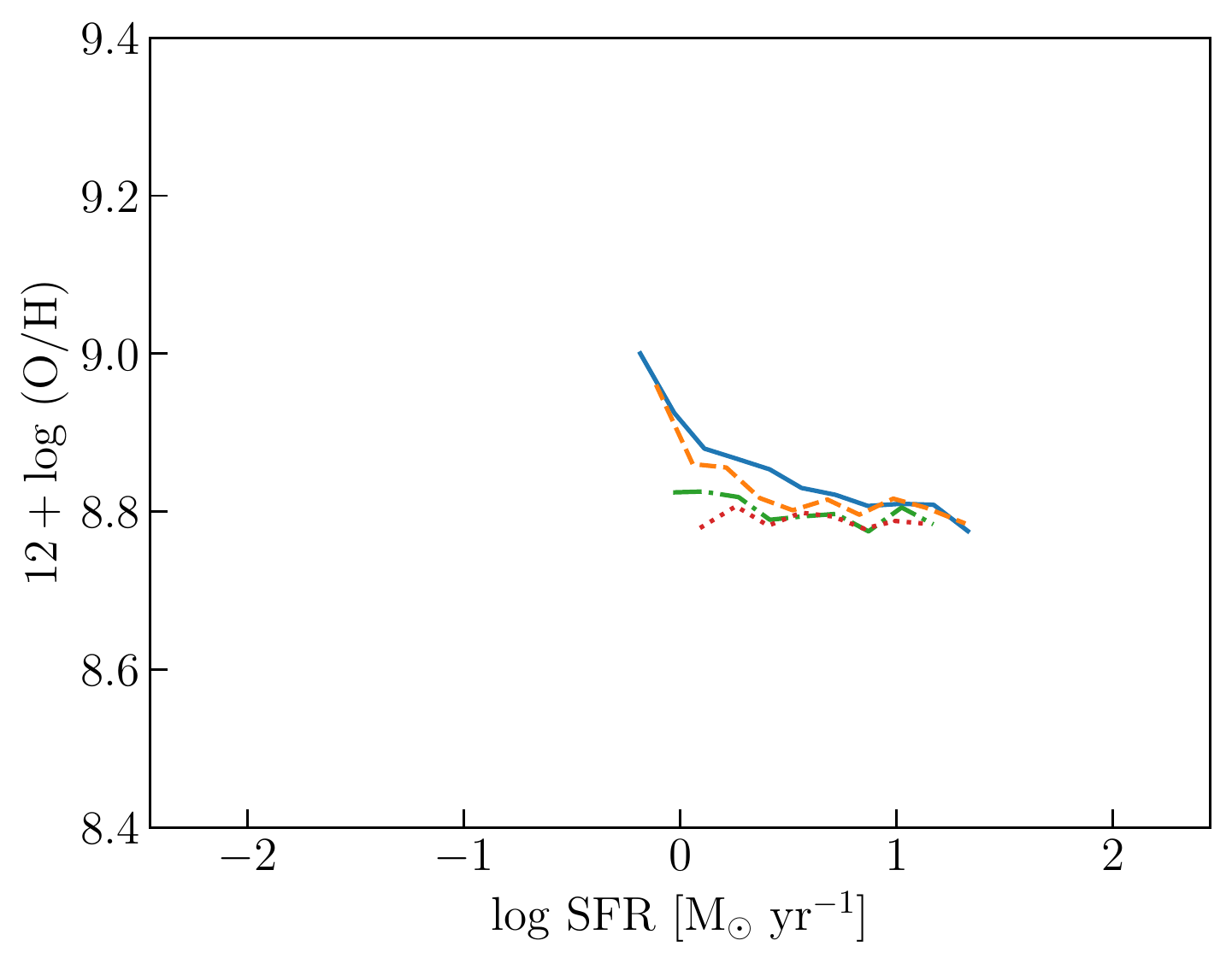}
    \includegraphics[width=8.5cm]{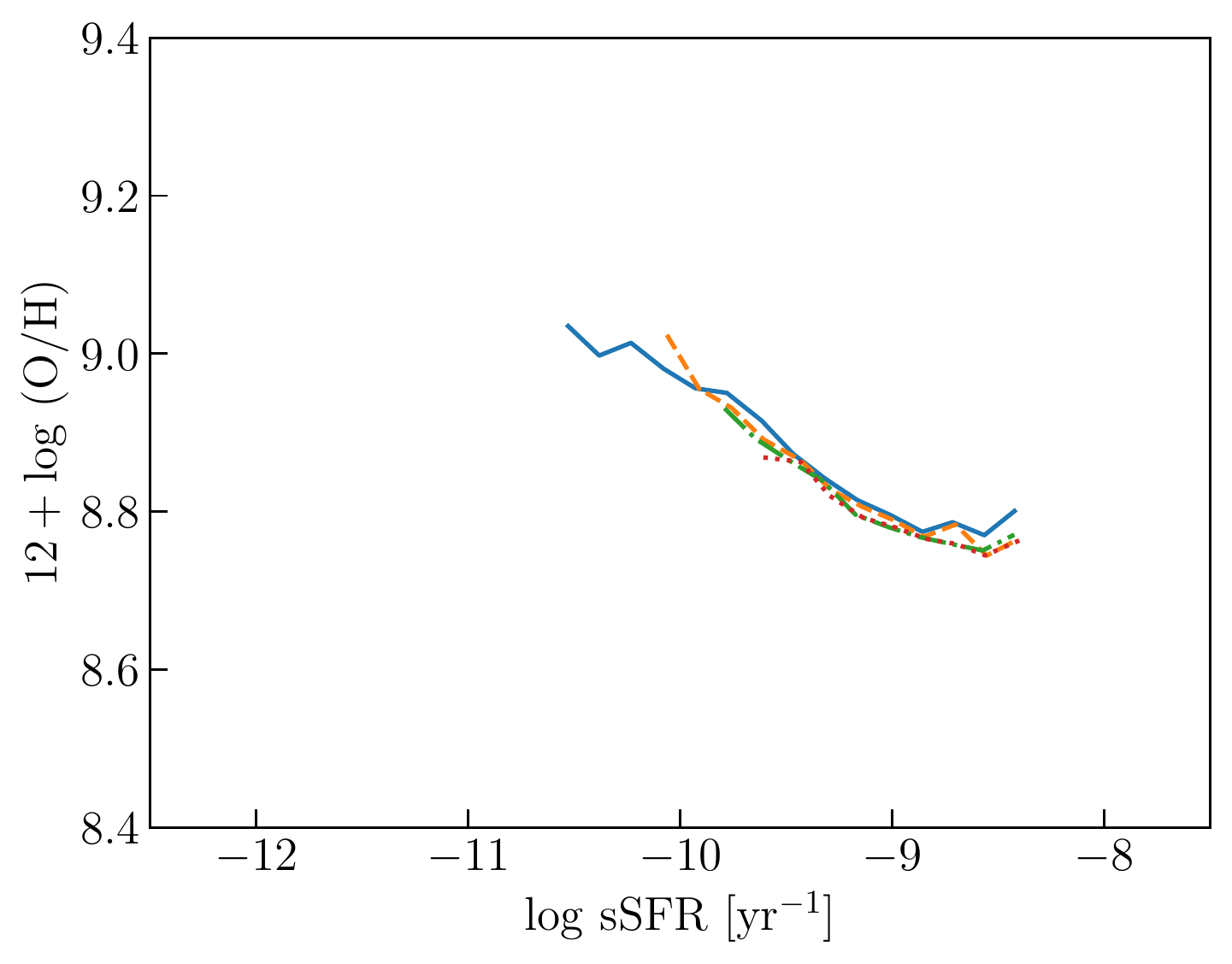}\includegraphics[width=8.5cm]{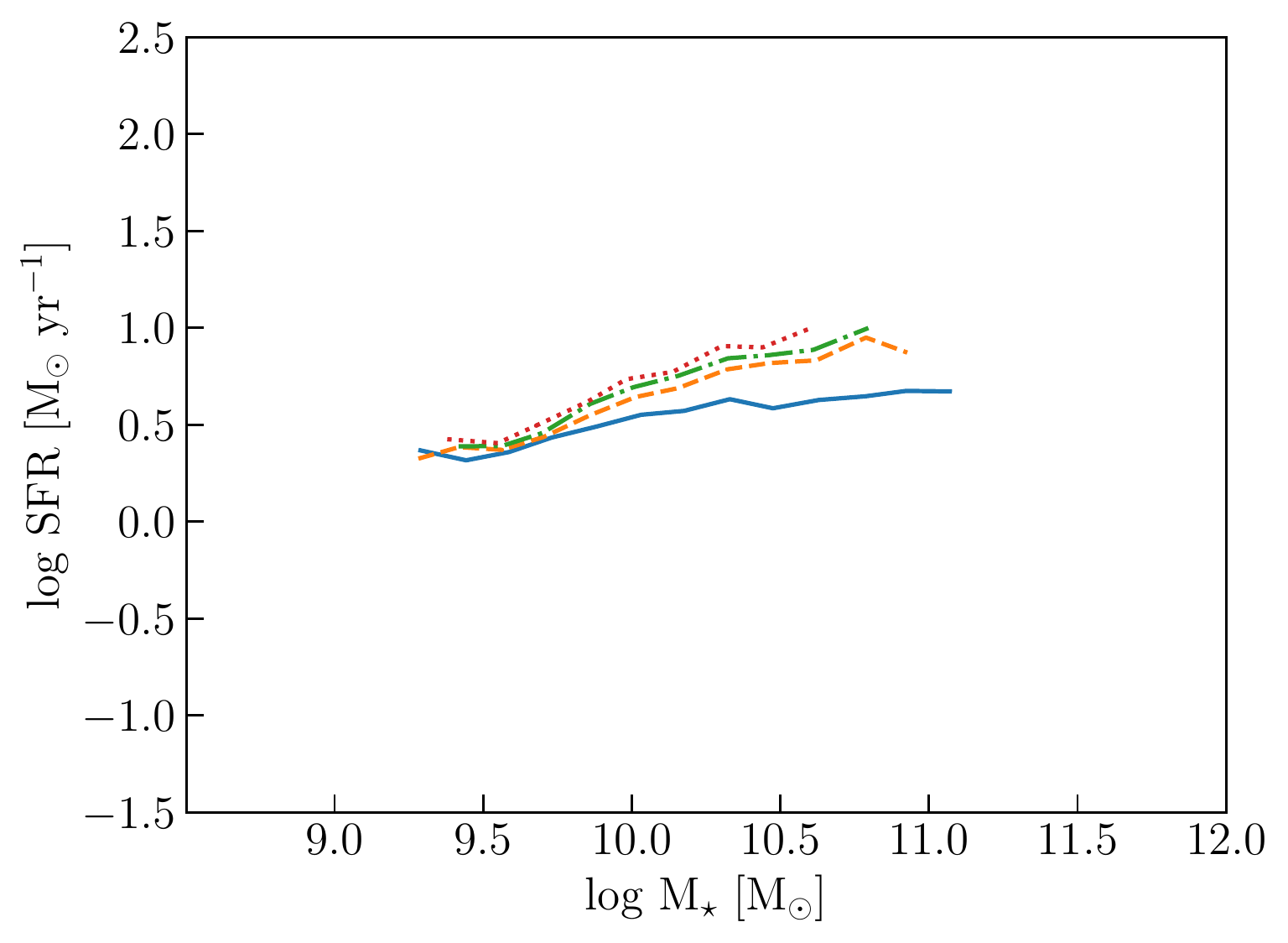}
    \caption{Effects of the flag selections on the projections of the FMR effects of flag selection (full sample: blue solid line; minimum flag equal to 1110: orange dashed line; minimum flag equal to 1111 dash-dotted line: green; minimum flag equal to 1112: red dotted line) for the VIPERS sample.}
    \label{fig:flagvipers}
\end{figure*}

Figure~\ref{fig:flagvipers} shows the effects on the FMR projections of the selections of the quality flag for the VIPERS sample. Each selection removes galaxies at high $\text{M}_\star$ and low SFR reducing the range explored in the corresponding projection. Another effect is to remove the metal-rich galaxies. This cut leads to flattening the MZR and the relation in the metallicity-SFR plane; while the relation in the metallicity-sSFR is not sensitive to these selections (it is only sensitive to the reduction of data at high $\text{M}_\star$ and low SFR). The main sequence (bottom right panel of Fig.~\ref{fig:flagvipers}) shows instead a shift towards the top of the diagram. The same shift is found in the case of the S/N selection (Fig.~\ref{fig:s/n_bias}). Although flattening of the relations is negligible compared to the error on metallicity ($\sim 0.1$ dex), the selection of the quality flag remains a possible source of a bias.

\begin{figure*}
    \centering
    \includegraphics[width=8.5cm]{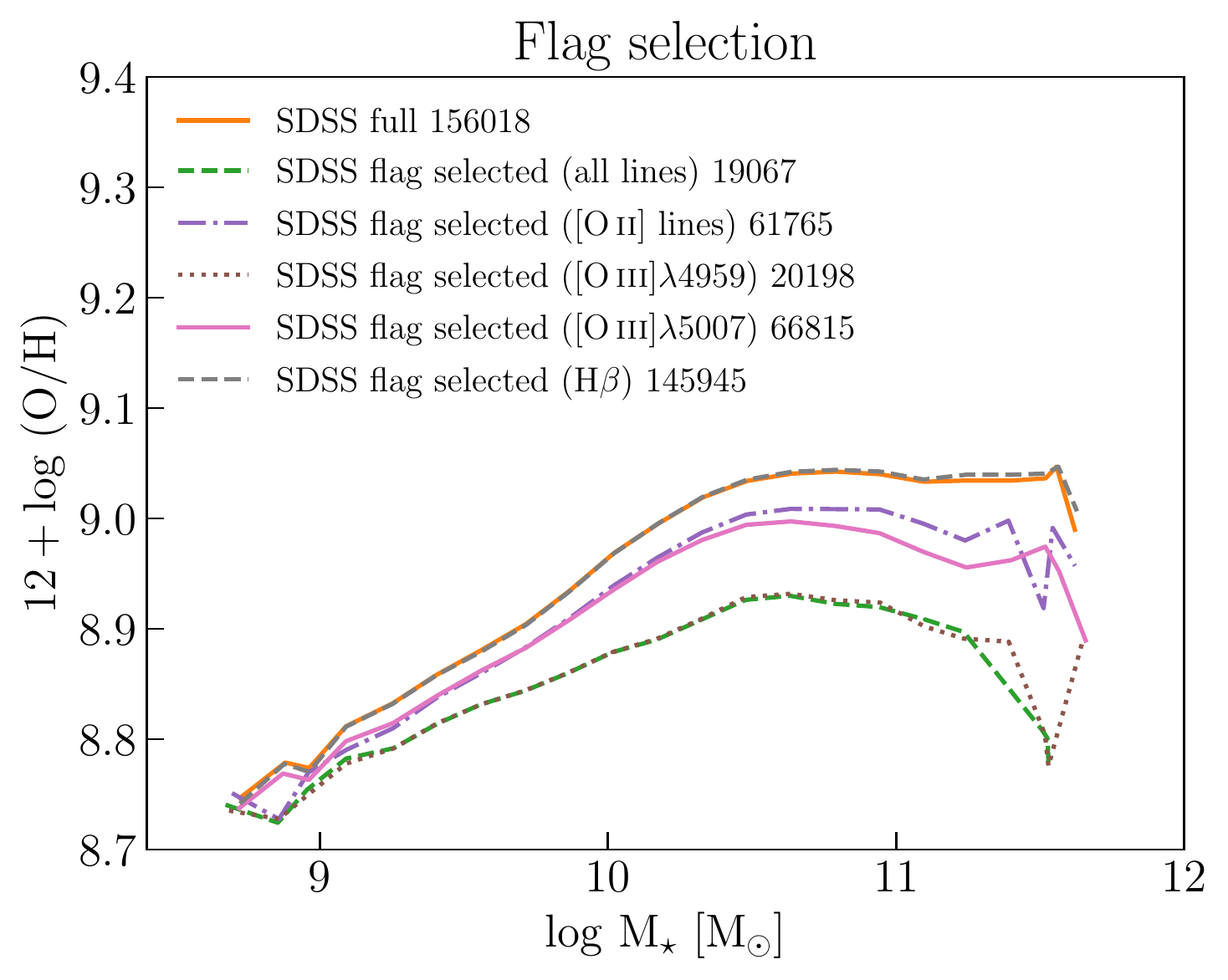}\includegraphics[width=8.5cm]{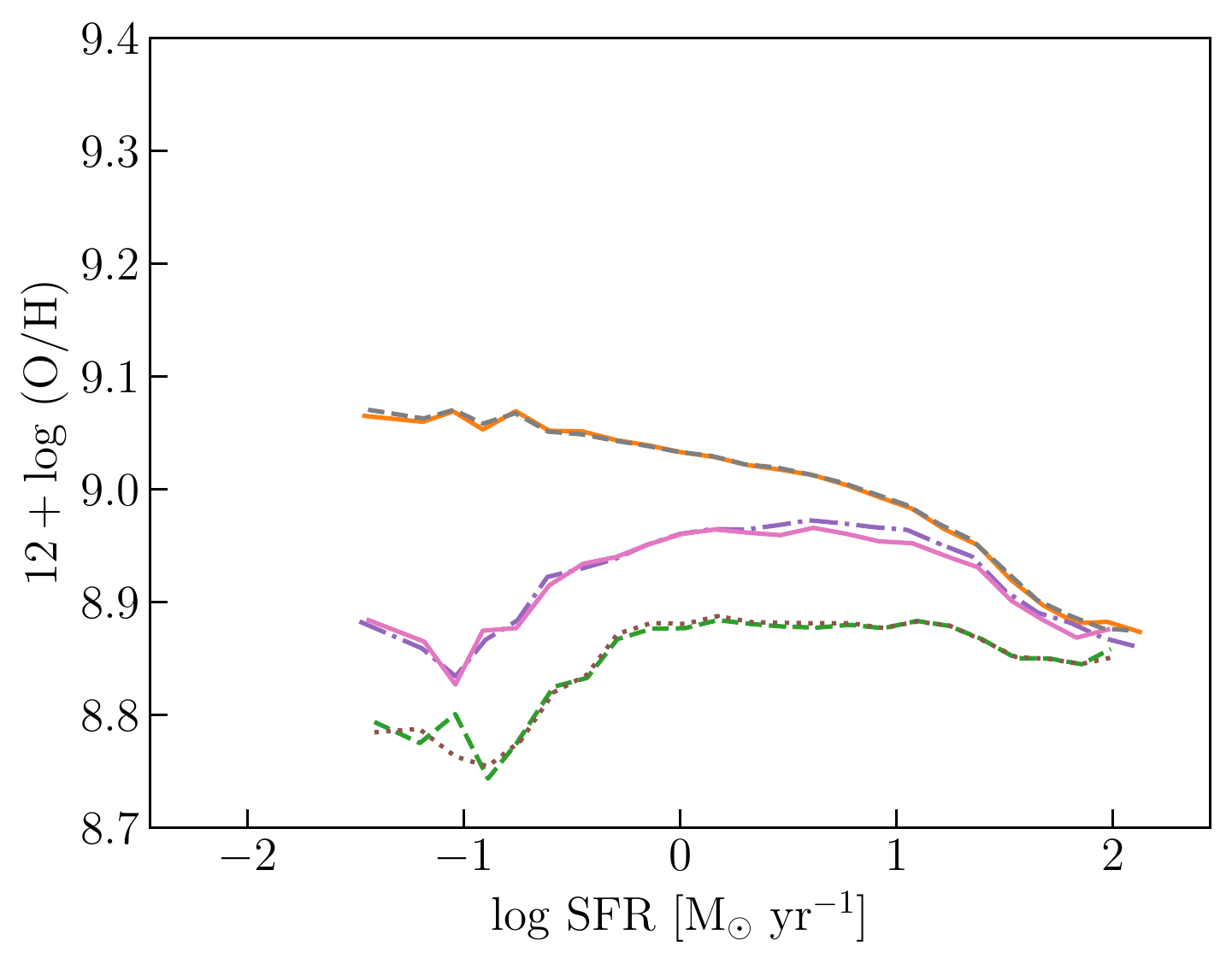}
    \includegraphics[width=8.5cm]{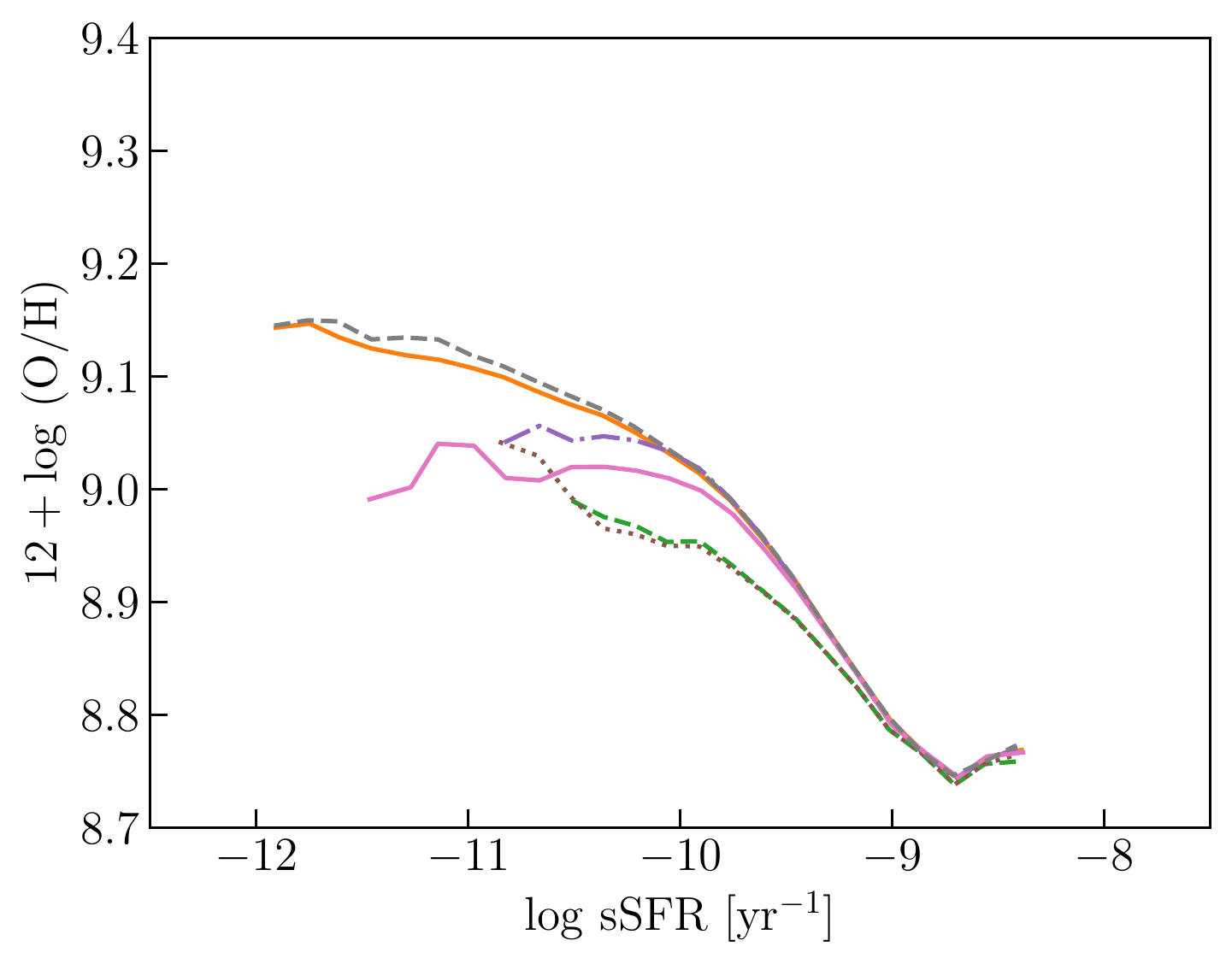}\includegraphics[width=8.5cm]{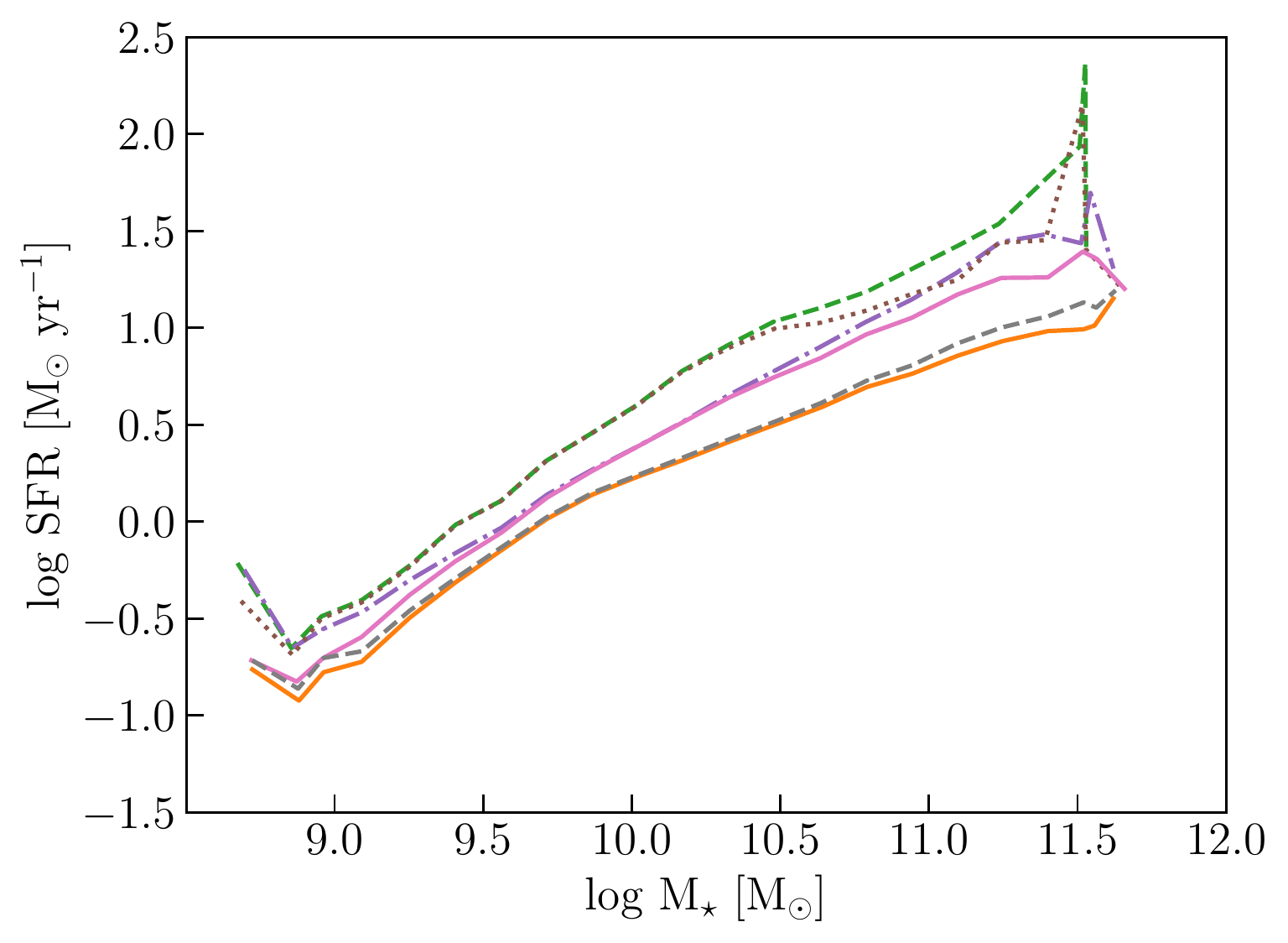}
    \caption{Effects of the selection on t-value flag for the emission lines (orange solid line: full sample; green dashed line: all lines; purple dash-dotted line: $\left[ \text{O{\,\sc{ii}}} \right]$; brown dotted line: $\left[ \text{O{\,\sc{iii}}} \right]\lambda 4959$; pink solid line: $\left[ \text{O{\,\sc{iii}}} \right]\lambda 5007$; grey dashed line: $\text{H}\beta$) on the projections for SDSS sample.}
    \label{fig:flag_bias}
\end{figure*}

Figure~\ref{fig:flag_bias} shows the projections of the FMR for the SDSS sample. The selection flattens the relations as already seen for the VIPERS sample. Thanks to the larger statistics of the SDSS sample and a wider interval in the SFR, it is possible to see that the selection on oxygen line flag leads to an inversion of the relation with the metallicity increasing at high SFR. The relation in the metallicity-sSFR seems also to be much more sensitive to this selection compared to the VIPERS sample.

The MZR is shifted toward lower metallicity when the selection is done on $\left[ \text{O{\,\sc{ii}}} \right]\lambda 3727$ and $\left[ \text{O{\,\sc{iii}}} \right]\lambda 5007$, namely, we mainly cut  the metal-rich galaxies at high $\text{M}_\star$es. The selection performed solely on $\left[ \text{O{\,\sc{iii}}} \right]\lambda 4959$ gives a further shift towards lower metallicities and a further flattening. This is the selection with the strongest effect. The selection on $\text{H}\beta$ only does not produce any difference. In this plane the median $\Delta \log \left( \text{O} / \text{H} \right) = 0.06$ dex and a maximum $\Delta \log \left( \text{O} / \text{H} \right) = 0.12$ dex in the case of the selection on all the emission lines together.

In the plane metallicity versus SFR, the bias affects the relation with a minimum around $\log \text{SFR}  \left[ \text{M}_\sun \text{ yr}^{-1} \right] \sim -1$ and then remains flat with the selection on $\left[ \text{O{\,\sc{ii}}} \right]\lambda 3727$ and $\left[ \text{O{\,\sc{iii}}} \right]\lambda 5007$, that is, we cut mainly metal-rich galaxies at lower SFRs. The selection on $\left[ \text{O{\,\sc{iii}}} \right]\lambda 4959$ shifts further the relation towards lower metallicity. Again, the selection on $\text{H}\beta$ only does not produce any difference. In this plane the median $\Delta \log \left( \text{O} / \text{H} \right) = 0.14$ dex and a maximum $\Delta \log \left( \text{O} / \text{H} \right) = 0.30$ dex in the case of the selection on all the emission lines together. Again, this plane is more sensitive to the selection than the MZR.

In this case, the metallicity versus the sSFR plane is not anymore insensitive but all the selections move in the same way as other projections with the selection on $\left[ \text{O{\,\sc{ii}}} \right]\lambda 3727$ and $\left[ \text{O{\,\sc{iii}}} \right]\lambda 5007$ that cuts mainly metal-rich galaxies at lower sSFR and the selection on $\left[ \text{O{\,\sc{iii}}} \right]\lambda 4959$ line is the strongest bias. The main sequence is shifted toward the top left part of the diagram when the selection on all lines is applied.

\begin{figure}
    \centering
    \resizebox{\hsize}{!}{\includegraphics{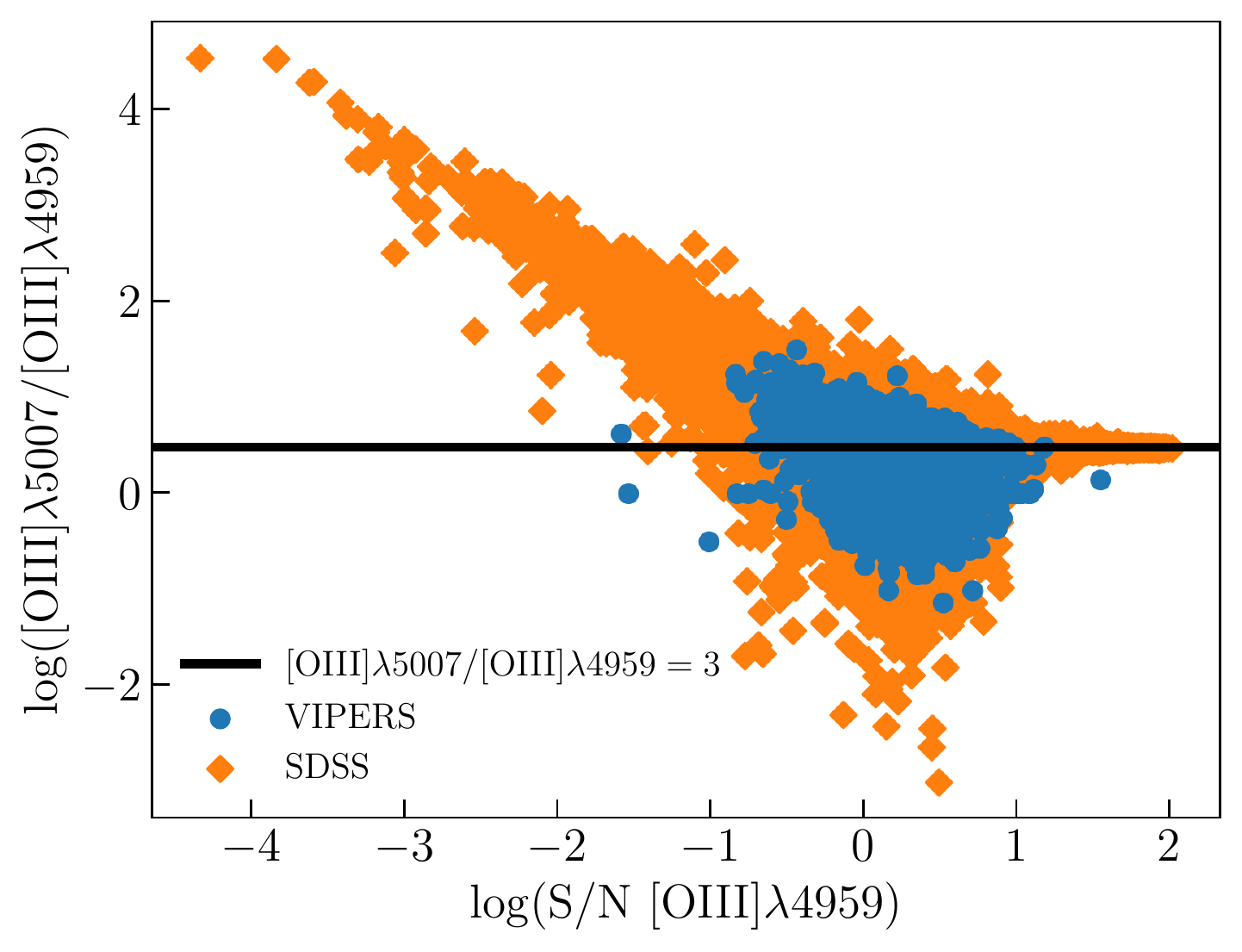}}
    \caption{Line ratio $\left[ \text{O{\,\sc{iii}}} \right]\lambda 5007 / \left[ \text{O{\,\sc{iii}}} \right]\lambda 4959$ vs S/N of the line $\left[ \text{O{\,\sc{iii}}} \right]\lambda 4959$ for VIPERS (blue dot) and SDSS (orange diamonds) samples. The solid black line show the intrinsic value of the line ratio.}
    \label{fig:study_oiii}
\end{figure}

To understand the reason why the effect is mainly due to the line $\left[ \text{O{\,\sc{iii}}} \right]\lambda 4959$, we studied the relation between the ratio of the $\left[ \text{O{\,\sc{iii}}} \right]$ doublet, $\left[ \text{O{\,\sc{iii}}} \right]\lambda 5007 / \left[ \text{O{\,\sc{iii}}} \right]\lambda 4959$, and the S/N of the line $\left[ \text{O{\,\sc{iii}}} \right]\lambda 4959$. The results are shown in Fig.~\ref{fig:study_oiii}. This plot shows that following the data selection used by \cite{curti2020mass} it is not possible to remove galaxies with very low S/N ($\log \text{S/N} < -1$) of $\left[ \text{O{\,\sc{iii}}} \right]\lambda 4959$.

\subsection{\texorpdfstring{$\text{B}-\text{B}^*$}{} selection effect}

The third bias analyzed here involves the range in absolute blue magnitude $\text{B}-\text{B}^*$, where $\text{B}^*$ is the characteristic magnitude at which the luminosity function changes dependence (from power law to exponential). In this way, we take into account the evolution of the luminosity function itself \citep[$\text{B}^*=-20.95$ for VIPERS and $\text{B}^* = -19.11$ for SDSS, these values are used for the whole redshift range,][]{ilbert2005vimos, ilbert2006vimos}. Then, we can observe the same luminosity interval of the distribution after taking into account its shift due to the redshift.

\begin{figure}
    \centering
    \resizebox{\hsize}{!}{\includegraphics{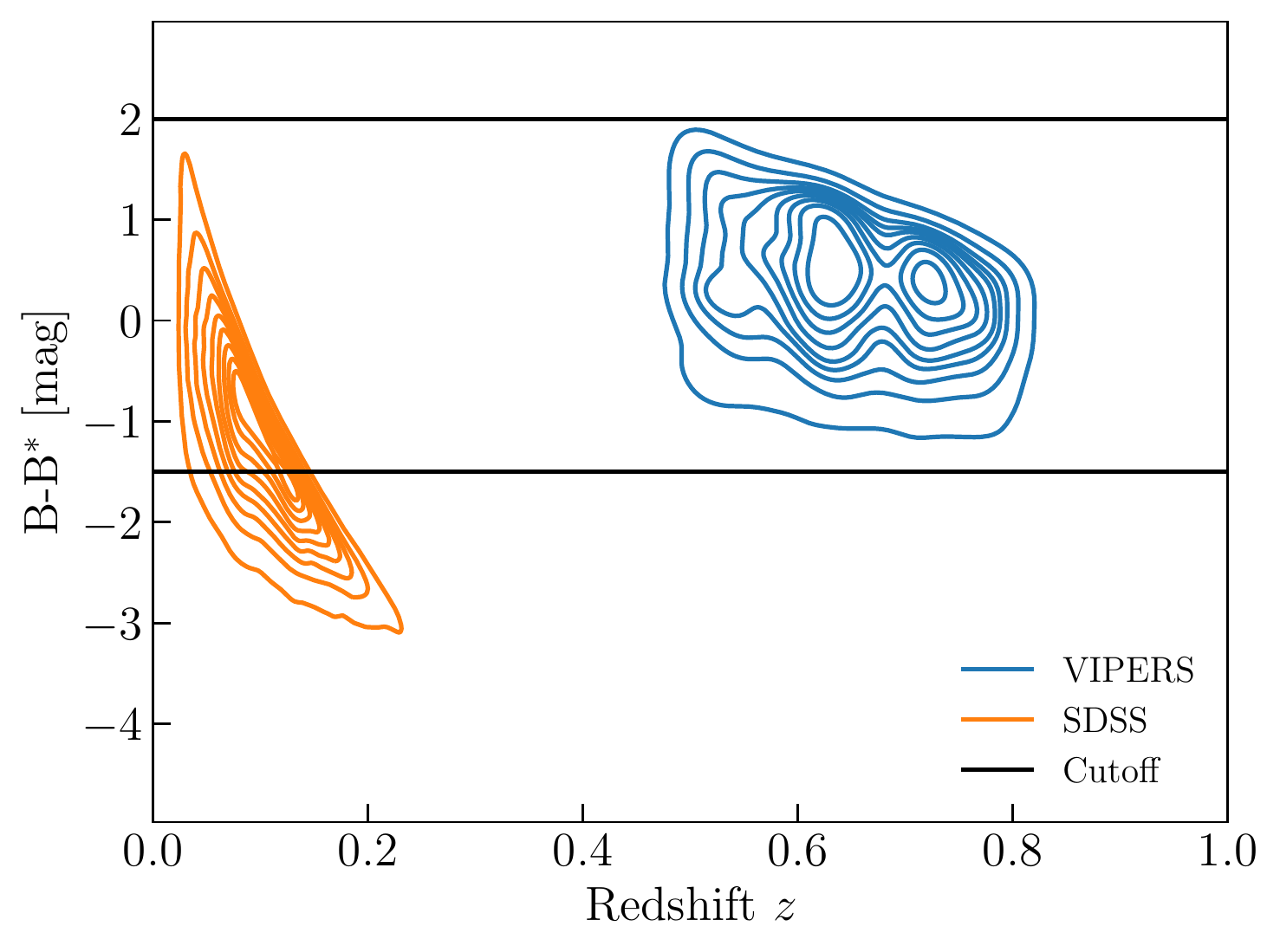}}
    \caption{$\text{B}-\text{B}^*$ vs redshift diagram and cutoff (black solid line) for the VIPERS (blue) and the SDSS (orange) samples.}
    \label{fig:bmagcut}
\end{figure}

Figure~\ref{fig:bmagcut} shows the distributions of both samples in $\text{B}-\text{B}^*$ with the cutoffs at $-1.5 \leq \text{B} - \text{B}^* \leq 2$ mag. $\text{B}^*$ is defined as the magnitude at which the luminosity function changes dependence (from power law to exponential) and it is redshift-dependent. In this way, we can analyze the same interval of the luminosity function for both samples.

\begin{figure*}
    \centering
    \includegraphics[width=8.5cm]{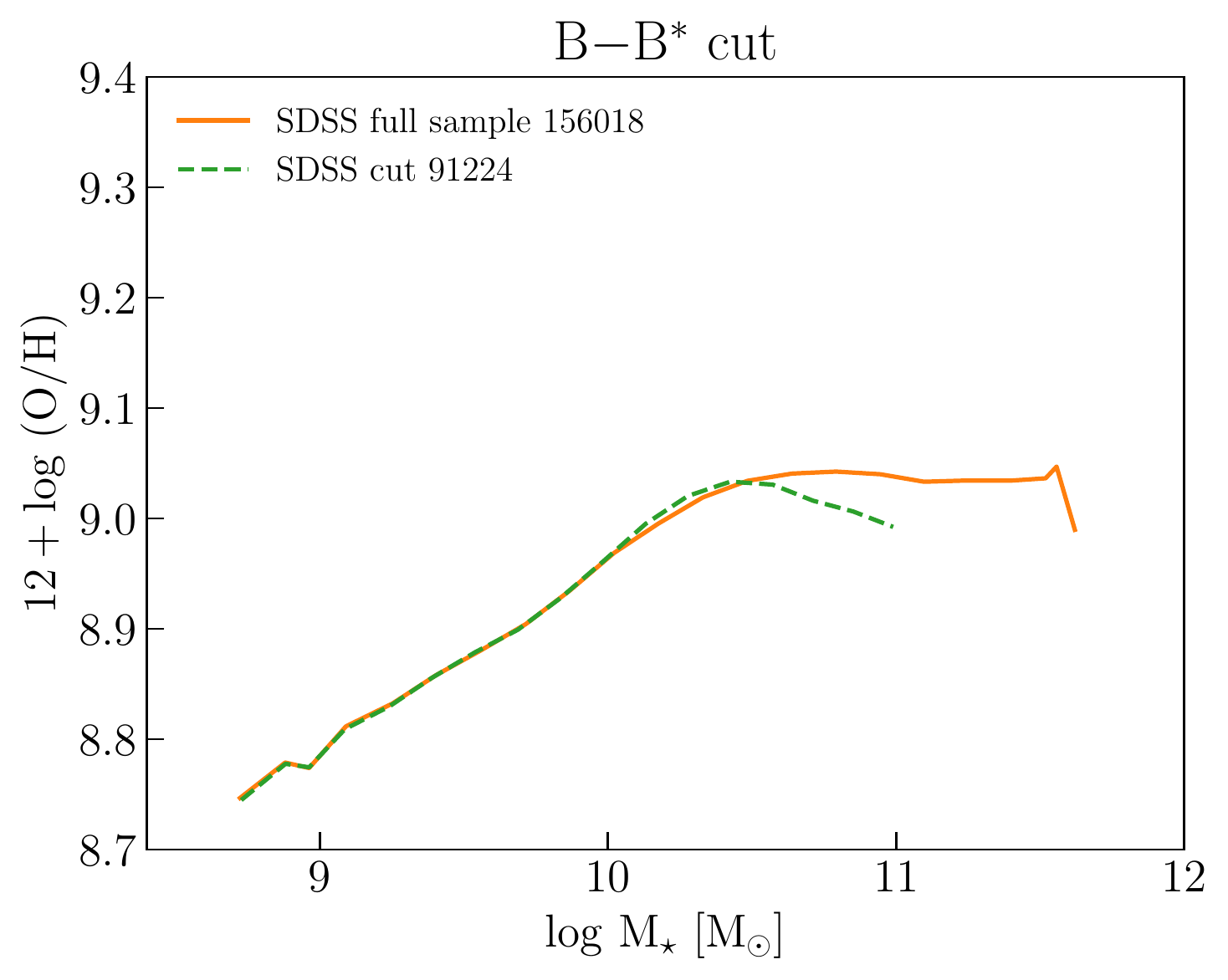}\includegraphics[width=8.5cm]{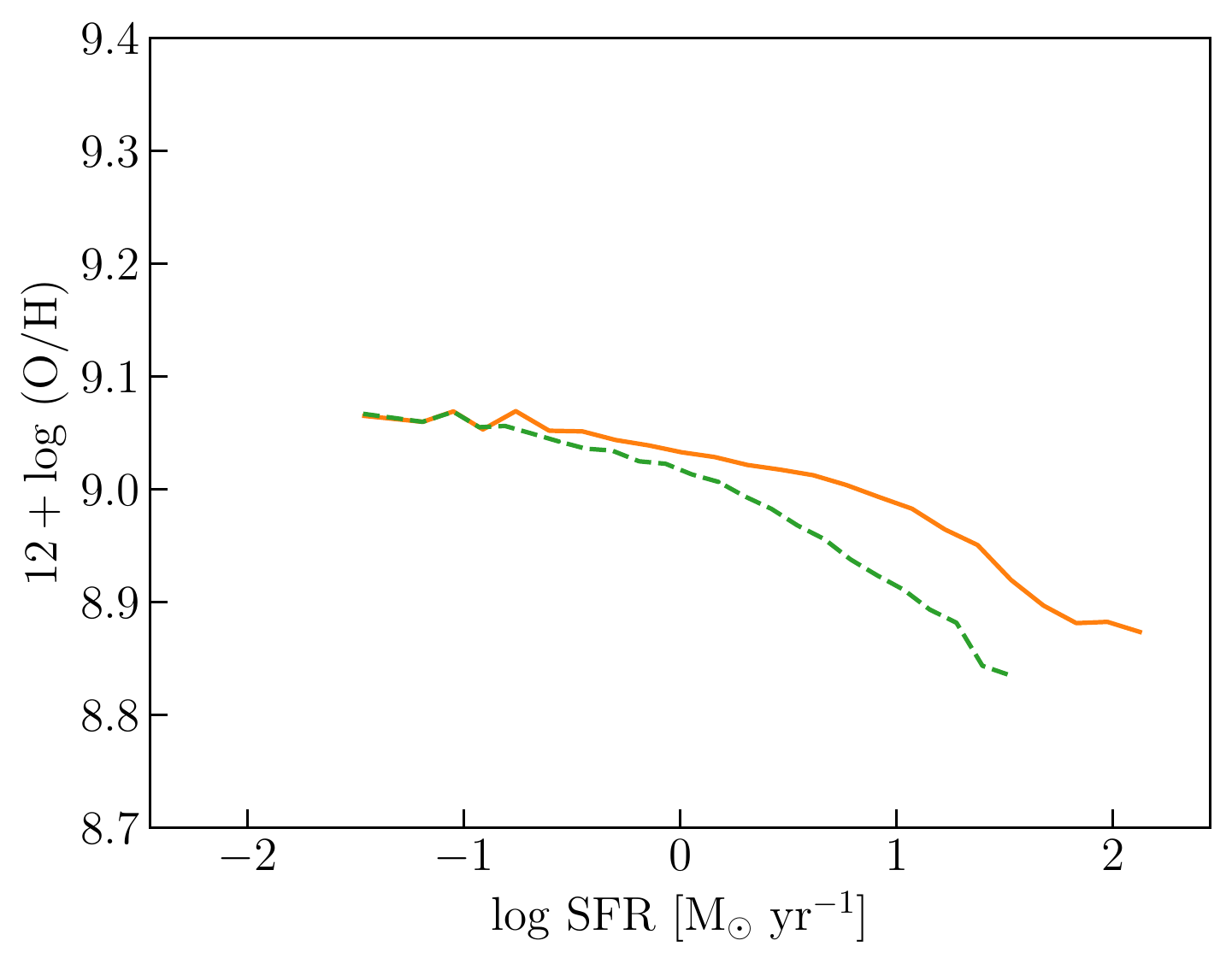}
    \includegraphics[width=8.5cm]{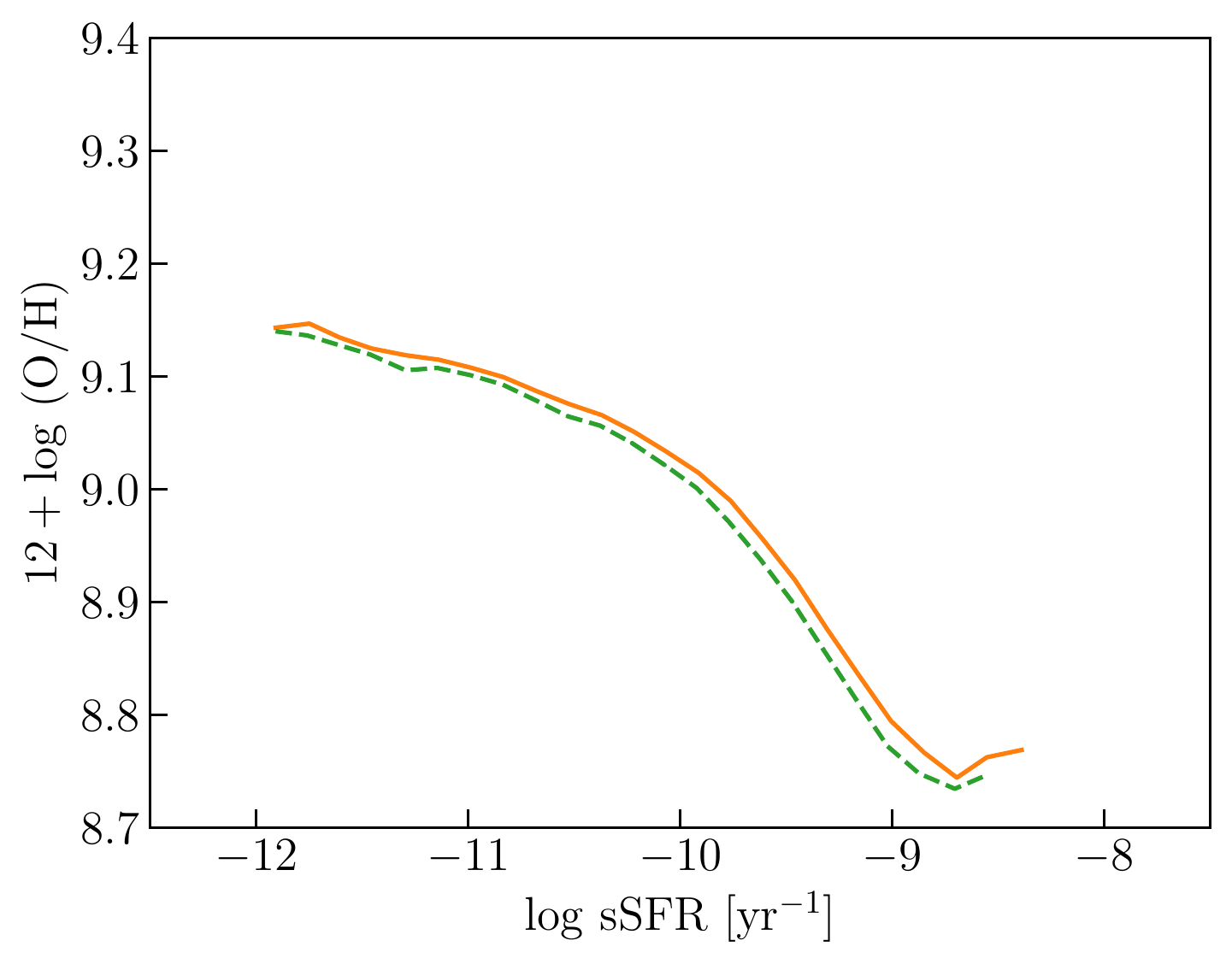}\includegraphics[width=8.5cm]{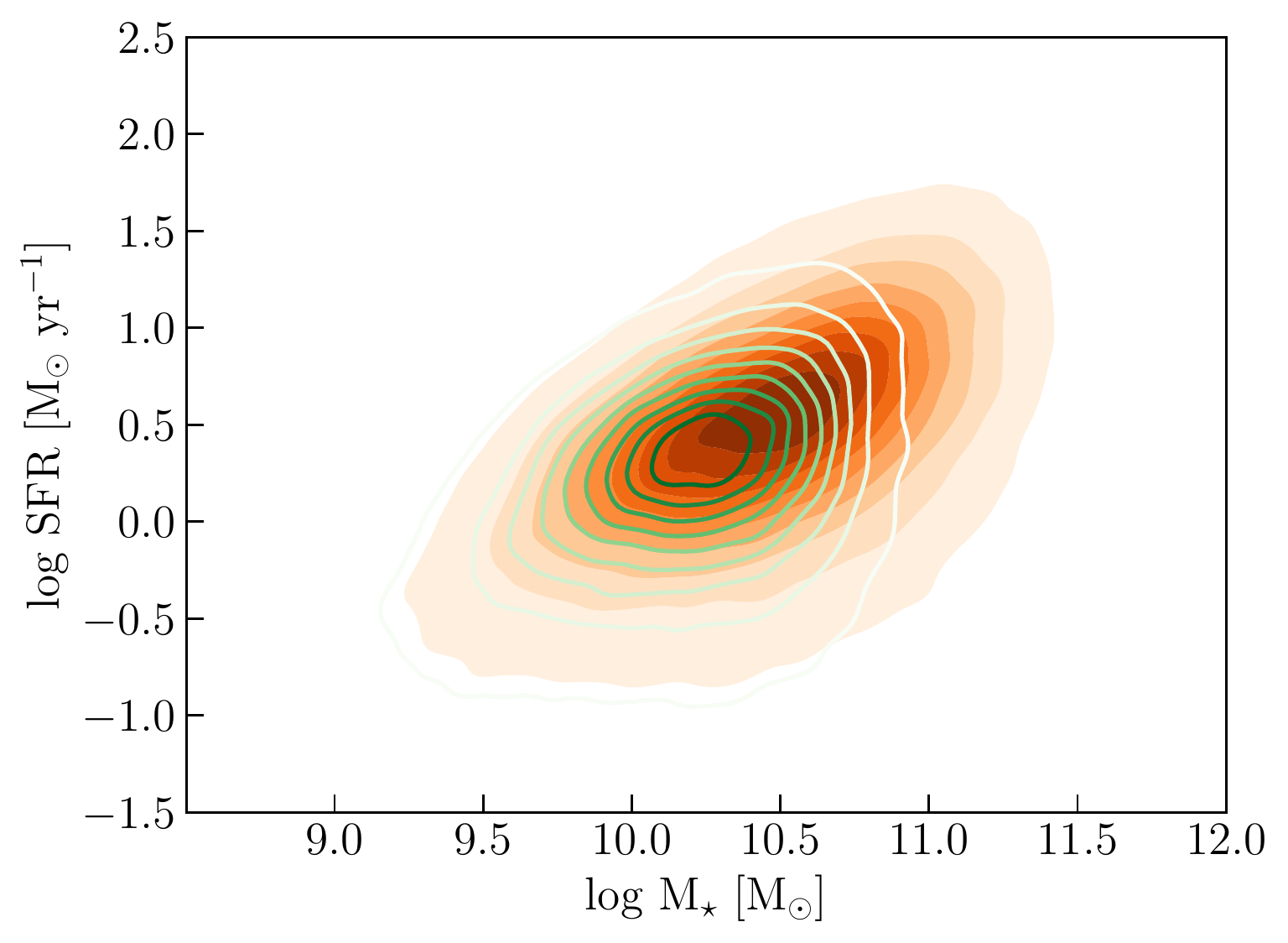}
    \caption{Effects of the selection on $\text{B}-\text{B}^*$ on the projections of the FMR (full sample: orange solid line; same luminosity volume than VIPERS: green dashed line).}
    \label{fig:bbstar_bias}
\end{figure*}

Figure~\ref{fig:bbstar_bias} shows the comparison of projections of the FMR between the full sample of SDSS data and after the cut. The MZR and the metallicity versus sSFR are insensitive to the selection on luminosity; while in the metallicity versus the SFR plane, the bias mainly cut the metal-rich galaxies at high SFR. In this plane the median $\Delta \log \left( \text{O} / \text{H} \right) = 0.01$ dex and a maximum $\Delta \log \left( \text{O} / \text{H} \right) = 0.03$ dex. The main sequence is moved towards the bottom left part of the diagram.

\subsection{Effect of the fraction of blue galaxies}

The last bias we analyze here is the fraction of blue galaxies selected on the distributions of the sSFR. There are two effects at work here: i) the observation of blue galaxies is reduced with the redshift since this type are mainly low-mass and are cut at high redshift because of the limited magnitude of the observations; ii) blue galaxies can be over-selected at high redshift --- namely, the VIPERS i-band selection translates to a B-band selection at high-z and for galaxies with bright emission lines is easier to estimate their metallicity. We wanted to check if the second point can introduce an observational bias in the studies on the FMR. We may also have introduced a bias with the redshift confidence level. This is because blue galaxies have brighter strong emission lines which are used to measure the spectroscopic redshift. We selected the blue galaxies via the sSFR distributions (VIPERS: ${-12 \leq \log \text{sSFR} \left[ \text{ yr}^{-1} \right] \leq -8}$; SDSS: ${-15 \leq \log \text{sSFR} \left[ \text{ yr}^{-1} \right] \leq -8}$).

We defined the fraction of blue galaxies as the ratio between the number of sources inside the sub-sample and the number of sources inside the full catalog:
\begin{equation}
    f_B = \frac{\text{N}_\text{sub-sample} \left( \text{M}_\star, \text{SFR} \right)}{\text{N}_\text{full catalog} \left( \text{M}_\star, \text{SFR} \right)}
.\end{equation}
We estimate the error on the fraction of blue galaxies as the propagation of the Poissonian errors of the counts. The relations between the fraction versus the $\text{M}_\star$ and SFR can be found in Appendix~\ref{app:fraction}.

We used the full catalog without doing any kind of selection (no selection on S/N of lines or selection based on BPT diagram) for VIPERS ($75\,369$ sources with a corresponding average confidence level between $50\%$ and $99\%$ for the redshift) and the full SDSS sample ($536\,140$ sources). We also cut the SDSS sample to have the same fraction of blue galaxies (in mass bin of the VIPERS sample). We are aware that we might introduce a bias on the possible evolution with redshift. However, we want to check if the ease of observing galaxies with brightest emission lines can produce an overselection of these kinds of sources.

\begin{figure*}
    \centering
    \includegraphics[width=8.5cm]{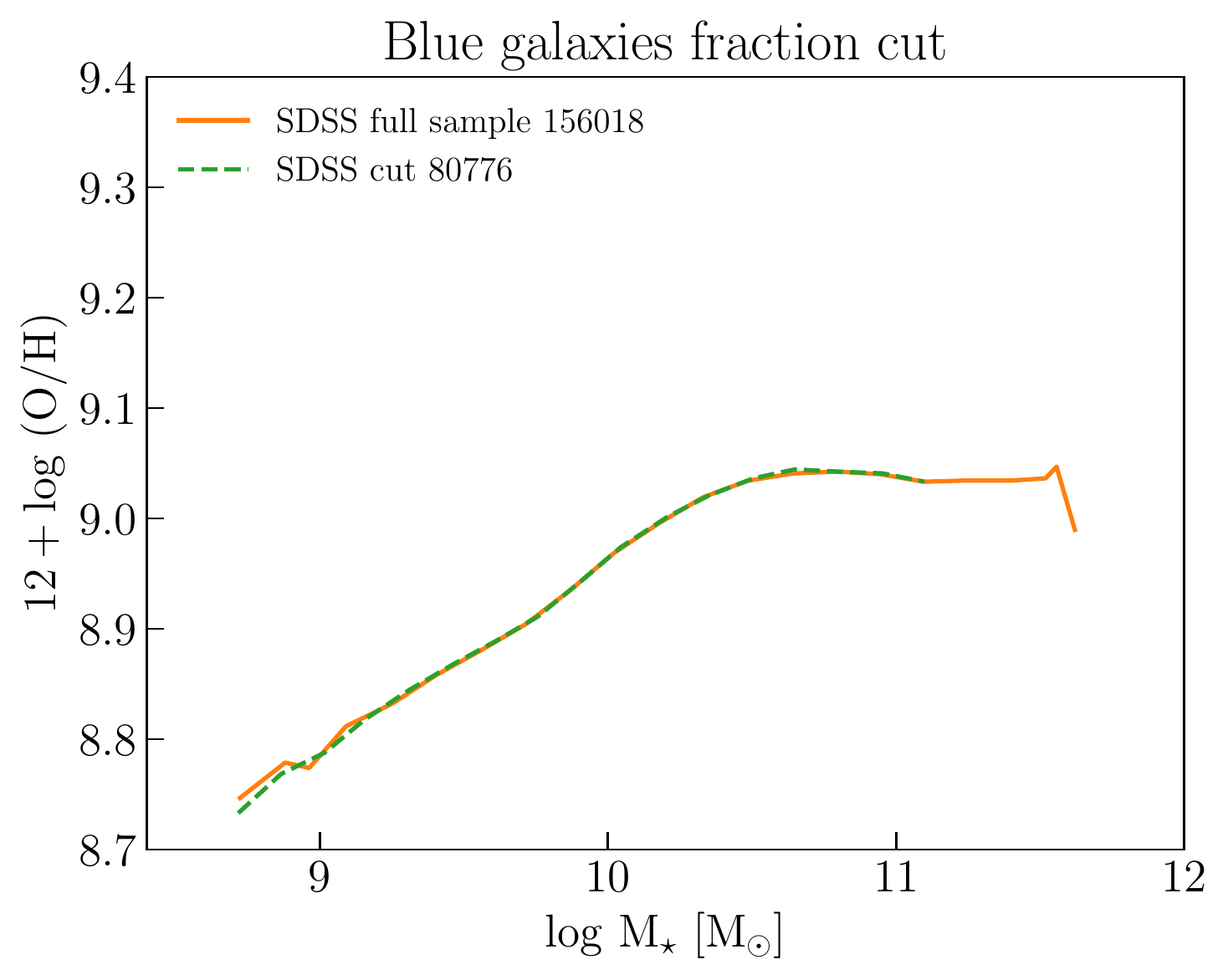}\includegraphics[width=8.5cm]{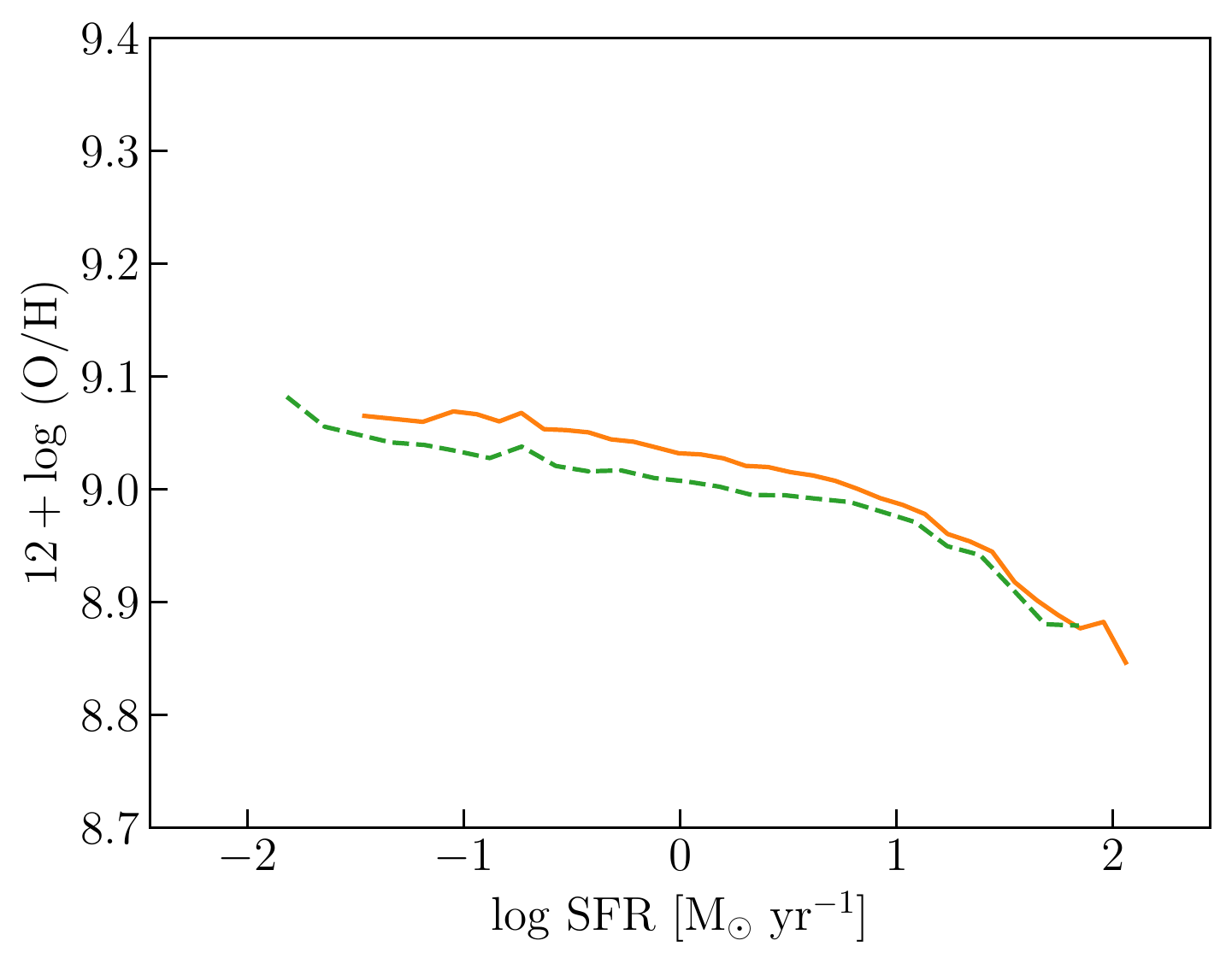}
    \includegraphics[width=8.5cm]{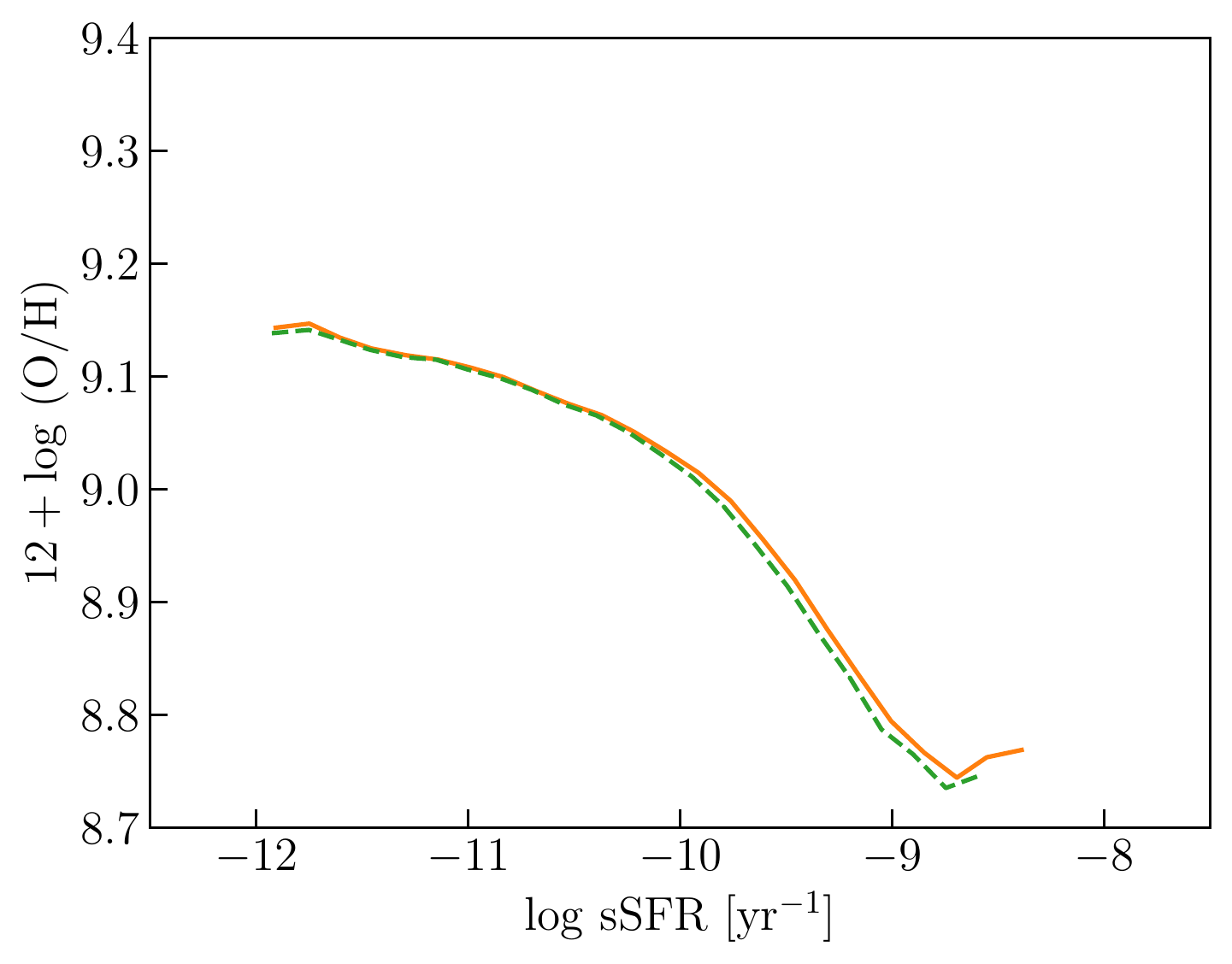}\includegraphics[width=8.5cm]{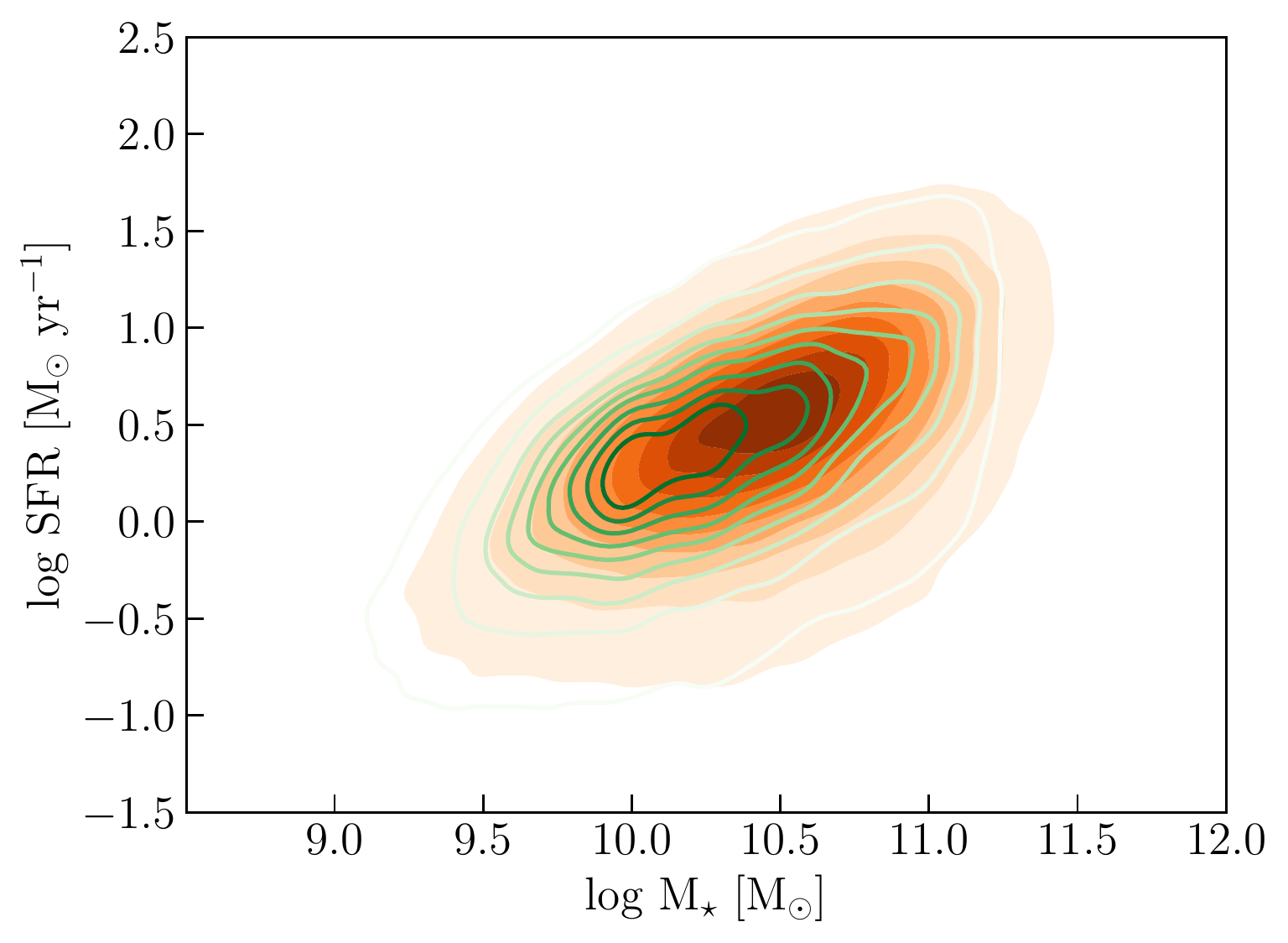}
    \caption{Effects of the fraction of blue galaxies on the projections of the FMR (full sample: orange solid line; same fraction on $\text{M}_\star$ than VIPERS: green dashed line).}
    \label{fig:bluefraction_bias}
\end{figure*}

Figure~\ref{fig:bluefraction_bias} shows the projections of the FMR. As for the luminosity selection, the MZR and the metallicity versus the sSFR are insensitive to the selection on the fraction of blue galaxies; while in the metallicity versus the SFR plane, the relation is shifted toward the bottom at $\log \text{SFR}  \left[ \text{M}_\sun \text{ yr}^{-1} \right] \leq 0$. The main sequence shows a significant shift towards the bottom left part of the diagram.
The independence on the fraction of blue galaxies can be due to the fact that the different fractions do not imply a redshift evolution or this effect is compensated by the observational effect.

\section{Comparison between VIPERS and SDSS samples}\label{sect:comparison}

Finally, we selected a sub-sample of the SDSS sample with the same characteristics as the VIPERS sample in terms of S/N, $\text{B}-\text{B}^*$ as well as the fraction of blue galaxies. These cuts are made in this order to avoid breaking the reconstruction of the VIPERS fraction of blue galaxies. In this case, the fraction of blue galaxies is just reduced, while the shape in function of $\text{M}_\star$ and SFR does not change. We did not apply any selection on the quality of the spectra because it would  completely change the shape of the relations, especially of the projections. The SDSS sub-sample equivalent to the VIPERS sample is composed of $60\,614$ SF galaxies ($\sim 39 \%$ of the full sample).

\begin{figure*}
    \centering
    \includegraphics[width=5.6cm]{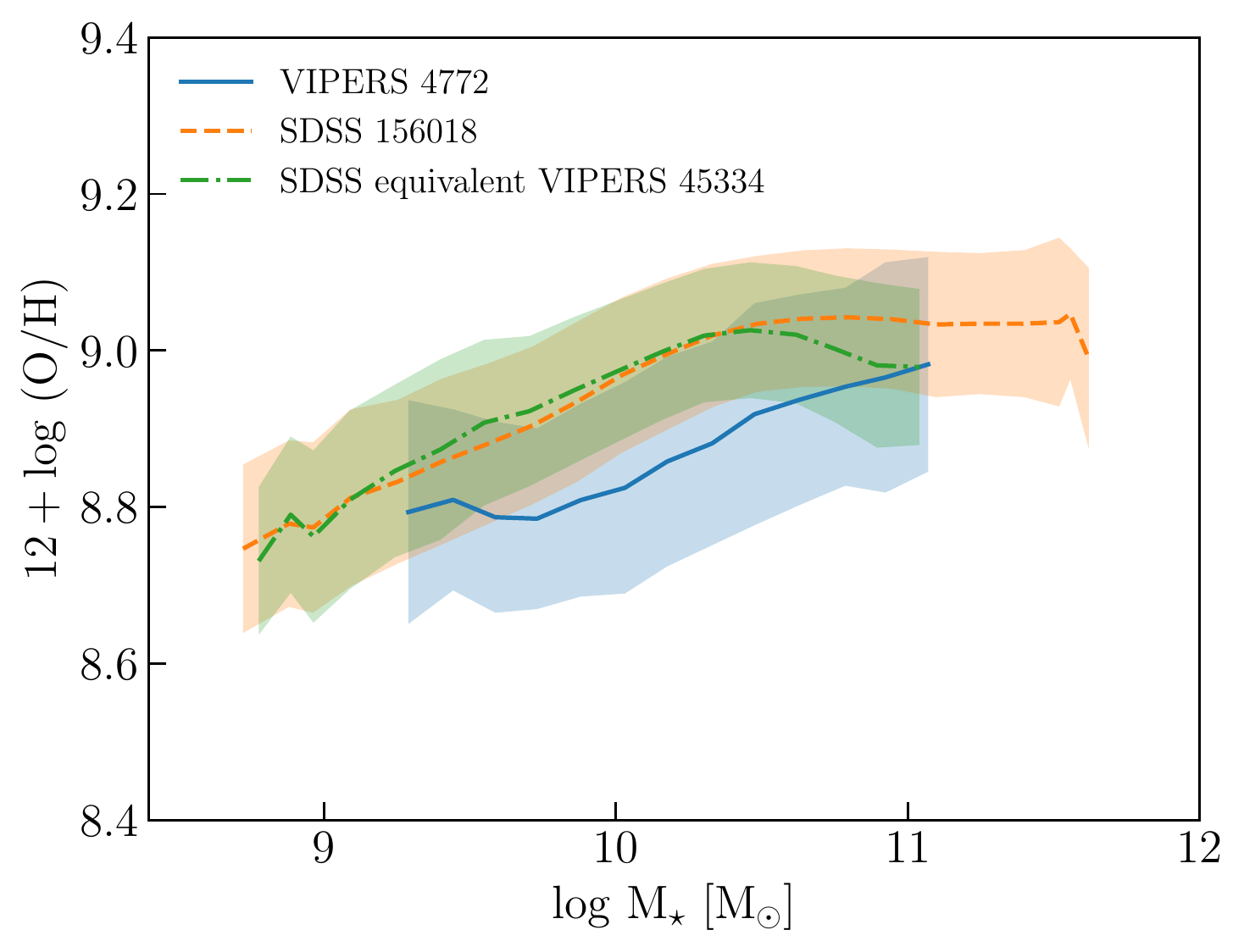}\includegraphics[width=5.6cm]{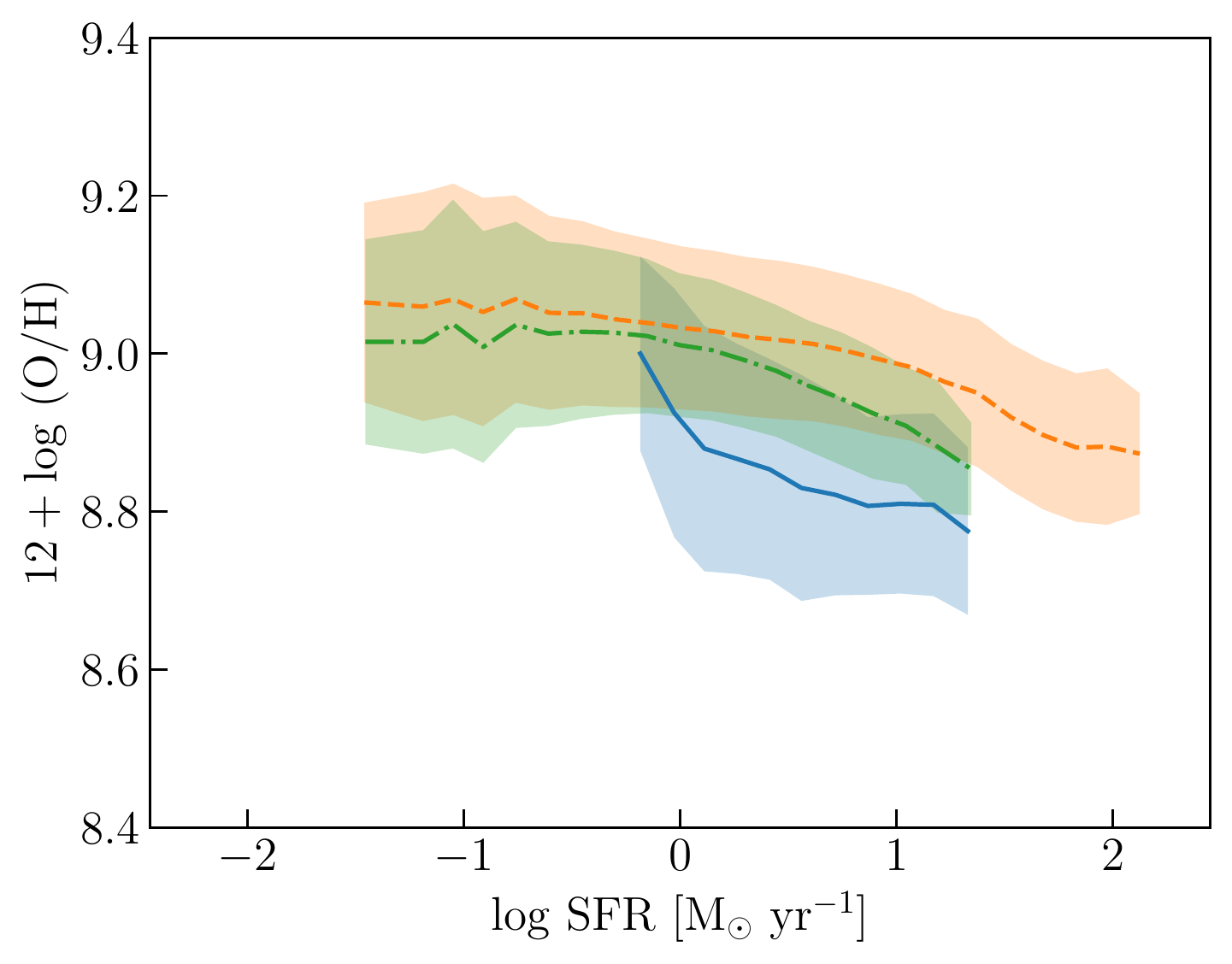}\includegraphics[width=5.6cm]{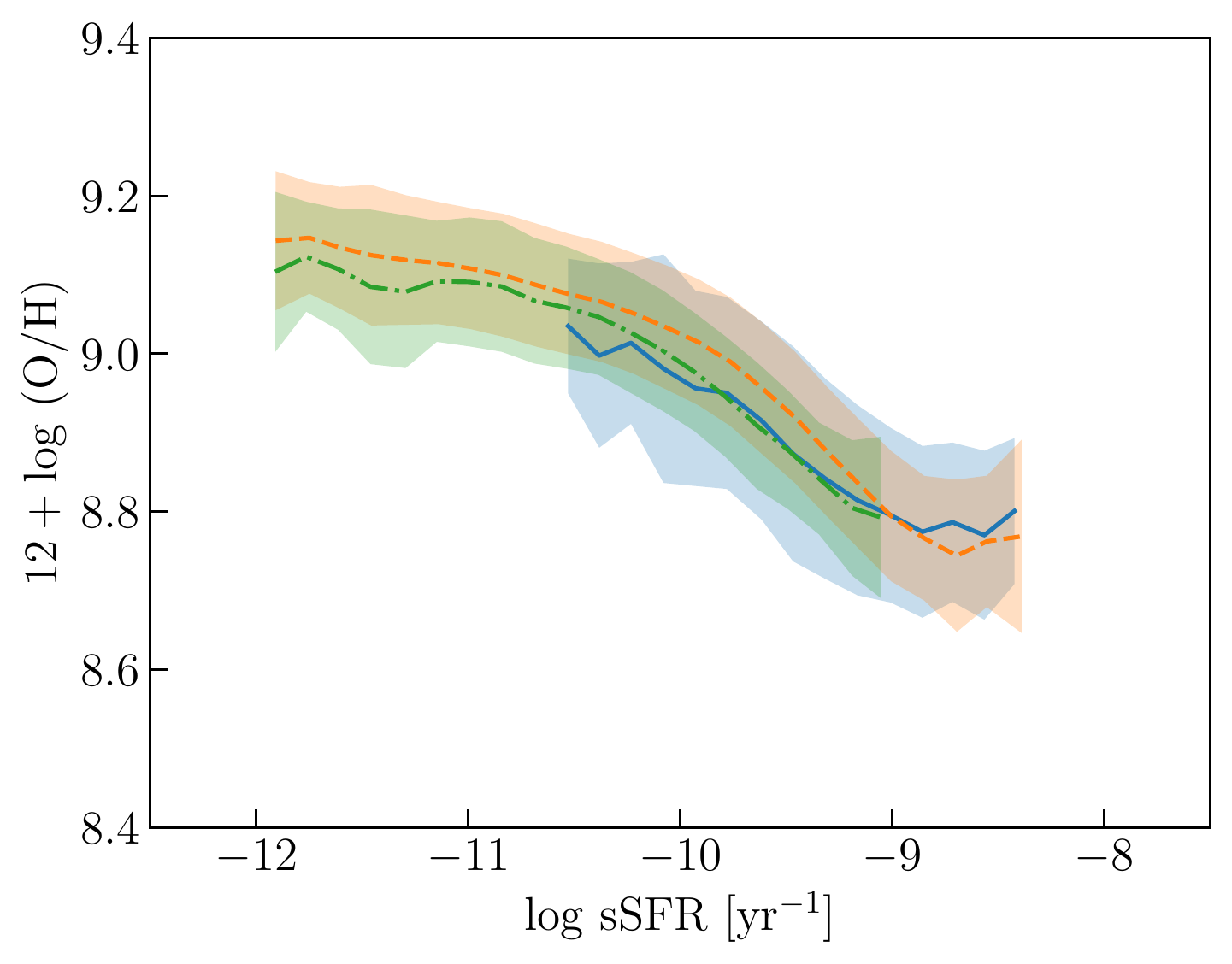}
    \caption{Three projections of the FMR: MZR (left), metallicity vs SFR (mid right), metallicity vs sSFR (right) for VIPERS (blue solid line), SDSS (orange dashed line), and SDSS equivalent to VIPERS (green dash-dotted line) samples.}
    \label{fig:comp_fmr_eqvipers}
\end{figure*}

Figure~\ref{fig:comp_fmr_eqvipers} shows the comparisons of the FMR projections: MZR, metallicity versus SFR, metallicity versus sSFR, and the main sequence for the VIPERS, the full SDSS, and SDSS after all the cuts to reproduce the characteristics of the VIPERS sample. The MZR of the VIPERS sample does not show the characteristic flattening at high $\text{M}_\star$, while it shows this behavior at low $\text{M}_\star$. In this projection, the biases, once taken into account, do not have any effects. In the metallicity versus SFR plane, the VIPERS sample has lower metallicity than SDSS at the same SFR. Once the biases taken into account, the relation for SDSS gets close to the one of VIPERS in the medium-high range covered from the latter. Finally, in the metallicity versus the sSFR plane, the two samples are in the closest agreement and the biases do not have any significant effect.

\begin{figure*}
    \centering
    \includegraphics[width=8.5cm]{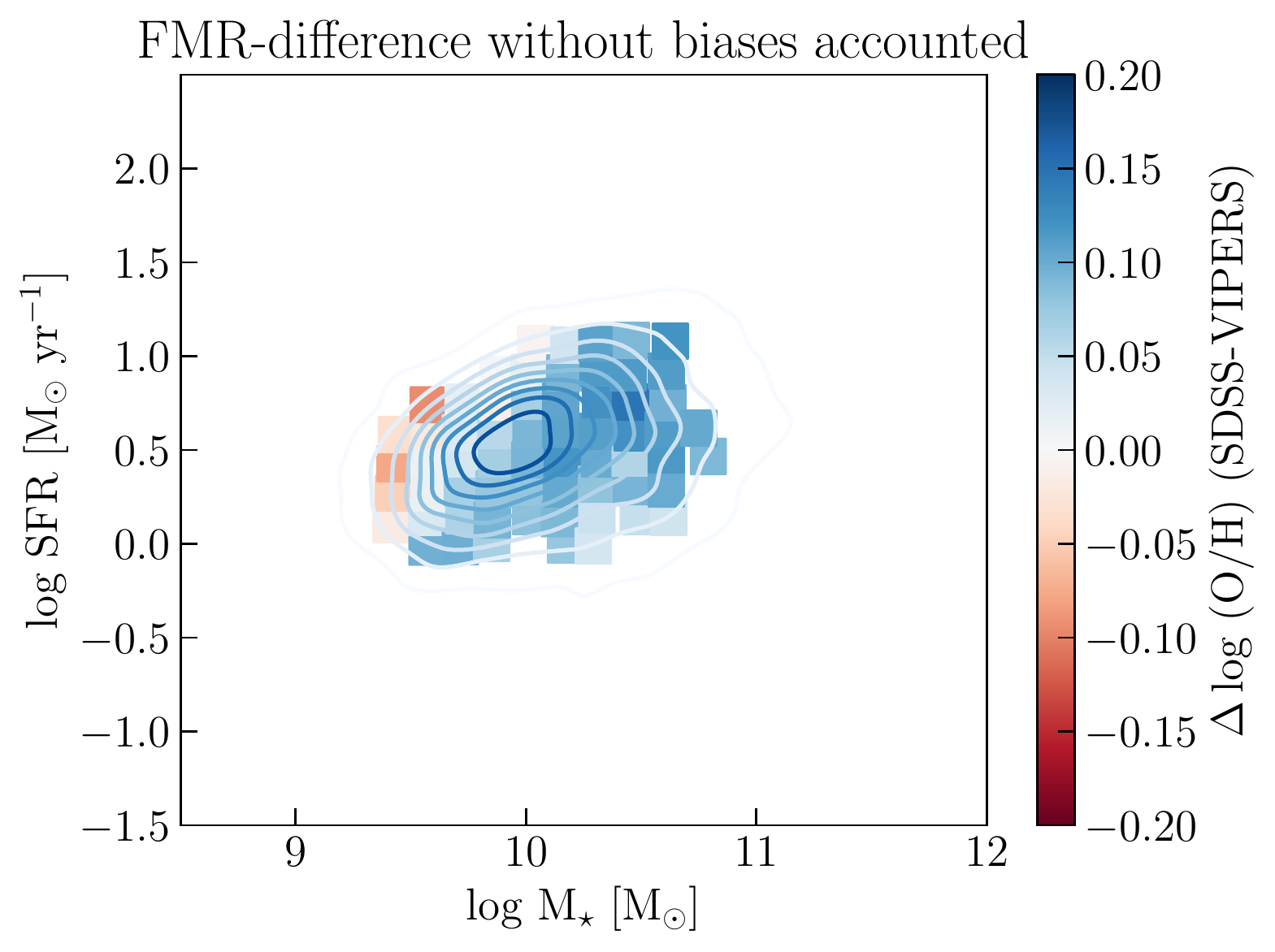}\includegraphics[width=8.5cm]{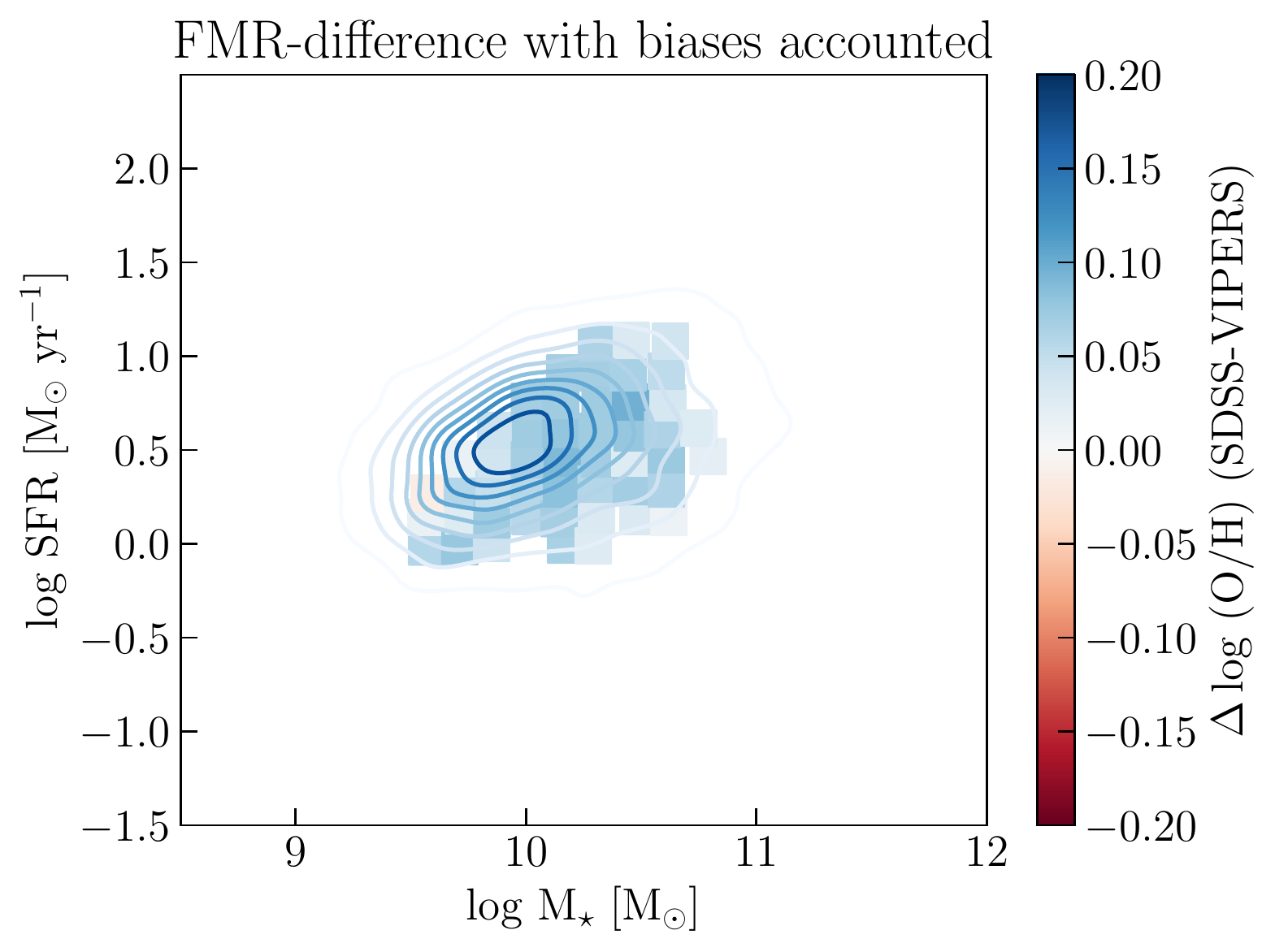}
    \caption{Difference in metallicity between SDSS without (left) and with (right) biases accounted and VIPERS projected on the main sequence of VIPERS sample (KDE contour plot).}
    \label{fig:ms_diff}
\end{figure*}

Figure~\ref{fig:ms_diff} represents the metallicity difference, color-coded according to the difference between SDSS and VIPERS sample, on the MS relation. This is the most direct comparison between the two samples. The difference between them increases towards higher $\text{M}_\star$ and SFR. The median difference is $\sim 0.5 \left< s_\text{VIPERS} \right>$, and $\sim 0.4 \left< s_\text{VIPERS} \right>$, with and without accounting biases, respectively. As it can be seen from Fig.~\ref{fig:ms_diff}, biases decrease the difference mainly at high $\text{M}_\star$ and SFR.

\section{Discussion}\label{sect:discussion}

\subsection{Biases analysis}

We analyzed the effects of four different biases introduced by observations (range in $\text{B}-\text{B}^*$ and fraction of blue galaxies) and data selection (S/N ratio and quality flags) on the SDSS sample to understand how they could affect the comparison between different samples. We find that the biggest bias is the data selection on quality flags of the spectra. The main result of this analysis is that the metallicity versus the sSFR  relation is the least sensitive to the biases analyzed here (completely independent on the selection on the range in $\text{B}-\text{B}^*$ and fraction of blue galaxies; dependent on the selection criteria on oxygen lines.) This adds value to the non-parametric framework described by \cite{salim2014critical, salim2015mass}, which allows for a generalization of the study neglecting the effects of these biases.

The biases occurring due to data selection (S/N selection and quality flag on the spectra of galaxy samples at different redshifts) can mimic an evolution mainly of the MZR and in the plane metallicity versus SFR. Restrictive sample cleaning, which requires galaxies with high S/N line detection, can lead to a non-physical MZR. It can result in a non-monotonic relation with a fall at high $\text{M}_\star$ or a complete cancellation of the anti-correlation  between metallicity and SFR. This nonphysical behavior is stronger when the cutoff is ``safer'' on oxygen lines, especially if applied to $\left[ \text{O{\,\sc{iii}}} \right]\lambda 4959$. This line is weaker than $\left[ \text{O{\,\sc{iii}}} \right]\lambda 5007$ and not always well measured. For this reason, it  will be particularly sensitive to S/N level leading to the selection of the most SF galaxies, in the case of VIPERS. This results in a distortion of the FMR projections, especially in the case of the MZR and metallicity versus the SFR plane.

\subsection{Comparison with the literature}

As we have shown above, a direct comparison between samples is not straightforward, but, nonetheless, this kind of comparison is often done in the literature. Therefore, it is interesting to perform a comparison taking into account all the limitations on the conclusions. For example, the VIPERS sample shows a flattening at $\log \text{M}_\star \left[ \text{M}_\sun \right] < 10$ in the MZR, which is not reproducible by any of the biases studied and a comparison with the literature can give some ideas on the reason for this behavior.

\begin{figure}
    \centering
    \resizebox{\hsize}{!}{\includegraphics{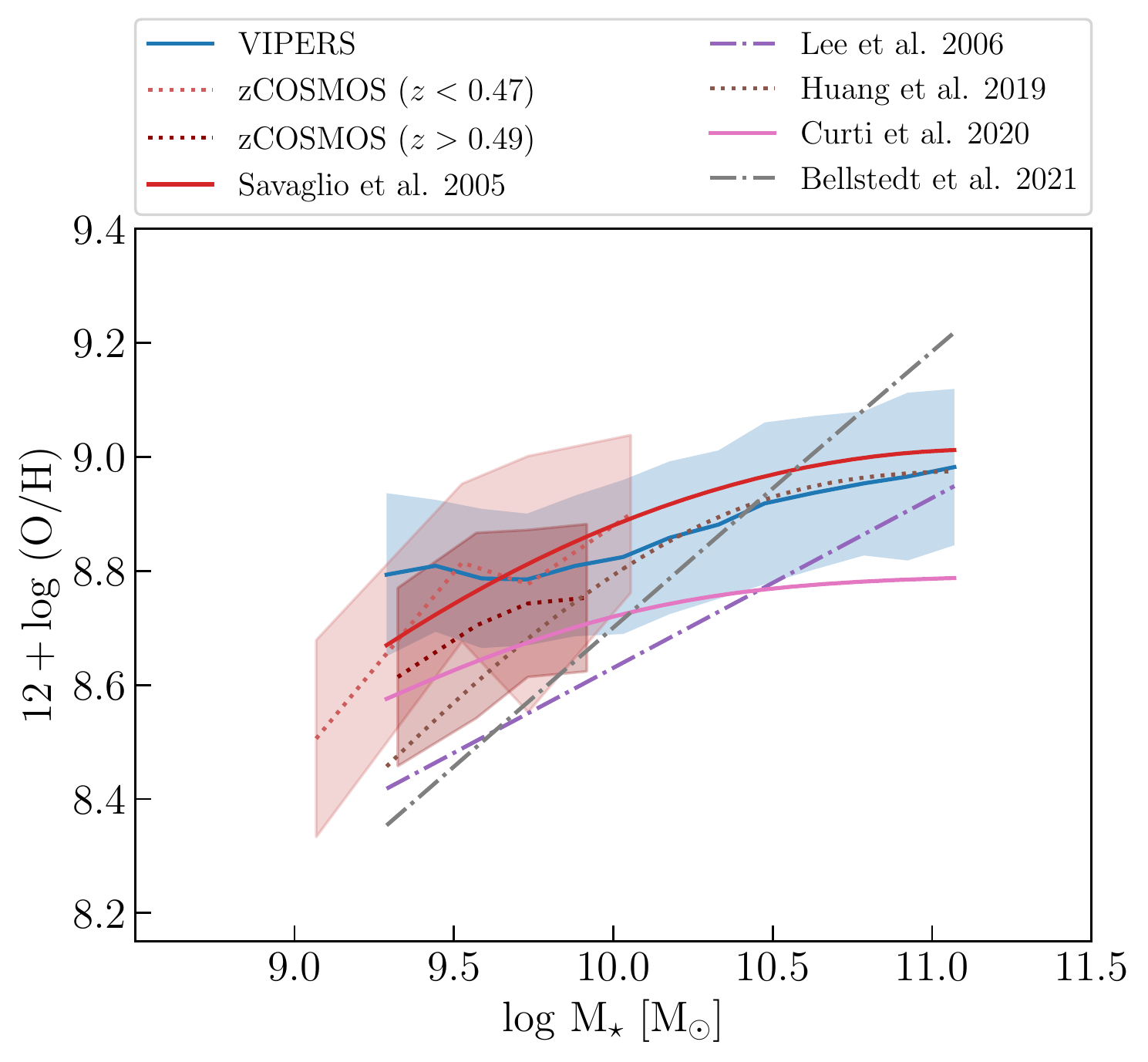}}
    \caption{Comparison between MZR of the VIPERS sample (blue solid line) and the relations from the literature (red solid line by \cite{savaglio2005gemini}; purple dash-dotted line by \cite{lee2006dwarf}; and brown dotted line by \cite{huang2019mass}; pink solid line by \cite{bellstedt2021gama}). In the same plot, we report the zCOSMOS data (red dotted lines) for both redshift bin from \cite{cresci2012zcosmos}.}
    \label{fig:mzr_lit}
\end{figure}

\cite{savaglio2005gemini} studied the MZR for galaxies at $0.4 < z < 1.0$ with data from the Gemini Deep Deep Survey (GDDS) and Canada-France Redshift Survey (CFRS) with a total of $\sim 60$ galaxies, finding a shift for this relation. They found a linear MZR described by
\begin{equation}
    12 + \log \left( \text{O}/\text{H} \right) = 0.478 \log \text{M}_\star + 4.062.
\end{equation}
These authors also compared their data with the ``converted'' (to take into account different IMF used) relation found by T04 
\begin{equation}
    12 + \log \left( \text{O}/\text{H} \right) =-2.4412 + 2.1026 \log \text{M}_\star  - 0.09649 \log^2 \text{M}_\star
.\end{equation}
They considered a different IMF as described by \cite{baldry2003imf};  but as shown in Fig. \ref{fig:mzr_lit}, after we convert the function to \cite{chabrier2003galactic} IMF, the VIPERS sample is in agreement with the quadratic function reported by \cite{savaglio2005gemini}.

\cite{lamareille2007vimos, lamareille2009vimos} reported the MZR for the VIMOS VLT Deep Survey (VVDS). They found a relation in agreement with the local MZR for galaxies at $z < 0.5$, while at higher redshift, the median relation starts to flatten, which is in disagreement with the SDSS data. This flattening and evolution compared to the local relation agrees with an open-closed model:
i) galaxies with small $\text{M}_\star$es evolve like open-boxes --- the metals produced by stellar evolution are ejected in the intergalactic medium by stellar winds and supernovae explosions;
ii) galaxies with high $\text{M}_\star$ evolve like closed-boxes --- the metals are kept in the galaxy thanks to a high gravitational potential.

\begin{figure}
    \centering
    \resizebox{\hsize}{!}{\includegraphics{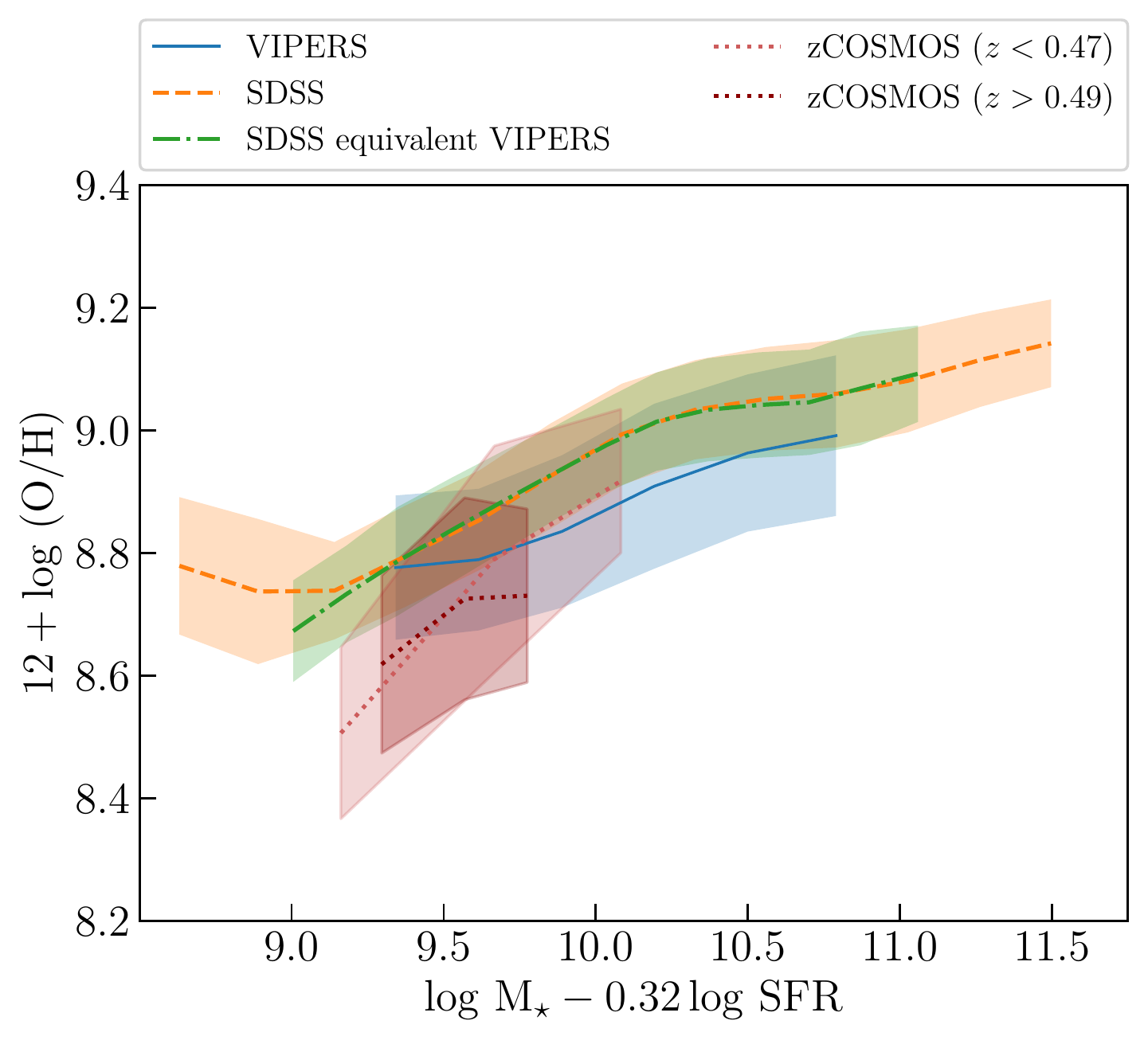}}
    \caption{Comparison between the projection on $\log \text{M}_\star - 0.32 \log \text{SFR}$ for SDSS (orange dashed line), VIPERS (blue solid line), SDSS equivalent to VIPERS (green dash-dotted line), and zCOSMOS (red dotted lines).}
    \label{fig:fmrproj}
\end{figure}

\cite{cresci2012zcosmos} analyzed the FMR and the MZR for galaxies in the zCOSMOS sample in the range $0.2 < z < 0.8$ divided in $169$ galaxies at $z < 0.47$ and $165$ at $z > 0.49$. They found an agreement with the relations in the local Universe with the SDSS data, but they used a different method to estimate the metallicity using also other lines as compared to \cite{mannucci2010fundamental}. Both studies used the calibration described by \cite{nagao2006gas}: for the local galaxies where more lines are available, the metallicity was estimated from the average between values obtained using $\left[ \text{N{\,\sc{ii}}} \right]\lambda 6584/\text{H}\alpha$ and $\text{R}_{23}$; the metallicity of zCOSMOS galaxies was instead estimated using the ratio $\left[ \text{N{\,\sc{ii}}} \right]\lambda 6584/\text{H}\alpha$ for $z<0.47$ and the ratio $\text{R}_{23}$ for $z> 0.49$. As shown in Fig. \ref{fig:mzr_lit} and \ref{fig:fmrproj}, the VIPERS and zCOSMOS data are in great agreement with each other.


\cite{huang2019mass} examined the composite spectra of galaxies from SDSS IV/eBOSS with a median redshift of $0.83$. They found a redshift evolution of the MZR described by the relation:
\begin{equation}
    12 + \log \left( \text{O}/\text{H} \right) = Z_0 + \log \left[ 1 - \exp \left( - \left[ \frac{\text{M}_\star}{\text{M}_0} \right]^\gamma \right) \right]
,\end{equation}
where $Z_0 = 8.977$, $\log \text{M}_0 = 9.961,$ and $\gamma=0.661$ for the redshift range $0.60$--$1.05$. The FMR approximately follow the local surface once the sample inhomogeneity and incompleteness are considered.

\cite{curti2020mass} used the calibration described in \cite{curti2016new} to describe the MZR via the equation:
\begin{equation}
    12 + \log \left( \text{O}/\text{H} \right) = Z_0 - \frac{\gamma}{\beta} \log \left( 1 + \left( \frac{\text{M}}{\text{M}_0} \right)^{- \beta} \right)
,\end{equation}
where $Z_0 = 8.793$ is the asymptotic metallicity at high $\text{M}_\star$, $\log \left(\text{M}_0/\text{M}_\sun \right) = 10.02$ is the turnover mass below which the MZR reduces to a power law of index $\gamma = 0.28$, and $\beta = 1.2$ is a measure of how fast the relation reach the asymptotic value.

\cite{bellstedt2021gama} studied the metallicity histories of $\sim 4500$ galaxies from the Galaxy And Mass Assembly (GAMA) survey with the SED fitting code ProSpect considering an evolving metallicity. With these metallicity histories, they can infer the MZR at different epochs. These authors described the evolution of the metallicity in function of $\text{M}_\star$ and look-back time ($t_\text{lb}$) as:
\begin{equation}
    \log \left( Z_\text{gas} \right) \left( \text{M}_\star,  t_\text{lb}\right) = \sum_{i=0}^3 f_i \left( t_\text{lb}\right) m^i
,\end{equation}
where
\begin{align}
    Z_\text{gas} &= Z_\sun \times 10^{\left[ 12 + \log \left( \text{O}/\text{H} \right) \right] - \left[ 12 + \log \left( \text{O}/\text{H} \right) \right]_\sun} ,\\
    Z_\sun &= 0.0142\footnotemark ,\\
    \left[ 12 + \log \left( \text{O}/\text{H} \right) \right]_\sun &= 8.69 ,\\
    m &= \log \left(\text{M}_\star \left[ \text{M}_\sun \right] \right) - 10 ,\\
    f_i \left( t_\text{lb}\right) &= \sum_{j=0}^5 a_{i,j} t_\text{lb}^j.
\end{align}
\footnotetext{This value differs from the commonly used value of $0.02$.}
In this way, they can reproduce the MZR from $z \sim 3.5$ to $z \sim 0$.

Figure~\ref{fig:mzr_lit} shows the MZRs of the VIPERS sample and other relations from the literature described above. At high $\text{M}_\star$, VIPERS data are in agreement with both \cite{lee2006dwarf}, \cite{huang2019mass}, and \cite{bellstedt2021gama} relations. The VIPERS sample follows the relation described by \cite{curti2020mass} quite well. From this comparison with the literature, we can conclude that an overestimation of the metallicity or an overselection of metal-rich galaxies at lower $\text{M}_\star$es seems to be present. In this plot are reported also the zCOSMOS data by \cite{cresci2012zcosmos}. These data are divided into four bins (in $\text{M}_\star$ or $\log \text{M}_\star - 0.32 \log \text{SFR}$) with the same number of galaxies per bin (to take into account the much lower statistics,$\sim 100$ galaxies per redshift bin, compared with SDSS or VIPERS sample) and the errorbars are estimated via the $16$th and $84$th percentile in each bin. The zCOSMOS data are in good agreement with the VIPERS sample.


Figure~\ref{fig:fmrproj} shows the projection on ${\log \text{M}_\star - 0.32 \log \text{SFR}}$. On this projection, the biases do not modify the relation for SDSS data. On this projection, the VIPERS sample is in agreement with the local data and it is in good agreement with the zCOSMOS data at a relative similar redshift.

\section{Conclusions}\label{sect:concl}

Since many studies of FMR are based on comparisons of differently selected samples, we have analyzed the effects of different biases that can be introduced during the sample selection on the relations between metallicity, $\text{M}_\star$, and SFR to assess how comparable in reality so constructed samples are. We studied biases introduced by physical constraints (evolution of the luminosity function and differences in the fraction of blue galaxies) or data selection (S/N and quality of the spectra).

The study of FMR projections is not the same comparison as for the full FMR. For example, the evolution of the MZR does not affect the FMR \citep{mannucci2010fundamental}. For this reason, the study of a more direct comparison from non-parametric analysis (being, indeed, a more reliable study of FMR) will be the subject of a subsequent separate paper.

The main conclusions reached in this paper can be summarised as follows.
\begin{itemize}
    \item The VIPERS sample is in good agreement with the SDSS sample with ``standard'' data selection, with an average metallicity difference of $\sim 0.6 \left< s_\text{VIPERS} \right>$. The biases taken into account can reduce the metallicity difference between these samples to $\sim 0.4 \left< s_\text{VIPERS} \right>$.
    \item Data selection based on S/N cutoff and flag quality of the lines affects the MZR and the metallicity versus the SFR plane. It leads to nonphysical relations (fall of the MZR at large $\text{M}_\star$ and hiding of the anti-correlation between the galactic properties in the plane metallicity versus the SFR), which can be misunderstood as evidence of evolution. These kinds of selections can introduce biases if applied, for instance, to the oxygen lines --- especially if applied to $\left[ \text{O{\,\sc{iii}}} \right]\lambda 4959$.
    \item The plane metallicity versus $\log \text{M}_\star - 0.32 \log \text{SFR}$ reduces the metallicity difference between the two samples. In this plane and in the MZR, the VIPERS sample is in good agreement with the zCOSMOS data \citep{cresci2012zcosmos}.
    \item VIPERS sample is in agreement with the relation found by \cite{savaglio2005gemini} in the whole range of $\text{M}_\star$ ($9.25 \leq \log \text{M}_\star \left[ \text{M}_\sun \right] \leq 11.0$) and with the relations found by \cite{lee2006dwarf}, \cite{huang2019mass}, and \cite{bellstedt2021gama} at high $\text{M}_\star$ ($\log \text{M}_\star \left[ \text{M}_\sun \right] > 10.0$). This comparison suggests that an over-selection of metal-rich galaxies or an overestimation of the metallicity at low $\text{M}_\star$ is still present in the VIPERS sample.
    \item The main bias is the selection of the flags of spectra quality which is not easily simulated by the selection of the S/N ratios of the emission lines. It shows that metal-poor SF galaxies have spectra with intrinsically better quality.
    \item S/N cutoffs affect the MZR and metallicity versus the SFR selectively cut the high metallicity at higher $\text{M}_\star$ and lower SFR flattening the curves. In the plane metallicity versus the sSFR, this cut has negligible effects within uncertainties.
    \item All the projections of the FMR are insensitive to the fraction of blue galaxies selection.
    \item The plane metallicity versus $\log \text{M}_\star - 0.32 \log \text{SFR}$ and metallicity versus sSFR are the least sensitive to observational biases among the 2D relations.
    \item When analyzing metallicity versus $\text{M}_\star$ or SFR, we have to be careful when carrying out the sample selection as this may introduce biases.
\end{itemize}

As demonstrated, a sample-selection-based comparison can be complicated to do even if often used in the literature \citep[e.g.][]{savaglio2005gemini, calabro2017vuds, huang2019mass}. In addition, the FMR projections do not fully describe the FMR itself. A more direct comparison of the FMR at different redshifts can be provided by a non-parametric framework \citep[e.g.][]{salim2014critical, salim2015mass} which will be the subject of our next paper.

\begin{acknowledgements}

This research was supported by the Polish National Science Centre grant UMO-2018/30/M/ST9/00757. KM is grateful for support from the Polish National Science Centre via grant UMO-2018/30/E/ST9/00082. MF acknowledges support from the First TEAM grant of the Foundation for Polish Science No. POIR.04.04.00-00-5D21/18-00 (PI: A. Karska). This paper uses data from the VIMOS Public Extragalactic Redshift Survey (VIPERS). VIPERS has been performed using the ESO Very Large Telescope, under the ``Large Programme'' 182.A-0886. The participating institutions and funding agencies are listed at \url{http://vipers.inaf.it}. We thank B. Garilli for the line measurements for the VIPERS sample. We thank F. Mannucci for the useful discussion. We thank G. Cresci for providing us with the zCOSMOS data. This research made use of Astropy,\footnote{\url{http://www.astropy.org}} a community-developed core Python package for Astronomy \citep{astropy2013, astropy2018}.
\end{acknowledgements}

\bibliographystyle{aa}
\bibliography{Bibliography3}

\begin{appendix}



\section{Conversion polynomials between different metallicity estimators}\label{app:conversion}

Following \cite{kewley2008metallicity}, we calculated new conversion polynomials from T04 to P01 and Z94 using both samples. These polynomials take the following form (see Fig.~\ref{fig:cal}):
\begin{equation}
\begin{split}
    \left[ 12 + \log \left( \text{O/H} \right) \right]_\text{P01} &=  0.94 \times \left[ 12 + \log \left( \text{O/H} \right) \right]_\text{T04} ^ 2 +\\
    &- 15.69 \times \left[ 12 + \log \left( \text{O/H} \right) \right]_\text{T04} + 74.07
\end{split}
,\end{equation}
and
\begin{equation}
\begin{split}
    \left[ 12 + \log \left( \text{O/H} \right) \right]_\text{Z94} &= 
    -0.33 \times \left[ 12 + \log \left( \text{O/H} \right) \right]_\text{T04} ^ 2 +\\
    &+6.92 \times \left[ 12 + \log \left( \text{O/H} \right) \right]_\text{T04} -26.56
\end{split}
.\end{equation}
The difference between our conversion polynomials and the ones described by \cite{kewley2008metallicity} is ascribable to the different sample selection and the different number of sources. They used a S/N of $8$ also for the oxygen and $\text{H}\beta$ lines and the total number of SF galaxies is $27\,730$ ($\sim 25\%$ of our sample). These S/N cutoffs result in a cut on metallicities. It is necessary to choose which conversion polynomials to use accordingly given the data selection performed.

\section{Check on the bi-modal redshift distribution of VIPERS samples}\label{app:red}

\begin{figure}[h!]
    \centering
     \resizebox{\hsize}{!}{\includegraphics{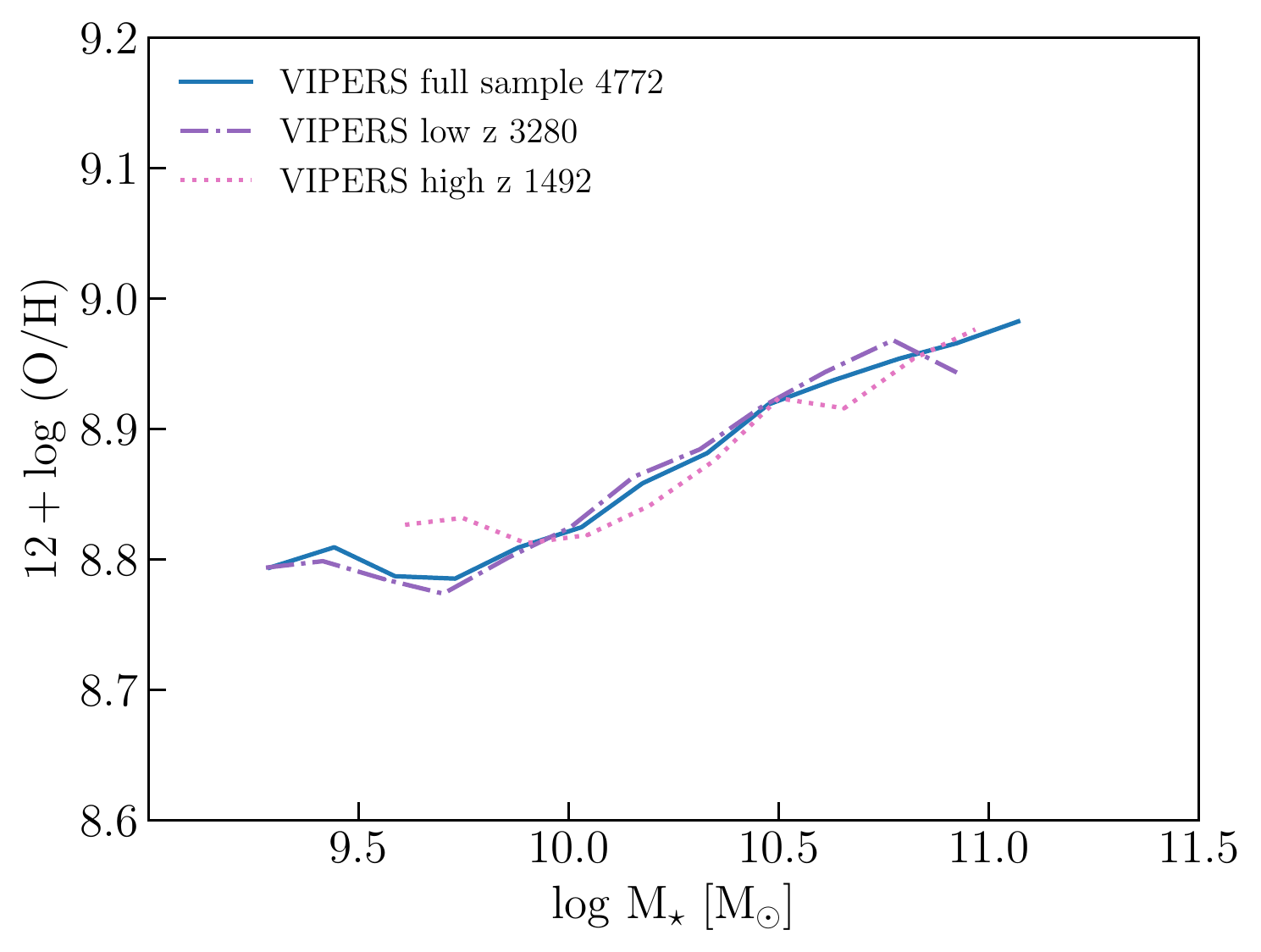}\includegraphics{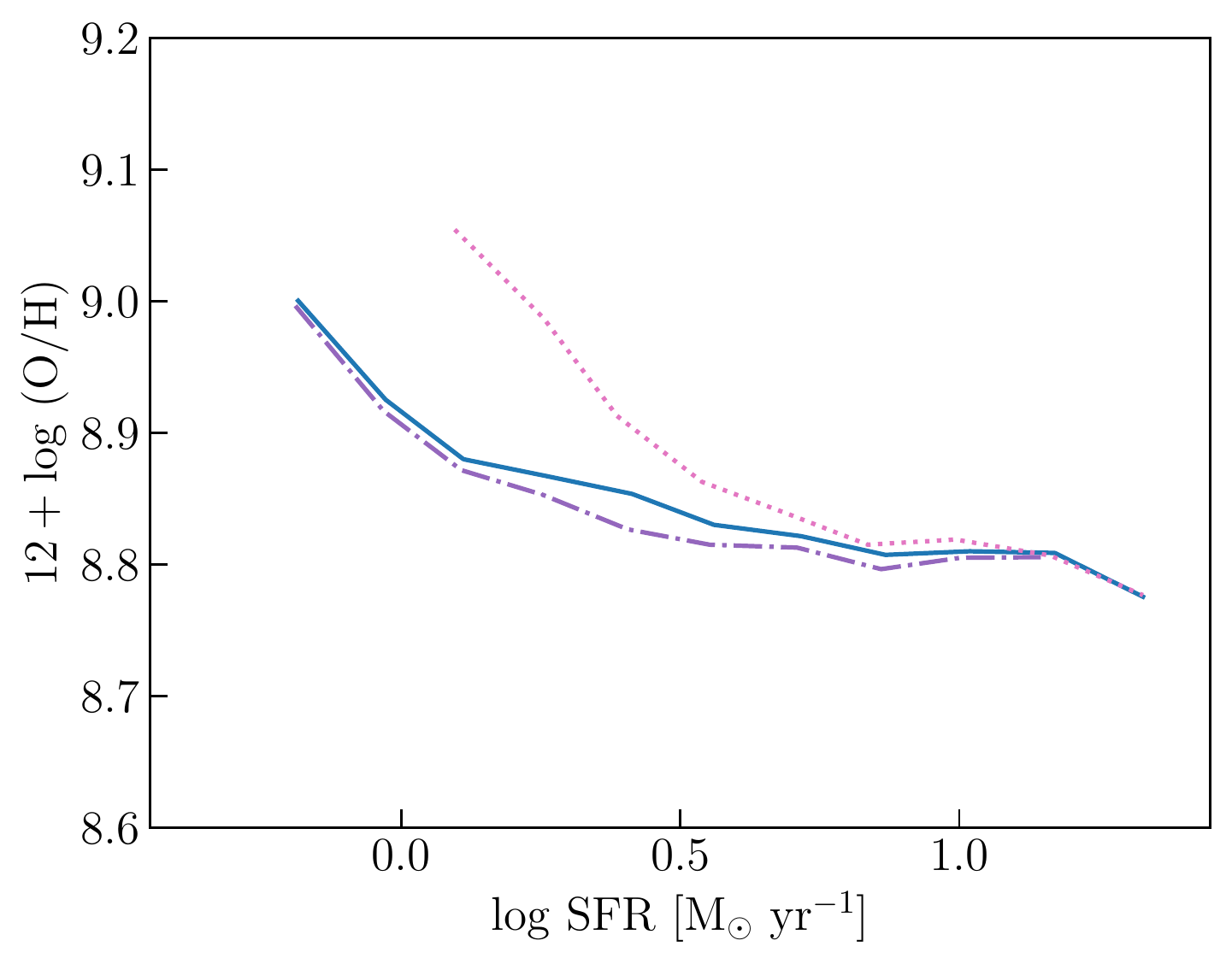}}
     \resizebox{\hsize}{!}{\includegraphics{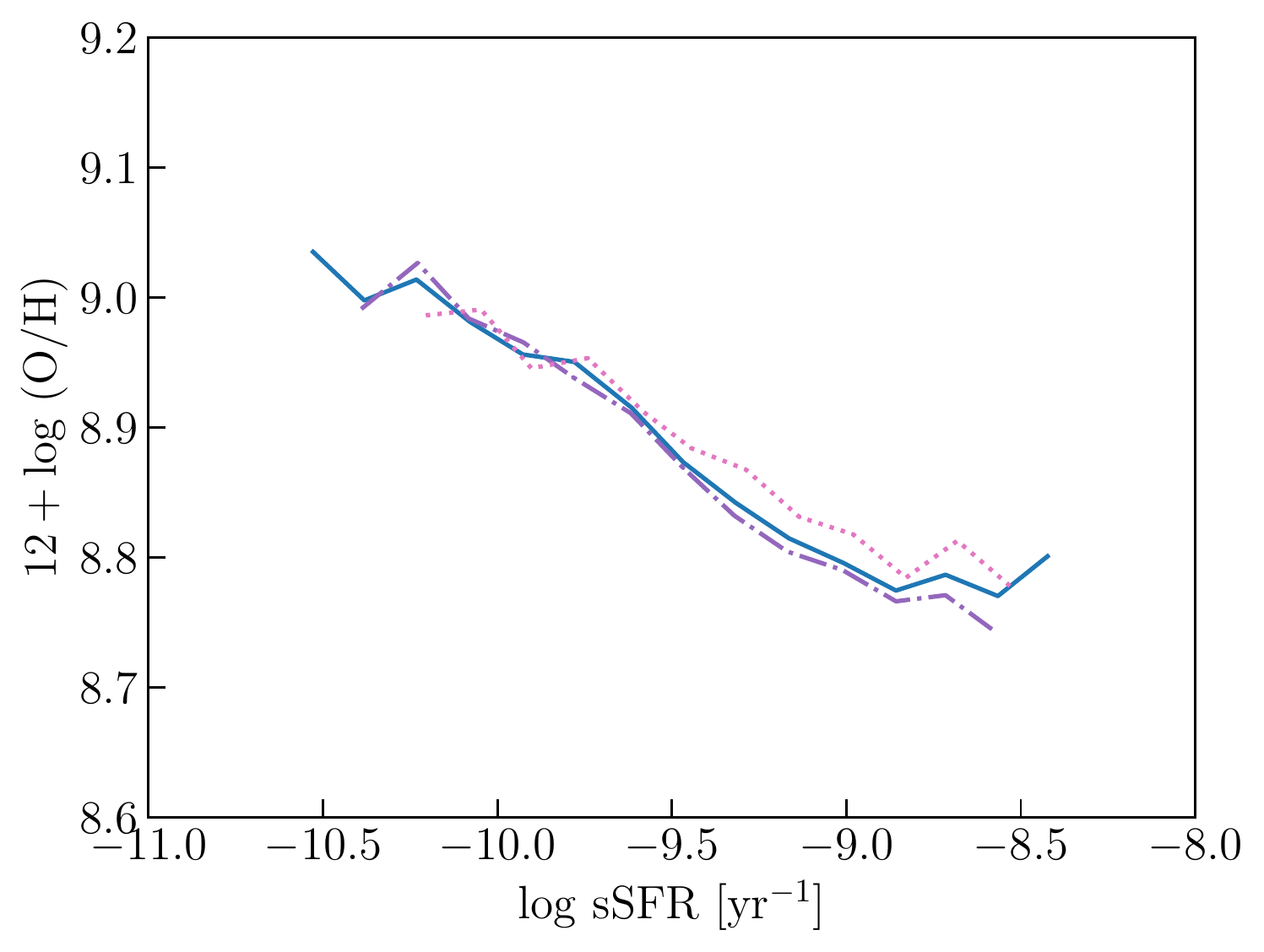}\includegraphics{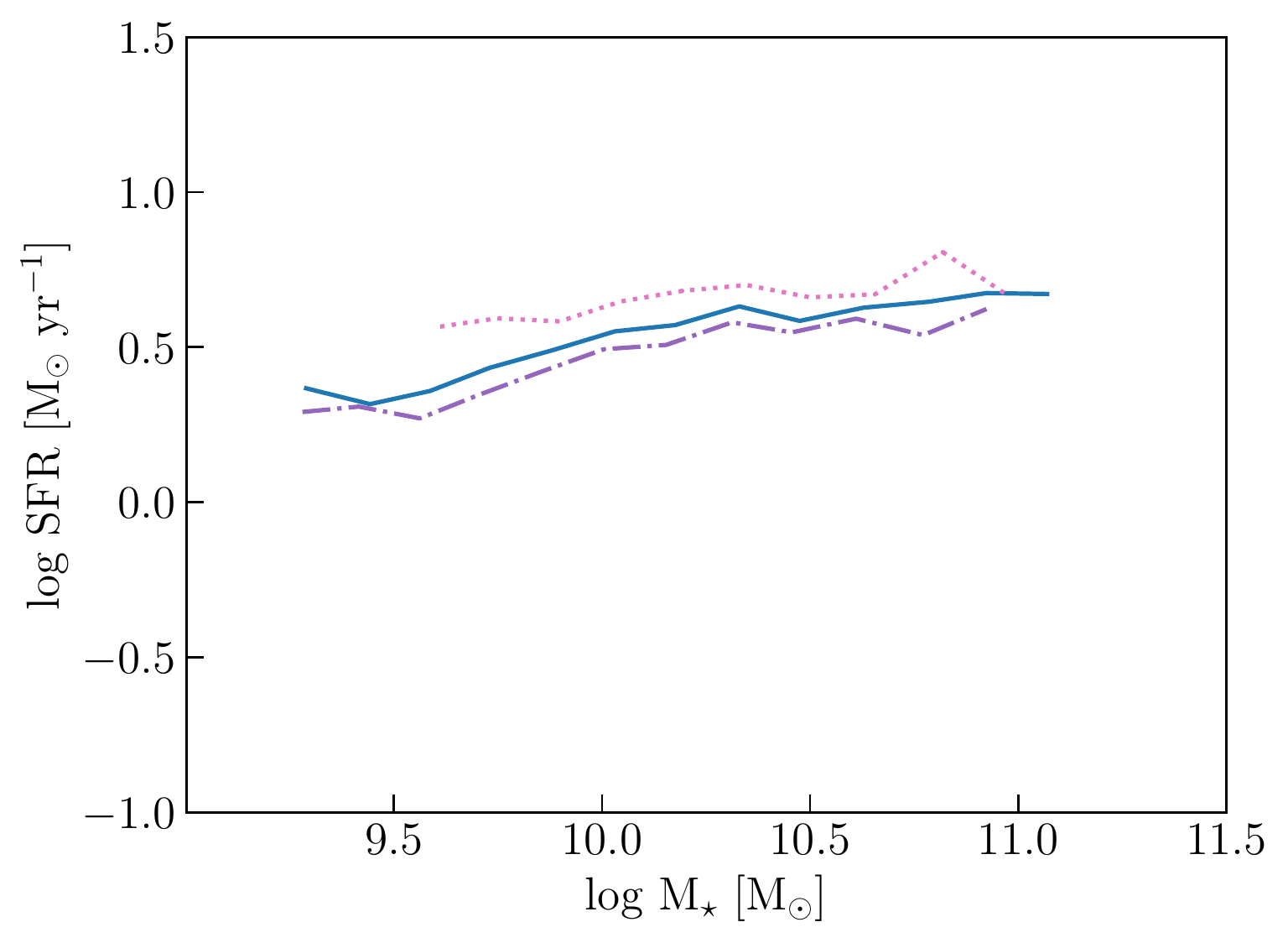}}
    \caption{Four projection of the FMR: (in order from the top) MZR, metallicity vs SFR, metallicity vs sSFR, and main sequence for VIPERS (blue solid line), VIPERS low-$z$ (purple dash-dotted line), and VIPERS high-$z$ (pink dotted line).}
    \label{fig:vipers_sepred}
\end{figure}

The VIPERS sample shows two separated populations in its redshift distribution with a gap at $z \sim 0.7$ which is a ``natural'' point to divide the sample into two sub-samples. Figure~\ref{fig:vipers_sepred} shows the projections of the FMR. The projections where the biggest difference is found between the sub-samples and the full sample is the metallicity versus SFR plane; here, the sub-sample at $z \geq 0.7$ shows higher metallicities at lower SFR. 
\FloatBarrier

\section{S/N distributions}\label{app:dist}
\begin{figure}[h!]
    \centering
    \resizebox{\hsize}{!}{\includegraphics{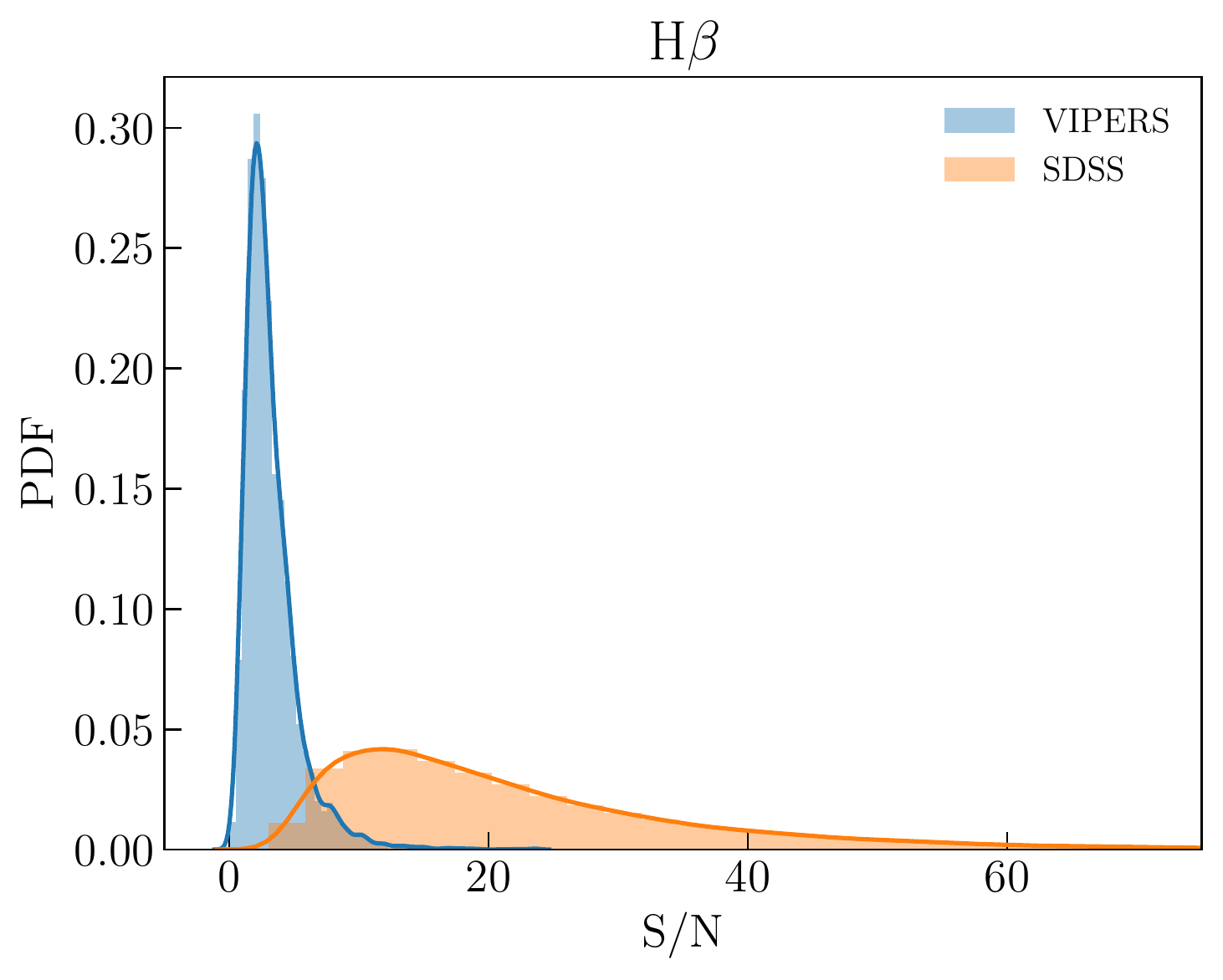}\includegraphics{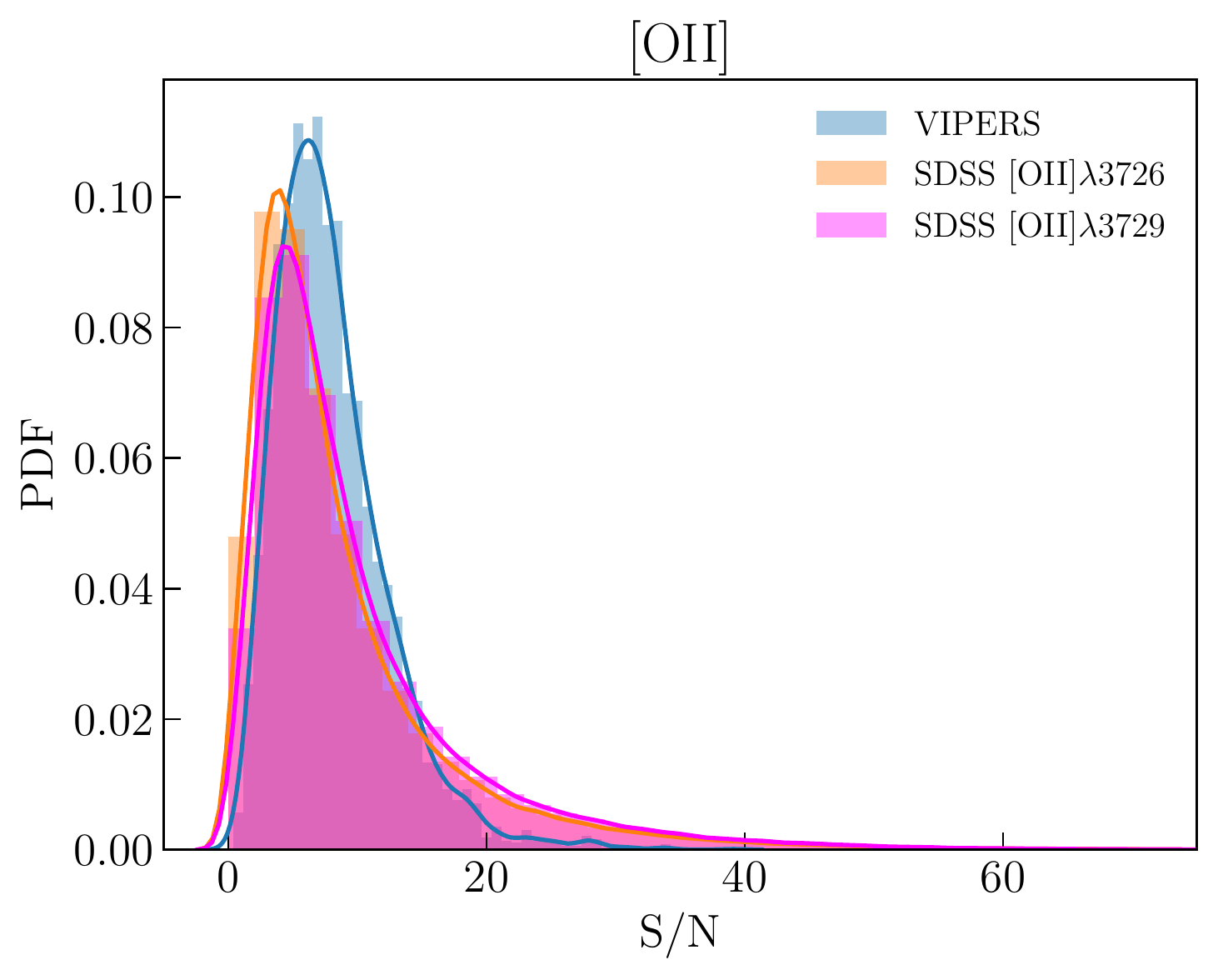}}
    \resizebox{\hsize}{!}{\includegraphics{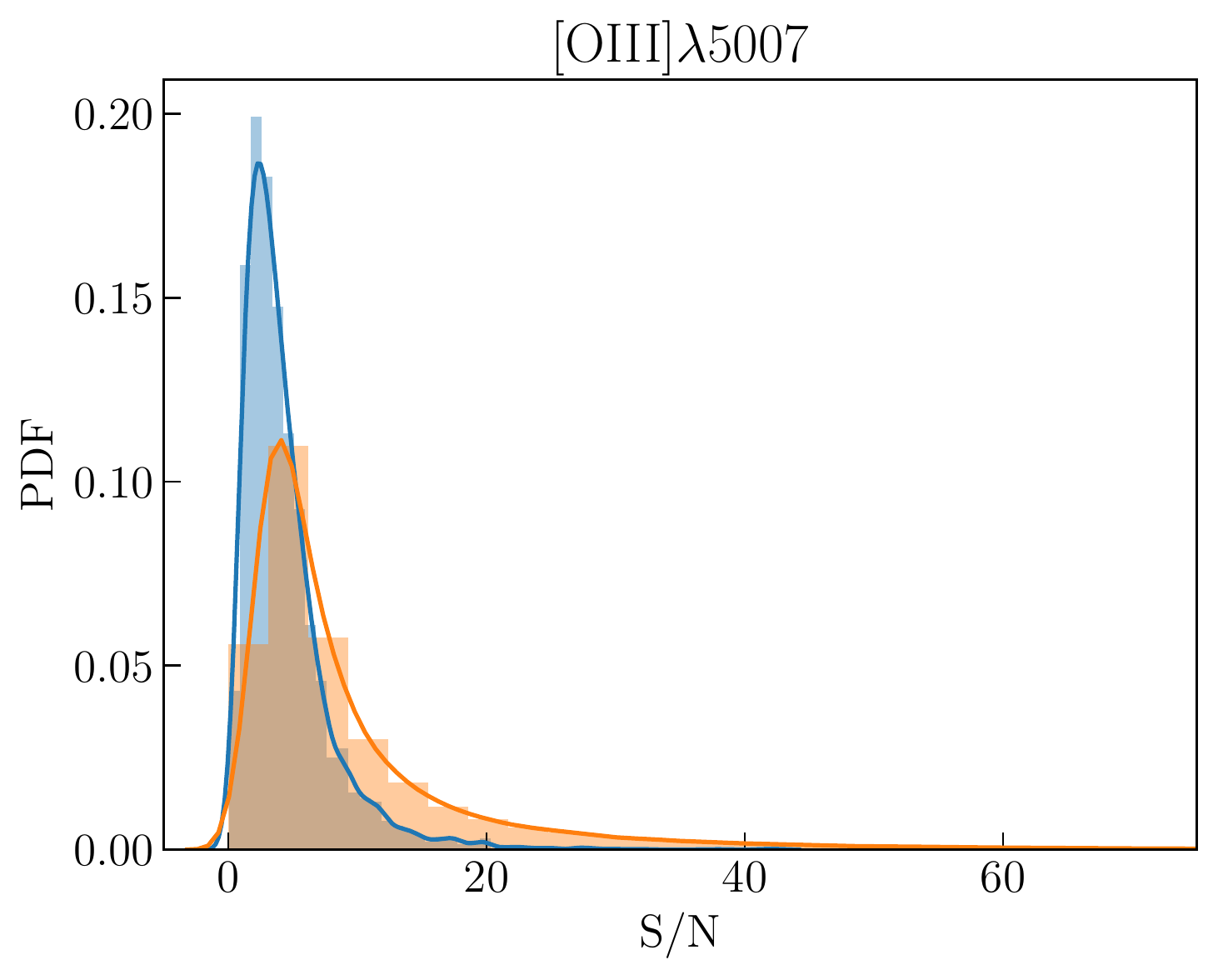}\includegraphics{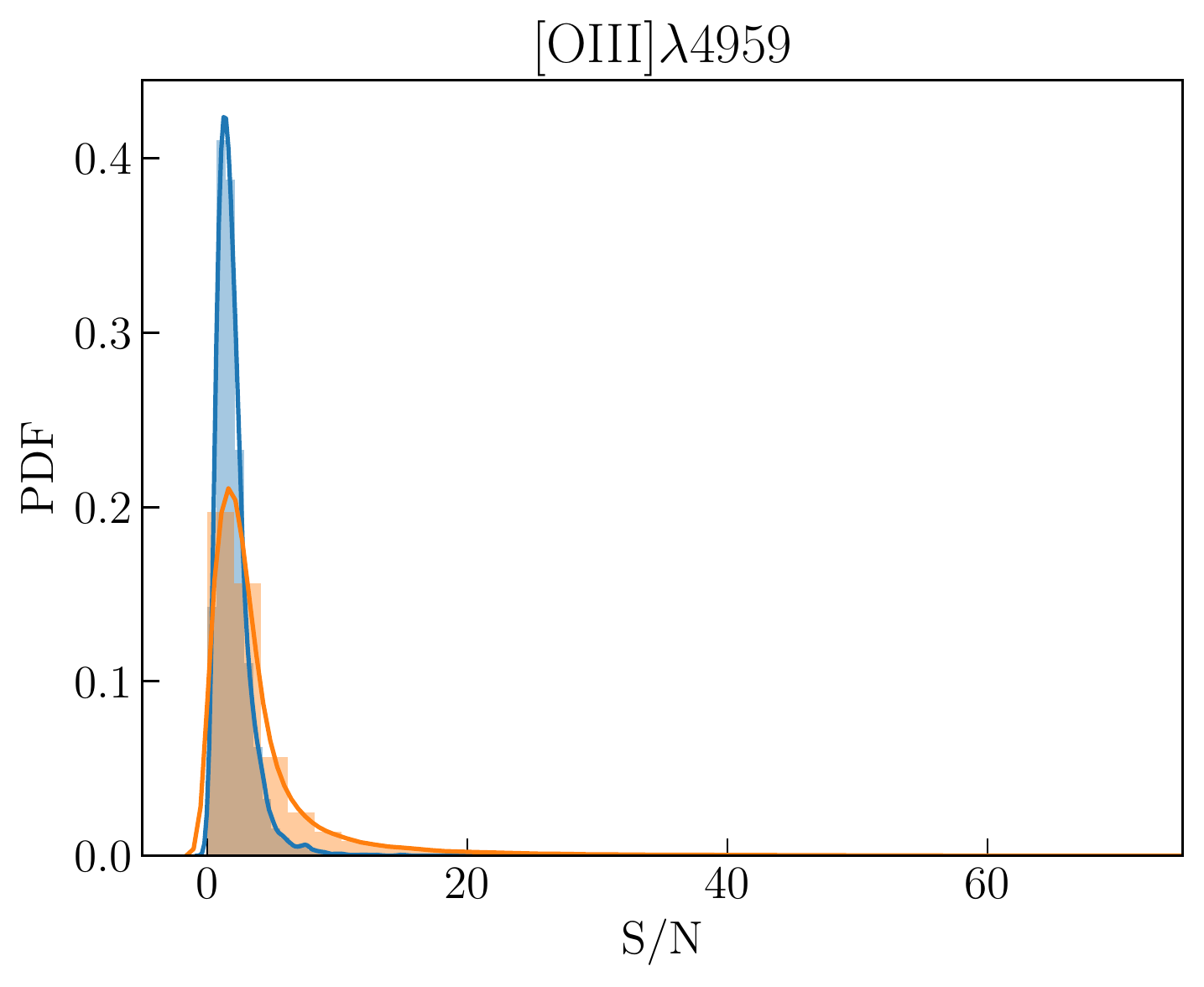}}
    \caption{Distribution in S/N for VIPERS (blue) and SDSS (orange) samples for (in order from the top) $\text{H}\beta$, $\left[ \text{O{\,\sc{ii}}} \right]\lambda 3727$, $\left[ \text{O{\,\sc{iii}}} \right]\lambda 5007$, and $\left[ \text{O{\,\sc{iii}}} \right]\lambda 4959$.}
    \label{fig:dists/n_bias}
\end{figure}

To check the intrinsic difference in the S/N of spectra observed by the two surveys, Fig.~\ref{fig:dists/n_bias} shows the global distributions in S/N for the emission lines ($\text{H}\beta$, $\left[ \text{O{\,\sc{ii}}} \right]\lambda 3727$, $\left[ \text{O{\,\sc{iii}}} \right]\lambda 4959$, and $\left[ \text{O{\,\sc{iii}}} \right]\lambda 5007$) for both samples.
\FloatBarrier

\section{Fraction of blue galaxies}\label{app:fraction}

Since blue galaxies are easier to observe at higher redshift, that is, the i-band VIPERS selection converts to a B-band selection at high-z, and for galaxies with bright emission lines is easier to estimate their redshift. This can introduce a bias in the study of the FMR.

\begin{figure}[h!]
    \centering
    \resizebox{\hsize}{!}{\includegraphics{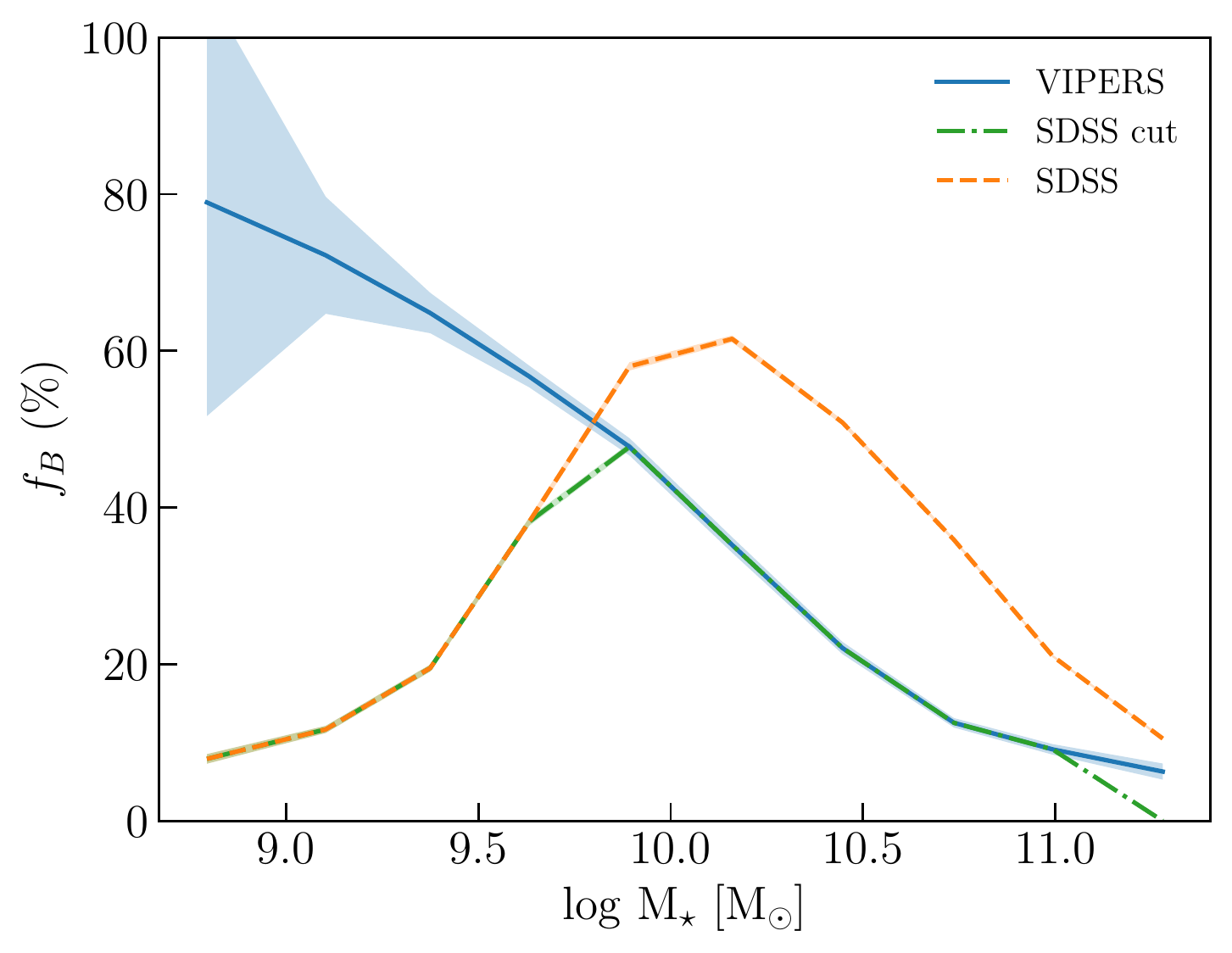}\includegraphics{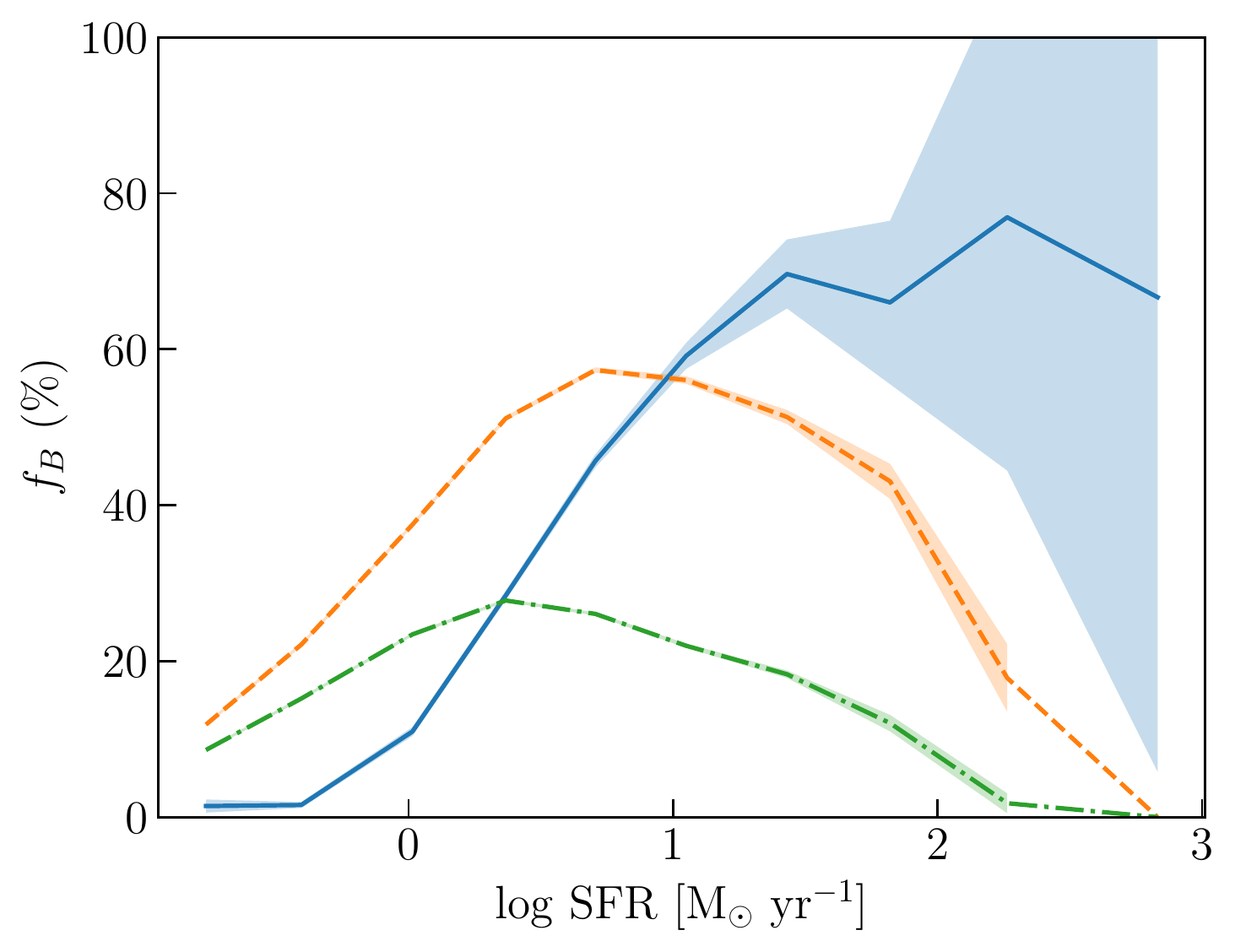}}
    \caption{Comparison of the fraction of blue galaxies of VIPERS (blue solid line) and SDSS (orange dashed line) samples in function of the $\text{M}_\star$ (left) and SFR (right). In green dash-dotted line is reported the fraction of blue galaxies of the SDSS sample after cut it to have the same fraction in function of the $\text{M}_\star$ than VIPERS sample.}
    \label{fig:comp_cuts}
\end{figure}

Figure~\ref{fig:comp_cuts} shows the fraction of blue galaxies in function of $\text{M}_\star$ and SFR. For VIPERS the fraction decreases with $\text{M}_\star$ while there is a maximum around $\log \text{M}_\star \left[ \text{M}_\sun \right] \sim 10.25$ for the SDSS sample. With the respect to the SFR, the two samples also showed different behavior: the fraction increases with SFR for VIPERS while it shows a maximum around $\log \text{SFR}  \left[ \text{M}_\sun \text{ yr}^{-1} \right] = 0.5$ for SDSS. We see that it is not possible to have the same fraction of blue galaxies for both samples simultaneously on $\text{M}_\star$ and SFR.
\FloatBarrier

\end{appendix}

\end{document}